# Improving the quasi-biennial oscillation
# via a surrogate-accelerated multi-objective optimization

**Luis Damiano[1], Walter Hannah[2], Chih-Chieh Chen[3], James J. Benedict[4], Khachik Sargsyan[5], Bert J. Debusschere[5], Michael S. Eldred[1]**

[1]Optimization & Uncertainty Quantification, Sandia National Laboratories, NM, USA
[2]Atmospheric, Earth & Energy Science, Lawrence Livermore National Laboratory, CA, USA
[3]Climate and Global Dynamics Laboratory, National Center for Atmospheric Research, CO, USA
[4]Fluid Dynamics and Solid Mechanics Group, Los Alamos National Laboratory, NM, USA
[5]Plasma & Reacting Flow Science, Sandia National Laboratories, CA, USA

**Key Points:**

- We developed an end-to-end workflow that calibrates gravity wave generation in E3SMv3, improving QBO realism.
- The fundamental frequency model compressed wind field data into physically interpretable quantities, isolated the QBO signal, and reduced dimensionality while retaining key QBO variability.
- Our surrogate-based multi-objective optimization quantified trade-offs in E3SM that limit QBO realism.

Corresponding author: Luis Damiano, `ladamia@sandia.gov`

**Abstract**

Accurate simulation of the quasi-biennial oscillation (QBO) remains a formidable challenge in climate modeling due in part to uncertainties in representing convectively generated gravity waves. We develop an end-to-end uncertainty quantification workflow that calibrates these gravity wave processes in the Energy Exascale Earth System Model to yield a more realistic QBO representation. Central to our approach is a domain knowledge-informed, compressed representation of high-dimensional spatio-temporal wind fields. By employing a parsimonious statistical model that learns the fundamental frequency of the underlying stochastic process from complex observations, we extract a concise set of interpretable and physically meaningful quantities of interest capturing key attributes, such as oscillation amplitude and period. Building on this, we train a probabilistic surrogate model that approximates the fundamental characteristics of the QBO as functions of critical physics parameters governing gravity wave generation. Leveraging the Karhunen–Loève decomposition, our surrogate efficiently represents these characteristics as a set of orthogonal features, thereby capturing the cross-correlations among multiple physics quantities evaluated at different stratospheric pressure levels, and enabling rapid surrogate-based inference at a fraction of the computational cost of inference reliant only on full-scale simulations. Finally, we analyze the inverse problem using a multi-objective approach. Our study reveals a tension between amplitude and period that constrains the QBO representation, precluding a single optimal solution that simultaneously satisfies both objectives. To navigate this challenge, we quantify the bi-criteria trade-off and generate a representative set of Pareto optimal physics parameter values that balance the conflicting objectives. This integrated workflow not only improves the fidelity of QBO simulations but also advances toward a practical framework for tuning modes of variability and quasi-periodic phenomena, offering a versatile template for uncertainty quantification in complex geophysical models.

**Plain Language Summary**

Simulating the quasi-biennial oscillation (QBO), a regular pattern of alternating winds high in the atmosphere, remains a major challenge for climate models. We developed an end-to-end workflow to calibrate gravity wave processes in the Energy Exascale Earth System Model, leading to more realistic simulations. We began by compressing complex spatio-temporal data into a few key, physically meaningful quantities, such as the oscillation's amplitude and period. This data reduction allowed us to isolate the QBO signal from noise and other atmospheric phenomena. Next, we built a fast statistical model that predicts QBO behavior based on critical physics parameters. This surrogate efficiently captures relationships among various atmospheric features, reducing the need for computationally expensive full-scale simulations. Our analysis revealed a trade-off between QBO amplitude and period, meaning that improving one aspect often worsened the other. Rather than finding a single perfect solution, we identified a range of balanced settings that offer the best compromise. This integrated approach not only leads to more realistic QBO simulation but also provides a practical framework for tuning other complex atmospheric phenomena.

# 1 Introduction

The process of tuning Earth system model components entails adjusting uncertain parameters within imperfect parameterizations designed to represent unresolved processes, such as cloud representations in conventional global atmospheric models used for future projections. The primary objective is to achieve a realistic time-mean state while maintaining balanced net top-of-atmosphere energy fluxes, ensuring model stability over 100 to 1000-year timescales in the absence of external forcing. In addition, tuning can affect the model's representation of variability across timescales ranging from hours to decades. However, efforts to tune modes of variability may inadvertently degrade the time-mean state, thereby relegating the faithful representation of variability to a secondary priority.

The tuning process is often labor-intensive and relies on subjective judgment in order to (a) anticipate how tuning might impact the time mean and/or variability and (b) assess whether a given modification yields adequate improvements. Although this appears to be a natural candidate for automated calibration techniques, the extensive range of attributes requiring optimization can be daunting. Recent studies have successfully employed automated calibration to optimize the time-mean state (Yarger et al., 2024); however, these techniques have been less frequently applied to tuning variability. While time-mean fields can be readily compared to satellite observations, performance metrics for modes of variability are more complex than time-mean assessments and must be tailored to each phenomenon. Furthermore, many modes of variability are episodic, allowing a model to simulate a phenomenon "correctly" even when the timing of events does not coincide with observations. Developing an automated calibration workflow to target episodic phenomena is challenging; however, overcoming this challenge could pave the way for broader adoption of auto-calibration in the Earth science modeling community.

In this study, we develop an end-to-end uncertainty quantification (UQ) workflow to calibrate the representation of convectively generated gravity waves in the U.S. Department of Energy's Energy Exascale Earth System Model (E3SM, Golaz et al., 2022) using surrogate-accelerated multi-objective optimization. We focus our efforts on convectively generated gravity waves for two reasons. First, these waves represent a large portion of the forcing for the quasi-biennial oscillation (QBO), a dominant mode of stratospheric variability that has global impacts, yet is poorly represented in many Earth system models. Second, both atmospheric moist convection and the gravity waves it generates are usually unresolved on the physical model grid and therefore must be parameterized. Because the characteristics of gravity wave generation are not well observed, their representation is calibrated through adjustment of several physical parameters in an attempt to obtain a more realistic QBO.

The end goal is to systematically adjust the parameters to minimize discrepancies, ensuring that its outputs align as closely as possible with observational or experimental data (Santner et al., 2018). This process improves the model's predictive capabilities and reduces epistemic uncertainty (Oberkampf & Roy, 2010). The fundamental statistical groundwork was established in the original works of Jones et al. (1998); Kennedy and O'Hagan (2001); Higdon et al. (2008). A thorough background on model calibration for climate models is provided by Hourdin et al. (2017). Applied scientists will find Mai (2023)'s ten strategies towards successful calibration of environmental models particularly useful. More generally, we direct the reader to R. C. Smith (2024) for a comprehensive exploration of UQ, including model calibration, in complex systems across various scientific and engineering disciplines.

Running E3SM to generate a single simulation with sufficient temporal duration and an acceptable level of spatial resolution can require the allocation of substantial computational resources on exascale computing platforms, rendering direct optimization approaches impractical. To mitigate the computational cost, we build a statistical surrogate based on Gaussian processes (GPs) to obtain a probabilistic approximation of the quantities of interest (QoIs) that, once trained, can be evaluated at a small fraction of the cost of a single simulation. GPs, as a computationally efficient alternative, have proven successful in multiple climate applications, including ice sheet modeling (Berdahl et al., 2021), greenhouse gas emission modeling (Beusch et al., 2022), and the calibration of atmospheric convection parameters in an idealized general circulation model (Dunbar et al., 2021). We direct the reader to Rasmussen and Williams (2005) for the most comprehensive self-contained treatment of GPs and to Gramacy (2020); Santner et al. (2018) for a modern overview of GPs in the context of design and analysis of computer experiments. For other constructions, Chowdhary et al. (2022) offers an extensive review of machine learning methods combined with projection-based model reduction techniques for data-driven surrogates, particularly for spatial or spatiotemporal field data.

As part of this work, we include a thorough discussion of the construction of the QoIs, along with an extensive analysis of the estimated quantities and sensitivity analysis. Al-though not strictly required for the calibration, sensitivity analysis yields two valuable complementary results: gaining insight into the effects of the parameterized convective gravity wave scheme on the QBO, and providing an opportunity for subject matter experts (SMEs) to assess the reasonableness of the surrogate predictions through the study of feature importance and global trends in the predictions. This enables a more transparent calibration approach, providing insight into how and why the climate model is tuned, mitigating the risk of misattribution of skillful predictions to data accommodation and vice versa (Schmidt et al., 2017). Among numerous techniques better suited for different model characteristics (Gan et al., 2014; Cheng et al., 2020), Sobol' indices for variance-based decomposition (Saltelli et al., 2006; Sobol , 2001) are particularly effective for gaining insight into the factors influencing the computer model. They can be estimated using random sampling approaches (Jansen, 1999; Saltelli et al., 2010) as well as efficient closed-form methods (Sudret, 2008; Crestaux et al., 2009; Sargsyan, 2015).

Next, we conduct a multi-objective optimization (MOO) to simultaneously optimize multiple conflicting objectives (Sharma & Kumar, 2022; Gunantara, 2018). In particular, we leverage our cost-efficient surrogate to find the Pareto frontier (Kang et al., 2024) and quantify an efficient trade-off between the QBO period and amplitude. MOO has gained significant traction in the calibration of climate models (Langenbrunner & Neelin, 2017) as well as adjacent fields of study such as hydrology (Efstratiadis & Koutsoyiannis, 2010), wind energy (Liu, Li, et al., 2020), geology (Gong et al., 2016), and environmental sciences (Sun et al., 2021). These advances reflect a growing recognition of the complexity of climate systems and the necessity of addressing diverse objectives in the calibration process.

The QBO is a regular variation of the winds that occurs in the tropical stratosphere (Baldwin et al., 2001). An alternating wind regime in the longitudinal (hereafter, "zonal") direction is primarily observed in the equatorial stratosphere, between approximately 10 km and 50 km altitude. The oscillation manifests as downward-propagating wind regimes, with westerly (west-to-east) and easterly (east-to-west) winds alternating in a quasi-periodic manner approximately every 28 to 30 months. The wind speeds can reach up to 40 m/s (Naujokat, 1986), and the regimes propagate from the upper stratosphere to the lower stratosphere at a rate of about 1 km per month (Baldwin & Dunkerton, 1998). To assess the quality of our results, we compare the physically interpretable QoIs, estimated from our domain knowledge-informed compressed representation of the high-dimensional spatio-temporal dense wind field, against these key attributes documented in the climate literature.

The QBO is a critical component of the Earth's atmospheric system, with far-reaching impacts on climate and weather. Although it manifests in the equatorial band, the QBO affects the global weather and climate via teleconnections (Anstey et al., 2021), defined as spatially remote responses to a local perturbation. It influences the frequency and intensity of tropical cyclones; for example, the QBO easterly phase (when lower stratospheric winds flow from east to west) is associated with reduced vertical wind shear, which can enhance cyclone development (Gray, 1984). The QBO affects the stratospheric circulation, which can in turn influence tropospheric weather patterns, including the jet streams and storm tracks (Baldwin & Dunkerton, 2001). QBO-related wind and temperature anomalies also modulate the distribution of ozone in the stratosphere, with implications for ultraviolet radiation reaching the Earth's surface (Hasebe, 1994). Additionally, the QBO impacts monsoon systems, particularly the Asian and African monsoons, by altering the large-scale atmospheric circulation (C. Li & Yanai, 1996). The QBO is also known to influence the boreal winter extratropical stratosphere (Naoe & Yoshida, 2019). Understanding and accurately simulating the oscillation remains a challenging but essential task for improving climate predictions and understanding atmospheric dynamics. Enhancing the representation of the QBO in E3SM is a crucial first step toward better representing teleconnected phenomena.

Mechanistically, the QBO is driven by the interaction of atmospheric waves with the mean flow in the stratosphere (Booker & Bretherton, 1967; Holton & Lindzen, 1972; Lindzen & Holton, 1968; Alexander & Holton, 1997). The primary waves involved are Kelvin waves, which are eastward-propagating equatorial waves that contribute to the westerly phase of the QBO; equatorial Rossby-gravity waves, which are westward-propagating waves that contribute to the easterly phase of the QBO; and gravity waves, generated by convection and other processes in the troposphere, which also play a crucial role in driving both phases of the QBO. In this article, we focus exclusively on the calibration of the parameterized convectively generated gravity waves via the deep convection scheme (G. Zhang & McFarlane, 1995). These waves propagate from the troposphere upward into the stratosphere. As the waves interact with the mean flow in the stratosphere, they can break and deposit momentum, which alters the wind patterns. Kelvin waves deposit eastward momentum, while equatorial Rossby-gravity waves deposit westward momentum. The deposition of momentum by these waves leads to the alternating easterly and westerly wind regimes. As one phase descends, it is eventually replaced by the opposite phase, completing the oscillation cycle (Wallace & Kousky, 1968; Plumb, 1977). These complex interactions result in a distinctive correlation structure over space and time that our workflow exploits to reduce the dimensionality of the data while guaranteeing spatial coherence across the atmosphere.

The QBO is well-known for being challenging to simulate (Anstey et al., 2020). In general, a realistic period can often be well represented by tuning the parameterized non-orographic gravity wave drag, but the QBO signal remains unrealistically weak in the lowermost tropical stratosphere, with a 50% underprediction in some key pressure levels (Anstey et al., 2022; Bushell et al., 2020). E3SM (Golaz et al., 2019, 2022) has also struggled to reproduce the observed QBO characteristics in version 1 (Y. Li et al., 2023; Richter et al., 2019) and version 2 (Golaz et al., 2022; Y. Li et al., 2025). The QBO involves small-scale wave processes that require high vertical and horizontal resolution (Giorgetta et al., 2002) and exhibit natural variability in its period and amplitude, which can be difficult to reproduce (Anstey & Shepherd, 2013). These challenges have been partially addressed by either recalibrating existing state-of-the-art parameterizations or developing yet more expressive parameterizations. Previous attempts at calibrating the representation of the QBO in E3SM versions 1 and 2 are documented in Richter et al. (2019) and Golaz et al. (2022), respectively. These attempts have been iterative, running the computer model using hand-picked values, which can result in a tedious, time-consuming, and subjective process. State-of-the-art automatic calibration workflows for E3SM, such as Yarger et al. (2024), do not include the QBO in their scope. To the best of our knowledge, our end-to-end UQ workflow is the first attempt at calibrating the QBO in E3SM by fully leveraging formal calibration tools. We successfully demonstrate the added value that statistical calibration has in improving the computer model. While our workflow currently employs a human in the loop, it builds the necessary components required to construct an automated pipeline.

The article is structured as follows. In Section 2, we introduce the observational and simulated data, discuss the structure of the wind fields, and present an exploratory analysis. In Section 3, we describe our end-to-end UQ workflow, including the fundamental frequency model that was developed to learn the QoIs from wind data (Section 3.1), a surrogate for highly correlated QoIs (Section 3.2), and a MOO strategy to efficiently search for physics parameter values (Section 3.3). In Section 4, we thoroughly examine the results from applying our workflow to the aforementioned data set, including a model-based characterization of the QBO (Section 4.1), an exploratory analysis of the simulations produced by the computer model (Section 4.2), a discussion on surrogate model selection and validation (Section 4.3), a global sensitivity analysis (Section 4.4), and a detailed account of the MOO results and limitations (Section 4.5). Finally, in Section 5, we summarize the most salient findings and allude to open questions for future research. Supporting Information S1 contains complementary details that are referenced throughout this manuscript.

## 2 Data

The purpose of our investigation is to calibrate E3SM to simulate wind fields showing a set of target characteristics learned from observational data. In this section, we describe the observations playing the role of reference data and an E3SM-generated simulations ensemble.

### 2.1 Reference data

Among the multiple atmospheric global reanalysis data sets (Wu et al., 2024), a particularly important observational data set is the ERA5 reanalysis (Hersbach et al., 2017). ERA5 is the fifth generation ECMWF atmospheric reanalysis of the global climate covering the period from 1940 to present, with a ~30 km horizontal grid resolution and 137 atmospheric levels from the surface up to a height of 80 km. Compared to its predecessors, such as ERA-Interim (Dee et al., 2011), ERA5 has significantly higher space-time resolution, a more sophisticated representation of physical processes in the atmosphere (including convection), and an improved representation of the stratosphere to better capture variability and extremes. Regarding the QBO, the main advancements include upgrades to the original parametrization of convection (Tiedtke, 1989) to improve the representation of mixed-phase clouds (Ahlgrimm & Forbes, 2014), tropical variability (Bechtold et al., 2008; Hirons et al., 2012), and the diurnal cycle of convection (Bechtold et al., 2014).

We subset the wind data from 1984 to 1993. This period excludes the anomalous behaviors associated with the more recent disruption events in 2015/2016 (Watanabe et al., 2018; Osprey et al., 2016) and 2019/2020 (Y. Wang et al., 2023). The comparatively short window is chosen to be consistent with the simulation lengths of the large E3SM ensemble to be described in Sec. 2.2, and to ensure that computational costs do not become prohibitively expensive. The data is structured as a 2D field with 6 pressure levels displayed on the vertical axis (7-70 hPa of atmospheric pressure, or approximately 18-33 km in elevation) and a complete sequence of 120 evenly spaced monthly time steps on the horizontal axis. Each horizontal slice corresponds to a time series for a fixed pressure level. The data set in Fig. 1 displays a characteristic alternating sequence of zonal wind. In 120 months of data, we observe a total of four approximately evenly spaced cycles comprised of two-thirds (right tail) of one cycle, three complete cycles, and one-third (left tail) of a fifth cycle. This indicates a reference period of approximately 30 months. The cycles are shifted over time as we descend through the stratosphere due to downward propagation, i.e., the movement of these alternating wind regimes from higher altitudes to lower altitudes in the stratosphere. Each phase of the oscillation starts at higher altitudes (around 30 km) and gradually descends to lower altitudes, creating alternating bands of easterly and westerly winds (Plumb, 1977).

### 2.2 Earth system model simulation ensemble

We generate simulations using a version forked from E3SMv2 (Golaz et al., 2022; E3SM Project, 2023) that implements the redesigned vertical grid introduced in Yu et al. (2024). The length and time span of the simulation are comparable to the reference data, despite an apparent phase shift attributed to minor differences in the atmospheric initial conditions. E3SM surface temperatures are prescribed from observations over regions of ocean and sea-ice during 1984-1993, while the land surface is fully prognostic. While surface temperatures are prescribed, the surface flux calculations still depend on the atmospheric state. Trace gases such as CO2 and CH4 are also prescribed consistent with observations.

E3SM simulates the signed wind speed over a 4D tensor indexed over latitude, longitude, pressure, and time. A single 10-year run typically requires 400 CPU core hours and writes to disk 17 billion wind daily averages over a 1°×1° global grid and 72 vertical levels. The vertical grid is a set of unevenly-distributed points ranging from 1000 to 0.1 hPa (~64 to 0.01 km) selected by the SMEs to satisfactorily resolve atmospheric behaviors and reduce model biases (Xie et al., 2018). Despite its massiveness, we exploit the nature of the QBO to reduce the data dimensionality. Because the QBO is in essence a tropical stratosphere phenomenon characterized by low-frequency dynamics (Baldwin et al., 2001), we constrain the geospatial and vertical domains as well as coarsen the time resolution. More specifically, we (i) only retain observations located in the 7-70 hPa pressure band and within the ±5° latitude band, (ii) spatially aggregate the data via averages over latitude and longitude, and (iii) temporally aggregate the data via monthly means. For rule (i), we build upon the empirical fact that the QBO is a lower stratosphere phenomenon and further trim the pressure band to exclude locations with overlapping atmospheric phenomena (e.g., semi-annual oscillation at altitudes above the 5 hPa pressure level (A. K. Smith et al., 2020)). All rules describing horizontal and temporal aggregation are consistent with the geotemporal aggregation criteria designed for the E3SM Diagnostics Package (C. Zhang et al., 2022), a comprehensive post-processing toolkit embedded in the E3SM process workflow. The resulting data set used in our study is reduced by these rules to 720 observations per simulation, structured as a 2D field over 120 months and 6 pressure levels (namely 7, 10, 20, 30, 50, and 70 hPa).

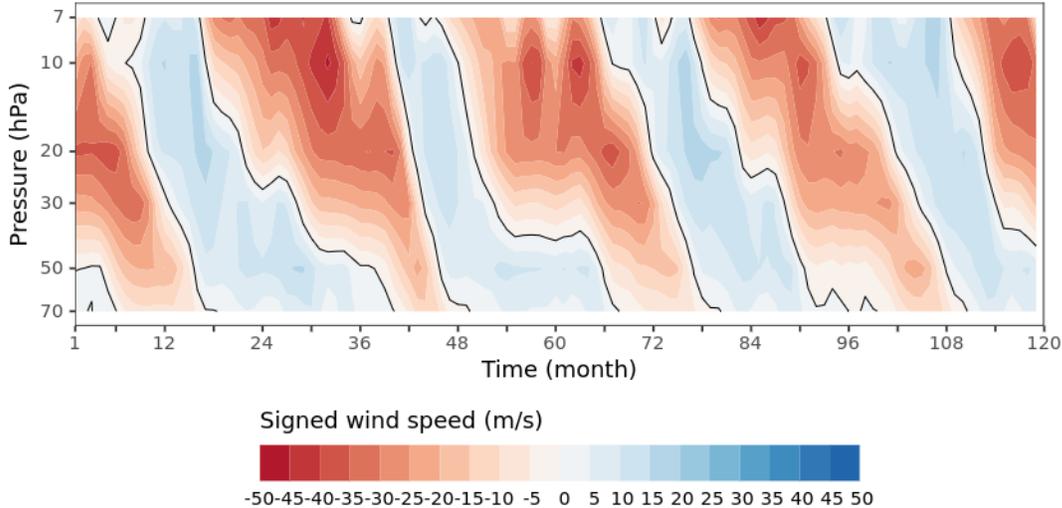

**Figure 1.** Cross-section of observed monthly mean and zonally averaged zonal winds in the stratosphere and the ±5° latitude band based on the ERA5 data set, spanning 1984-1993. Positive (blue) and negative (red) shading corresponds to westerlies and easterlies, respectively. The black boundaries delineate a change in wind direction (i.e., zero wind speed).

E3SM is controlled by numerous parameters, including both user-facing and internal variables. The key mechanistic QBO driver is the forcing generated by the breaking of vertically propagating and convectively generated gravity waves (Booker & Bretherton, 1967; Lindzen & Holton, 1968; Lindzen, 1987; Baldwin et al., 2001), and a correct QBO representation requires realistically represented large-scale (grid-resolved) atmospheric waves as well as small-scale (parameterized) gravity waves generated by convection (Richter et al., 2020). The oscillation generated in E3SMv2 is characterized by a shorter period and stronger amplitude compared with its predecessor due to updates to the deep convection parameterization, which makes convection more intense but less frequent while leaving the time-mean convective heating tendency almost unchanged (Y. Li et al., 2025). Accordingly, we have selected three physics parameters closely linked to the deep convection scheme and gravity wave generation whose ranges and default values are reported in Table 1.

First, we consider the efficiency (EF) of convection in generating gravity waves from the Beres scheme (Beres et al., 2004). When the Zhang-McFarlane deep convection scheme (G. Zhang & McFarlane, 1995) activates convection, the EF range from 0 to 1 indicates the proportion of time that gravity waves are generated. The higher the EF value, the higher the number of gravity waves generated over time. Second, we control the convective fraction (CF), which ranges between 0 and 1 and scales the grid-box-averaged heating rate resolved by E3SM. A higher value, indicative of more intense convection, results in a greater magnitude of localized heating. Treating the parameterization of convectively generated gravity waves as a black box, CF and EF are tightly coupled parameters scaling the input and output of the black box, respectively. The former scales the convective (condensational) heating that is provided to the scheme, while the latter scales the amount of gravity wave activity generated by the scheme. Third, we adjust the heating depth (HD) multiplier, which ranges from 0.25 to 1.50. HD is an empirical parameter determined heuristically rather than theoretically that scales the resolved heating depth, allowing for adjustments that increase or decrease the prescribed value. Both the vertical extent of positive convective heating values and the maximum amplitude of the heating vertical profile influence the spectrum of gravity waves that will be generated.

| Name | Lower | Upper | Default (v2) | Default (v1) | E3SM parameter name |
|------|-------|-------|--------------|--------------|---------------------|
| EF | 0.00 | 1.00 | 0.35 | 0.40 | `effgw_beres` |
| CF | 0.00 | 1.00 | 0.10 | 0.08 | `1 / gw_convect_hcf` |
| HD | 0.25 | 1.50 | 1.00 | 1.00 | `hdepth_scaling_factor` |

**Table 1.** Parameter ranges, default values, and corresponding definitions for E3SM version 2 (v2) and version 1 (v1). The parameters include EF (efficiency factor), CF (convective fraction), and HD (heating depth multiplier), which are integral to the convective gravity wave generation scheme.

## 3 Methods

We now present our end-to-end UQ workflow for improving the QBO, which consists of domain-knowledge informed dimension reduction, surrogate modeling, and MOO.

### 3.1 Fundamental frequency model

The QBO is succinctly defined as the leading mode of variability in the tropical stratosphere (Baldwin et al., 2001). It is not a directly observable quantity but serves as a framework for characterizing key patterns identified from observational data, which should be approximately replicated by the simulated wind field to be considered realistic. From a modeling perspective, we identify three key challenges to a successful formulation. First, wind fields, influenced by multiple chaotic and complex atmospheric phenomena, represent a rich but intricate data set. As not every identifiable source of variability is of interest to our calibration efforts, we require a clear definition of the relevant signal to mitigate overloading the calibration process with information that has no mechanistic link to the ability to accurately simulate the QBO and, eventually, its teleconnections. Second, calibrating the model to simulate a realistic QBO requires more than simply matching simulated fields to observations. The objective is to find optimal configurations that reproduce wind field patterns consistent with observations and grounded in physical principles, avoiding overly restrictive field-to-field comparisons that do not address the underlying scientific questions. As one simple example, an element-wise comparison between a wind field and a time-shifted copy would yield the same result, making frequency-domain calibration more suitable in this context. Third, since E3SM is a high-resolution model that runs on exascale computers and generates high-dimensional data, dimensionality reduction is crucial for streamlining the analysis and improving computational efficiency. In the remainder of this subsection, we introduce the fundamental frequency model (FFM) to translate established principles from atmospheric science into QoIs amenable to UQ, isolate the QBO signal from other oscillations and random error, and reduce data dimensionality.

### 3.1.1 Formulation

We formulate a set of equations characterizing the essential dynamics of the QBO coupled with a minimal set of constraints. At the most basic level, the FFM learns a single frequency that optimally explains the cyclical wind patterns while ensuring an empirically motivated phase coherence in the vertical dimension.

Let $y_{tk} \in \mathbb{R}$ be the signed wind speed in m/s at month $t$ and the $k$-th pressure level for $t = 1, \ldots, T \in \mathbb{N}$ and $k = 1, \ldots, K \in \mathbb{N}$. Positive and negative speed values correspond to eastward and westward winds, respectively. The wind field is driven by the following spatio-temporal set of equations,

$$y_{tk} = \beta_{0k} + \beta_{1k} \sin\left(2\pi t/\tau - \phi_k\right) + \varepsilon_{tk} \qquad (1)$$

$$\phi_k = \alpha_0 + \alpha_1 \log_{10}\left(\text{pressure}_k\right) \qquad (2)$$

where $\beta_{0k} \in \mathbb{R}$ is the QBO E-W (zonal wind) bias at the $k$-th pressure level, $\beta_{1k} \in \mathbb{R}^+$ is the QBO amplitude at the $k$-th pressure level, $\tau \in \mathbb{R}^+$ is the QBO period shared across all the pressure levels, $\phi_k \in [0, 2\pi]$ is the phase shift at the $k$-th pressure level, $\varepsilon_{tk} \sim \mathcal{N}(0, \sigma_k^2)$ is the error variance at the $k$-th pressure level, and $\alpha_0, \alpha_1 \in \mathbb{R}$ are the linear propagation coefficients. The E-W bias coefficients capture the mean wind speed over time, where positive values indicate the dominance of easterlies over westerlies. The amplitude coefficients measure the half-range of the signed wind speed or, more intuitively, half the distance between peaks and troughs. The period and the phase shift characterize the wave cycle length and starting point. Visually, $y_{tk}$ for a fixed $k$ behaves like a wave over time. The parameters $\beta_{0k}$ and $\phi_k$ are associated with vertical and horizontal shifts, respectively, while $\beta_{1k}$ and $\tau$ are associated with vertical and horizontal dilations.

In this parametrization, we allow the E-W biases $\beta_{0k}$ and amplitudes $\beta_{1k}$ to vary freely, the phase shift $\phi_k$ to propagate linearly, and the period $\tau$ is held constant over all pressure levels in the atmosphere. These three distinct levels of freedom are informed by previous empirical studies. The period $\tau$ is held constant as a function of pressure to enforce that the physical process remains coupled: even though it is possible to observe multiple waves with seemingly different periods in a finite sample, were the QBO period truly free over the atmosphere, the wind cycles at different pressure levels would eventually desynchronize. As for the phase shift $\phi_k$, a downward propagation from the top of the troposphere until the signal dissipates near the tropopause has been widely documented observationally (Baldwin et al., 2001). The waves are allowed to be coherently off-phase across the multiple pressure levels and, since they share the same period in our parametrization, the phase shift translates directly into a time shift equal to $\tau \times \phi_k/2\pi$ months. The phase offset $\phi_k$, however, is not erratic but follows a spatial progression. As a first-order approximation to the true relationship, in Eq. (2) we incorporate a propagation equation that is log-linear in pressure and, thus, inversely and approximately linear to the geopotential altitude in the stratosphere per the barometric formula pressure $\propto e^{-\text{altitude}}$ (United States Committee on Extension to the Standard Atmosphere, 1976). This simple parametric model mimics a descending wind regime that reaches the lower levels as a new regime begins to form in the upper levels. Finally, we allow the E-W wind bias and amplitude to vary freely throughout the atmosphere. Although there is no reason to expect that they form a rough function over pressure levels or exhibit discontinuities, the limited number of pressure levels and the strength of the signal in the data rarely necessitate regularization to prevent unphysical fits. Enforcing smoothness, for example by incorporating a penalization or regularization term, may be warranted in other applications.

A particularly appealing aspect of the FFM is that every QoI has a clear visual counterpart. Fig. 1 essentially resembles a noisy sequence of diagonal stripes in alternating colors:

(i) the linear propagation rate $\alpha_1$ captures the stripes' inclination or angle; (ii) the period $\tau$ captures the distance between two consecutive stripes of the same color; (iii) the amplitudes $\beta_{1k}$ determine the contrast between the darkest shades of red and blue (i.e., the local minima and maxima); and (iv) the E-W wind biases $\beta_{0k}$ capture the dominating color.

We present two additional minor considerations regarding Eq. (1). First, because the FFM is designed to capture the leading mode of variability in signed wind speed, we incorporate an additive error term to account for potential secondary sources of variability, such as the well-known semi-annual oscillation in the upper stratosphere (A. K. Smith et al., 2020) and wind variations associated with the El Niño-Southern Oscillation (Timmermann et al., 2018). While assuming an independent and identically distributed normal error simplifies parameter estimation considerably, time-aware alternatives, such as an autoregressive process, could yield more accurate estimates of the QoIs. Second, although periodic waves are customarily parameterized as functions of frequency in the engineering and digital signal processing communities, we express the equation explicitly as a function of period to facilitate the direct estimation of the QoI and its associated uncertainty. Parameterizations in terms of both frequencies and periods are equivalent.

### 3.1.2 Estimation

Fitting the FFM to a data set involves learning $2K + 3$ parameters for the mean and $K$ parameters for the variance from $T \times K$ data points. The log-likelihood is given by the expression

$$L(\boldsymbol{\theta}|\mathbf{Y}) = \sum_{k=1}^{K} \left[ -\frac{1}{2}(\mathbf{y}_k - \mathbf{m}_k)^\top \boldsymbol{S}_k^{-1}(\mathbf{y}_k - \mathbf{m}_k) - \frac{1}{2}\log|\boldsymbol{S}_k| - \frac{T}{2}\log 2\pi \right] \tag{3}$$

$$\mathbf{m}_k = \beta_{0k} + \beta_{1k}\sin\left(2\pi\mathbf{t}/\tau - \phi_k\right) \tag{4}$$

where $\mathbf{m}_k$ is a vector of length $T$ with the mean at time snapshots for the $k$-th pressure level, $\boldsymbol{S}_k = \sigma_k^2\mathbf{I}$ is a diagonal matrix with a time-homogeneous variance at pressure level $k$, and $\boldsymbol{\theta}$ denotes the parameter vector. The log-likelihood, which is readily found by noticing that Eq. (1) is a non-linear regression model with Gaussian additive error, reduces to a sum of $T \times K$ terms that can be computed efficiently.

There are several methods to estimate the parameters of the FFM and quantify their uncertainty. Maximum likelihood estimation (MLE) can be conducted by optimizing the log-likelihood in Eq. (3). Note that, conditional on $(\tau, \alpha_0, \alpha_1)$, the MLE and ordinary least-squares estimates for $(\boldsymbol{\beta}_0, \boldsymbol{\beta}_1, \boldsymbol{\sigma}^2)$ are equivalent. This effectively reduces the numerical optimization input dimension down to three free parameters facilitating a computationally efficient search even for large $K$. The parameter uncertainty matrix can be approximated by evaluating the Hessian at the MLE.

Alternatively, Bayesian estimation may be performed by updating Eq. (3) with possibly non-uniform priors. When fitting the model to the reference data, in particular, the following set of informative priors can be elicited from published empirical studies (Baldwin et al., 2001): $\tau \sim \mathcal{N}(28, 10^2)$ to center the mode around the widely accepted value of 28 months and place 80% of the density in the 20-40 month window, $\alpha_1 \sim \mathcal{U}(0, \infty)$ to force downward propagation, and $\beta_{1k} \sim \text{Gamma}(2, 0.1)$ to place the central 80% of the density for wind speed amplitude in 5-40 m/s. Numerical optimization over the posterior surface can be performed using standard Bayesian inference methods (Gelman et al., 2013), such as the maximum a posteriori probability (MAP) estimate via derivative-based maximization algorithms or full posterior distribution estimates via Markov chain Monte Carlo (Brooks et al., 2011). Since the MLE and MAP under uniform priors are numerically equivalent, a computationally efficient plug-in estimate may be constructed by optimizing numerically over $(\tau, \alpha_0, \alpha_1)$ and interpreting the point estimate under the Bayesian framework. The plug-in estimate is likely to result in the underestimation of the uncertainties for certain QoIs (White, 1982), which may not be critical if these uncertainties are not propagated through downstream analyses.

The FFM can also be framed as a regression problem. Nonlinear least squares can be applied to Eqs. (1) and (2) to estimate the set of waves that best fit the data without assuming normal errors (D. M. Bates & Watts, 1988; Nocedal & Wright, 2006). Given the varying scale of residuals across the atmosphere, it may be beneficial to minimize a weighted sum of squares with weights inversely proportional to the wind speed variance: $w_k^{-1} = (T-1)^{-1} \sum_t (y_{tk} - \bar{y}_k)^2$ for $\bar{y}_k = T^{-1} \sum_t y_{tk}$. Alternatively, cross-validation is often argued to be more robust against model misspecification (Wahba, 1990). For fixed parameters $(\tau, \alpha_0, \alpha_1)$, the leave-one-out predictive mean and variance can be computed analytically, reducing the computational burden of performing leave-one-out cross-validation,

$$\mathrm{E}\left\langle \hat{y}_{-tk} | \tau, \phi_k \right\rangle = y_{tk} - \frac{\left[\mathbf{S}_k^{-1}\mathbf{y}\right]_t}{\left[\mathbf{S}_k^{-1}\right]_t} \quad \text{and} \quad \mathrm{V}\left\langle \hat{y}_{-tk} | \tau, \phi_k \right\rangle = 1 / \left[\mathbf{S}_k^{-1}\right]_t \qquad (5)$$

where $\mathbf{S}_k = \mathbf{b}_k \cdot \mathbf{b}_k^\top$, $\mathbf{b}_k = \sin\left(2\pi\mathbf{t}/\tau - \phi_k\right)$ is the dot product matrix of the sine basis vector.

### 3.1.3 An analogy to dimension reduction techniques

Setting aside the physics motivating its formulation, the FFM can also be viewed as a tool for reducing the dimensionality of the wind field. A single model run typically produces $T \times K$ values arranged in a two-dimensional map over time and atmospheric pressure, whereas Eqs. (1) and (2) only require $2K + 3$ parameters, yielding a reduction factor of approximately $T/2$ except for very small $K$. The percentage of variance explained by the FFM can be estimated as

$$R^2 = 1 - \frac{\sum_k \left(\mathbf{y}_k^\top \mathbf{y}_k - \mathbf{y}_k^\top \mathbf{B}(\mathbf{B}^\top \mathbf{B})^{-1} \mathbf{B}^\top \mathbf{y}_k\right)}{\sum_k \mathbf{y}_k^\top \mathbf{y}_k}, \qquad (6)$$

where $\mathbf{B}$ is the sine basis matrix.

If dimension reduction were the sole intent, one might consider applying off-the-shelf data reduction techniques to the wind fields, such as the Karunen-Loève expansion (Ghanem & Spanos, 2003), functional principal component analysis (Shang, 2013), or t-distributed stochastic neighbor embedding (van der Maaten & Hinton, 2008). For example, using the first technique, (Mueller et al., 2025) recently extended conventional polynomial chaos surrogates to optimally represent random fields through a generative construction that captures the first- and second-order two-point statistics of a random process. However, the FFM has at least two added values: parsimony and scientific interpretability.

First, since the size of the low-dimensional representation is selected on conceptual grounds rather than based on goodness of fit or cross-validation, the FFM has a built-in Occam's razor mechanism that safeguards against overfitting. Although preserving a large percentage of the variance (e.g., 99%) is often preferred in many learning tasks, in line with the definition of the QBO, we focus on capturing only the primary mode of variability. This approach is analogous to tasks like sound processing, where the fundamental frequency corresponds to the first harmonic and captures the identity of the signal but not its full complexity, which arises from the combination of all the harmonics. In Section 4.2, we argue that the FFM captures a substantial portion of the wind field variability, though not all of it. However, this is intentional, as the goal is to avoid misattributing other atmospheric oscillations to the QBO.

Second, because the reduced quantities have an intrinsic physical meaning derived from a model, they can be subjected to evaluation by SMEs. Not only does this facilitate interdisciplinary collaboration, but it also integrates our results with the extensive theoretical and empirical knowledge in climate science. In contrast, low-dimensional variables such as spectral features or principal component scores lack tangible physical meaning, and any data-driven interpretation is fully dependent on the correlation structure realized in the training data, which may shift as new data are collected.

### 3.2 Probabilistic surrogate

We construct a statistical model to approximate the QoIs as they would appear if E3SM were run with an arbitrary set of physics parameter values. Our discussion focuses on the QBO period and amplitude, noting that incorporating additional quantities into the proposed workflow is straightforward. At the core of the surrogate model are multiple independent GP regressions, each representing an unknown function that maps the physics parameter space to a truncated set of noisy spectral features associated with the QoIs. From these regressions, the predictive distribution of the QoIs is analytically reconstructed from the predictive distribution of the spectral features.

#### 3.2.1 Formulation

Let $\mathbf{Q}$ be the $N \times J$ matrix containing the QoIs, where each row corresponds to a vector $\mathbf{q}_n$ for the $n$-th E3SM ensemble member simulation. Since the QoIs are derived from noisy data rather than directly observed, $\{\mathbf{q}_n\}_{n=1}^{N}$ represents a collection of random vectors, each with a mean vector and a covariance matrix as discussed in Section 3.1.2. For the purpose of this analysis, however, we consider the vectors fixed and known and set them equal to the estimated mean.

Although the FFM does not impose an explicit functional structure on the QoIs, these vector elements are often highly correlated as they describe a coherent set of physical characteristics from a single atmospheric phenomenon. It is therefore essential that our surrogate, when evaluated at a new location in the design space, generates joint predictions with an internal consistency similar to the E3SM simulations. To achieve this, we apply the Karhunen-Loève expansion (KLE) (Karhunen, 1946; Loève, 1963; Ghanem & Spanos, 2003) to represent the vector $\mathbf{q}_n$ in terms of $J$ zero-mean uncorrelated random variables $\mathbf{z}_n$ referred to as *spectral features*. Let $\mathbf{Z} = \mathbf{Q}_0 \mathbf{V}$, where $\mathbf{Z}$ is the matrix containing the spectral features, $\mathbf{Q}_0$ is the standardized version of $\mathbf{Q}$ (centered by subtracting the sample mean and scaled by dividing by the sample standard deviation along each column), and $\mathbf{V}$ consists of the right singular vectors of $\mathbf{Q}_0$, or equivalently, the eigenvectors of the sample correlation matrix $\mathbf{Q}_0 \mathbf{Q}_0^{\top}$.

Let $\mathcal{X}$ and $\mathcal{Z}$ be the physics parameter and the spectral feature spaces, respectively. We model the unknown mapping $f_j : \mathcal{X} \to \mathcal{Z}_j$ via a GP with mean zero and positive definite correlation function $r_j : \mathcal{X}^2 \to [0,1]$,

$$\mathbf{z}_j = f_j(\mathbf{X}) \qquad\qquad \text{unknown function} \qquad (7)$$

$$f_j \sim \text{GP}(0, \sigma_{f_j}^2 r_j(\mathbf{x}, \mathbf{x}') + \sigma_{\varepsilon_j}^2 \delta(\mathbf{x}, \mathbf{x}')) \qquad\qquad \text{function prior} \qquad (8)$$

where $\sigma_{f_j}^2 > 0$ is the signal variance for the $j$-th spectral feature, $\sigma_{\varepsilon_j}^2 > 0$ is the error variance for the $j$-th spectral feature, and $\delta$ is the delta function. Implicit in the prior is the assumption of homogeneous signal variance and the inclusion of a nugget to model the response as a noisy, smooth function. The correlation, whose specification is discussed in Section S1 of Supporting Information S1, is often parameterized by $\{\sigma_{x_{pj}}^2 > 0\}$, where $\sigma_{x_{pj}}^2$ represents the length scale for the $p$-th physics parameter and the $j$-th spectral feature.

#### 3.2.2 Training

The GP models are trained separately for each of the $J$ spectral features. By setting up $J$ separate non-parametric regressions, each targeting one dimension of the orthogonal space, and reconstructing the physical predictions from the predicted spectral features, the surrogate effectively preserves the correlation across different physical quantities (most notably, the negative correlation between QBO period and amplitude discussed in Section 4.2) as well as the spatial coherence in amplitude at multiple pressure levels. Each spectral dimension has $2 + P$ hyperparameters, leading to a total of $(2 + P) \times J$ tuning parameters for the surrogate. The separate length scales allow the surrogate to learn different degrees of variability in $z_j$ with respect to $x_p$: the partial function $z_j = f_j(x_p)$ can become constant as $\sigma^2_{x_{pj}} \to \infty$, linear for moderate values of $\sigma^2_{x_{pj}}$, and highly non-linear as $\sigma^2_{x_{pj}} \to 0$ (Piironen & Vehtari, 2016). The distinct signal and error variances for each spectral feature provide individual signal-to-noise ratios (Ameli & Shadden, 2022).

Learning for GP regression models has been studied extensively, including exact (K. A. Wang et al., 2019) and approximate (Liu, Ong, et al., 2020) solutions. A brief discussion contrasting full Bayesian estimates and MLE plug-in estimates can be found in Bayarri et al. (2007). Type-II MLE (Rasmussen & Williams, 2005, sec. 5.4.1) is the most frequent exact training rule. Noting that the unknown function prior $p(\mathbf{f}|\mathbf{X})$ and the conditional distribution of the noisy observations given the function values $p(\mathbf{z}|\mathbf{f})$ are multivariate Gaussians, one can show (Rasmussen & Williams, 2005, eq. 5.8) that the observation marginal distribution is given by

$$p(\mathbf{z}|\mathbf{X}) = \int p(\mathbf{z}|\mathbf{f}, \mathbf{X})\, d\mathbf{f} = -\frac{1}{2}\mathbf{z}^\top \mathbf{S}_z^{-1} \mathbf{z} - \frac{1}{2}\log|\mathbf{S}_z| - \frac{T}{2}\log 2\pi, \tag{9}$$

where $\mathbf{S}_z$ is the observation covariance matrix. It is well-established that evaluating Eq. (9) at a large scale can be prohibitive due to the cubic time complexity $O(N^3)$ in computing the inverse and determinant of the covariance matrix (Liu, Ong, et al., 2020). Nonetheless, the computational cost of training the surrogate remains manageable given the limited number of simulations allowed by the comparatively costly Earth system model, the efficient dimension reduction enabled by the FFM, and the perfectly parallelizable nature of the separable GPs.

### 3.2.3 Selection

The strong correlation among the QoIs allows the data analyst to introduce regularization into the surrogate by truncating the KLE to the top $\tilde{J} \leq J$ modes, particularly when the number of perturbed parameters is much smaller than the number of QoIs ($P \ll J$). While the effective dimensionality of the spectral space can be determined automatically, for example, as part of the Bayesian inference procedure (Bishop, 1998), the analysis of exascale computer experiments often benefits from a human-in-the-loop approach due to relatively small sample sizes and highly complex, coupled structures. For instance, the following training statistics can be used to monitor the retained information. First, the distribution of energy across the modes can be examined to identify a sharp decrease (or "elbow") in the percentage of variance explained, given by $v_j / \sum_{j=1}^{J} v_j$, where $v_j$ is the $j$-th eigenvalue (Cattell, 1966). Second, because the spectral features are ordered by decreasing variance, higher-order modes tend to be relatively noisier, i.e., $\frac{\sigma_{f_1}}{\sigma_{\epsilon_1}} > \cdots > \frac{\sigma_{f_J}}{\sigma_{\epsilon_J}}$. Thus, the learned signal-to-noise ratio can serve as a guideline for selecting the truncation constant (Gramacy, 2020, sec. 5.3.4).

Alternatively, when the focus is on the surrogate's predictive capabilities, the optimal truncation can be determined via cross-validation. Under a Gaussian likelihood, the exact mean and variance for the leave-one-out prediction $\hat{z}_{jn}$ are available analytically through two numerically distinct but mathematically equivalent expressions (Vehtari et al., 2016; Sundararajan & Keerthi, 2001). The coefficient of determination, $R_j^2 = 1 - \sum_{n=1}^{N}(z_{jn} - \hat{z}_{jn})^2 / \sum_{n=1}^{N}(z_{jn} - \bar{z}_j)^2$ quantifies the proportion of variance in the actual values of the $j$-th spectral feature that is predictable by the surrogate. The posterior predictive log-density, $\text{PPLD}_j = \log p(\mathbf{z}_j \mid \text{E}\langle \hat{\mathbf{z}}_j \rangle, \text{V}\langle \hat{\mathbf{z}}_j \rangle)$, represents the log probability of the observed values under the model's predictive distribution, given that the model is trained on all other data points. A higher PPLD indicates more consistent predictability of a mode. Finally, the log-likelihood ratio, $\text{LLR}_j = -2\left[\text{PPLD}_j - \log p(\mathbf{z}_j \mid 0, 1)\right]$, provides a scaled version of the PPLD relative to a baseline model in which the spectral features follow a standard normal distribution. In other words, it quantifies the improvement gained by conditioning on the E3SM simulation ensemble rather than assuming their marginal distribution. A corresponding set of statistics can be defined analogously for the QoIs.

### 3.2.4 Prediction

Once the model is trained, the predictive distribution of the QoIs at a new location in the design space is readily available. The predicted QoIs follow a multivariate normal distribution, with a mean vector and covariance matrix obtained by applying standard linear algebra manipulation to the covariance in Eq. (8). This logic can be summarized in a three-step sequence: (1) calculate the prediction mean and variance for the spectral features, (2) derive the prediction mean vector and covariance matrix for the zero-mean, unit-variance quantities by evaluating the KLE, and (3) obtain the prediction mean vector and covariance matrix for the QoIs by reversing the scaling across columns. Step-by-step derivations and resulting expressions are provided in Section S2 of Supporting Information S1.

## 3.3 Multi-objective optimization

Our goal is to find physics parameter values that, based on the surrogate, are likely to generate wind fields with a period and amplitude similar to those estimated from the reference data. We set up a MOO, a decision-making method for optimizing under conflicting criteria supported by extensive literature offering theoretical guidance and algorithmic solutions (Collette & Siarry, 2004; Ruzika & Wiecek, 2005; Chinchuluun & Pardalos, 2007).

To define an appropriate set of objective functions, we note that the QoIs correspond to two physical quantities with markedly different characteristics. The period $\tau$ is measured in time units and quantifies the duration of one full cycle. The amplitude $\beta_{1k}$ is a speed measured in m/s and represents the maximum displacement of the wind speed from its time mean at the $k$-th pressure level. Moreover, the amplitude is a heterogeneous quantity whose scale does not remain constant throughout the depth of the atmospheric layer examined. Period and amplitude exhibit competing behaviors due to a set of complex and unknown dynamics simulated by the E3SM. Multiple strategies could be employed in our application. We find scalarization of vector optimization (Gunantara, 2018) unappealing due to the heterogeneity in their physical meanings, measurement units, and scales. Alternatively, minimizing spectral discrepancy, as proposed by (Mueller et al., 2025), would be immediate since the surrogate operates in the spectral domain. However, we decide against this approach to preserve the interpretability of the results introduced by the FFM.

We design the following pair of objective functions to capture the two most essential characteristics of the QBO, period and amplitude, as well as the trade-off between them:

$$\arg\min_{x^* \in \mathcal{X}} \begin{cases} (\hat{\tau}(\mathbf{x}^*) - \tau^{\text{REF}})^2 & \text{(period)} \\ \sum_k w_k \left( \hat{\beta}_{1k}(\mathbf{x}^*) - \beta_{1k}^{\text{REF}} \right)^2 & \text{(amplitude.)} \end{cases} \tag{10}$$

Here, $\hat{\tau}(\mathbf{x}^*)$ and $\hat{\beta}_{1k}(\mathbf{x}^*)$ are the QBO period and amplitude predicted by the surrogate at a new design location $\mathbf{x}^*$ in the design space $\mathcal{X}$, $\tau^{\text{REF}}$ and $\beta_{1k}^{\text{REF}}$ are the QBO period and amplitude estimated from the reference data, and $w_k > 0$ are weights. Considering that amplitude varies in scale throughout the depth of the atmospheric layer, heteroskedasticity is mitigated by weighting observations inversely proportional to their variance.

Multiple mathematical programming approaches have been proposed to search for the Pareto efficient solutions. However, exploiting our highly economical surrogate, we conduct a simple grid search over a low-dimensional design space. To find the physics parameter values on the Pareto frontier, we evaluate the surrogate over a fine tensor product grid covering the physics parameter space, plug the predicted mean of the QoIs into the objective functions in Eq. (10), and subset the dominant solutions. The resulting set is a discrete approximation of the infinitely many solutions for which the QBO period cannot be improved without deteriorating the amplitude-weighted sum, and vice versa.

## 4 Results

In this section, we apply the workflow developed in Section 3 to the data described in Section 2 to achieve a more realistic representation of the QBO in E3SM. We center our attention on the period $\tau$ and the amplitudes $\beta_{1k}$ over $K = 6$ pressure levels, working with a total of $J = K + 1 = 7$ QoIs.

### 4.1 Reference data

We fit the FFM in Eqs. (1) and (2) to the ERA5 data and estimate, by numerically optimizing Eq. (3), the reference period and amplitudes that will be used as target values in our optimization. We now assess the appropriateness of the model fitted to the observational data by visually inspecting the extracted signal as well as the unexplained component in the time-pressure physical space, and comparing the estimated QoIs to statistics reported in the climate literature.

***Time-pressure cross-sections***

Figure 2 shows lower stratospheric equatorial monthly mean and zonally averaged zonal winds in the time-pressure domain for the reference data, the FFM-based mean (signal), and the FFM-based deviations (residuals). The raw data manifest as a sequence of irregular stripes, whereas the FFM signal has perfectly aligned, identical stripes. Both data and signal show five diagonal stripes with similar inclination ($\hat{\alpha}_1 > 0$), red peaks at 10 and 20 hPa (large positive coefficients $\hat{\beta}_{1k} : k = 2, 3$), and red stripes that are more prominent than the blue stripes ($\hat{\beta}_{0k} > 0$). Overall, the signal captures the essence of the data while also discarding high-frequency oscillations, and the FFM signal is consistent with the stylized understanding of the QBO in the climate literature.

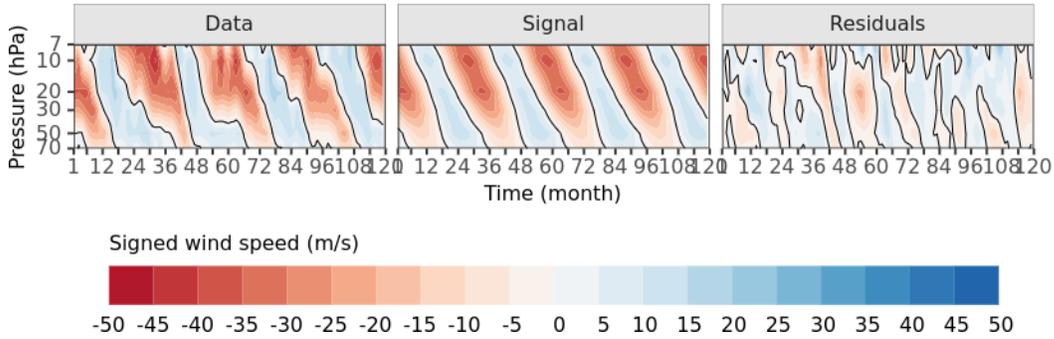

**Figure 2.** Monthly mean and zonally averaged zonal winds in the lower stratosphere based on (top) the reference data set (ERA5), (middle) the FFM (model mean), and (bottom) the ERA5-FFM difference (model residual or unexplained component). The time window, color shading, and line contours are identical to Fig. 1.

The residuals exhibit an overall lighter shade accounting for approximately 20% of the variance in the raw data. The residual map is a composite of two patterns. First, we observe short wavelength dynamics over time at 7-10 hPa, indicating that the model effectively distinguishes the QBO from the semi-annual oscillation "leaking" from the upper stratosphere. Second, we note a weaker oscillation pattern with a periodicity of 12-16 months, explained by extracting a periodic signal from quasi-periodic data. Even though this mild periodic component could motivate a higher-order approximation, the empirical residuals support our claim that it is reasonable to use a periodic model for a 10-year time series because we observe as few as approximately four cycles. Contrary to the common practice of preferring residuals with a random unstructured pattern, the error need not be fully random as long as the residual structure is not attributable to the QBO. The irregularity in the residual stripes suggests that the constant period and linear propagation constraints are a reasonable approximation for the data.

Additional empirical results are reported in Section S3 of Supporting Information S1. A spectral analysis provides further evidence that the FFM learns the most relevant frequency and the residuals have active frequencies not associated with the QBO. Moreover, a time series analysis shows that the sinusoidal wave is sufficient to capture the global pattern and the local deviations from the signal belong to high-frequency oscillations that we want to separate from the QBO.

### *QBO period and amplitude*

Figure 3 shows the QBO period and amplitude learned from the reference data set. The FFM captures 79.7% of the variance, achieving a 4:1 signal-to-noise ratio while reducing the data set dimensionality by a factor of 50 with as few as a single harmonic. The retained variability is comparable to the maximum QBO contribution of 81% at 20 hPa reported for the ERA-40 data set in Pascoe et al. (2005, sec. 4.1). The remarkably high information content in one single mode is the data-driven counterpart of the definition of the QBO as the main variation in the winds.

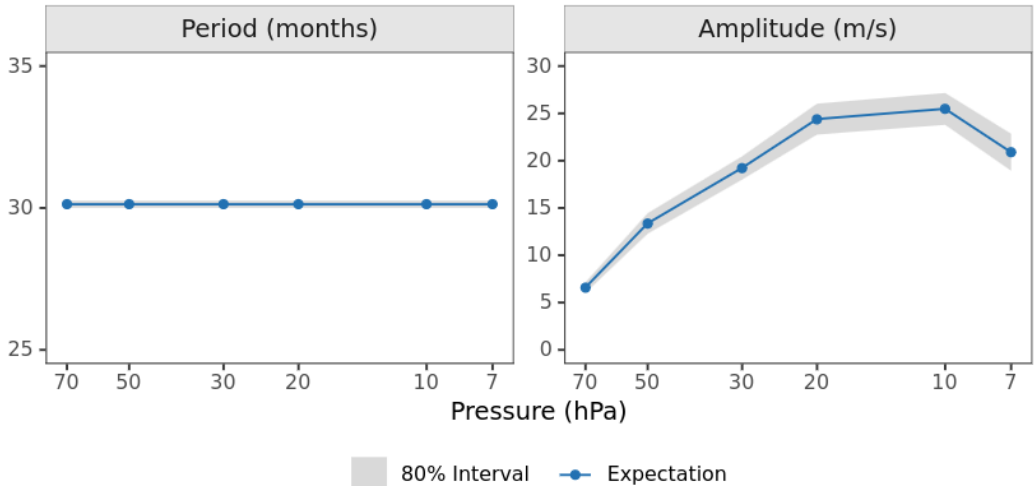

**Figure 3.** QoIs learned from the reference data set. These correspond to the period (left) and amplitude (right) of the sine waves in Figure S1 of Supporting Information S1. Expectation and interval found via maximum likelihood as discussed in Section 3.1.

Because our dimension reduction technique produces highly interpretable coefficients, we can establish a direct comparison to the substantial amount of empirical findings reported in the climate literature. As our analyses do not rely on the same data set and methodology as the other studies, we expect our estimates to be congruent but not necessarily statistically identical. We estimate the QBO period to be between 29.9 and 30.2 months, consistent with the 26.4 to 30.4 months reported in Pascoe et al. (2005, sec. 4.1), and marginally larger than the previously reported 27.7 months (Naujokat, 1986, tab. 1), 28.2 months (Baldwin et al., 2001), and 29.4 months (Coy et al., 2020). We estimate the QBO peak amplitude to be between 23.8 and 27.2 m/s at 10 hPa. The estimate is nominally close to 23.1 m/s at 20 hPa (Naujokat, 1986, reconstructed from tab. 3) and 25 m/s at 10 and 20 hPa (Coy et al., 2020, reconstructed from subfig. 2b and 2c). The peak vertical location is similar to the peak shown between 10-11 hPa in Pascoe et al. (2005, fig. 4d), although our amplitude is larger in magnitude, possibly due to their band filtering. Although we did not enforce smoothness or regularization on the amplitude estimates across pressure, we observe no sudden jumps in the estimated quantities.

### The FFM satisfactorily characterizes the QBO

In summary, we find that the goodness of fit is reasonable, the constraints are appropriate, the estimates are consistent with other studies, the free amplitudes display no sudden jumps despite the lack of a built-in smoothness mechanism, and the signal's local deviations from the data show a successful separation of the QBO from other high-frequency phenomena. On the downside, the standard error for the QBO period is optimistically narrow due to the simplistic nature of the FFM, leaving opportunities for future research.

## 4.2 E3SM simulation ensemble

In the same manner as done directly above with the reference data, we fit the FFM in Eqs. (1) and (2) separately to each wind field generated by the E3SM simulation ensemble. To assess the goodness of fit, we compute the percentage of variance explained by the QBO, defined in Eq. (6). The coefficients, which differ across the ensemble members, range from 11% to 91% with an ensemble median of 71%. Since the FFM explains 80% of the reference data and more than 71% in half of the ensemble members, we affirm that the single, most important frequency provides a sufficient degree of compression. The small but significant unexplained portions indicate that the QBO is not the only source of variability, but it is indeed the dominant mode for every single ensemble member.

Not all of the attempted simulations are carried forward for the surrogate-accelerated MOO. A total of 61 instances were attempted to explore the physics parameter space, as shown in Fig. 4. An initial space-filling design was constructed with a Latin hypercube sample (McKay et al., 1979; Stein, 1987) from the region defined in Table 1 and was sequentially augmented through manual inspection of the posterior predictive surface. E3SM did not terminate successfully for 10 parameter combinations with relatively large values of CF, suggesting that there is an implicit upper bound on the convective fraction beyond which the model becomes numerically unstable. The model terminated successfully but did not generate a pattern resembling the QBO for another 5 parameter sets located close to the sampling space boundaries. Formally, we deem the QBO non-existent in a simulation if the estimated signal has an excessively fast periodic component ($\hat{\tau} < 6$) or a periodic component slower than the Nyquist frequency ($\hat{\tau} > T/2$). The remaining 46 simulations are retained for downstream analysis. The ensemble member with the closest period is 3.3 months faster than the reference. There is at least one ensemble member that individually approximates the reference amplitude at one pressure level, but no simulation approximates the reference amplitude at all pressure levels simultaneously.

### Tension between the simulated period and amplitudes

The FFM does not establish any explicit dependence between the QBO period and amplitude. Nonetheless, the QoIs learned from the ensemble show a strong correlation structure with a profound impact on the optimization results. The pairwise analyses in Fig. 5 highlight the two core empirical dynamics driving the three-block structure in the correlation. The relationship between period and amplitude at 20 hPa (left) is nonlinear, with an overall downward curve such that longer periods are associated with lower amplitudes. On the other hand, the relationship between period and amplitude at 70 hPa (right) has an approximately negative log-linear trend. The strength and direction observed in both curves exhibit an evident tension between the simulated period and the amplitudes. More subtly, the dissimilarity in the shape of these associations implies that increasing amplitude at 20 hPa will increase amplitude at 70 hPa disproportionately, which conveys some of the challenge in matching the full set of amplitudes.

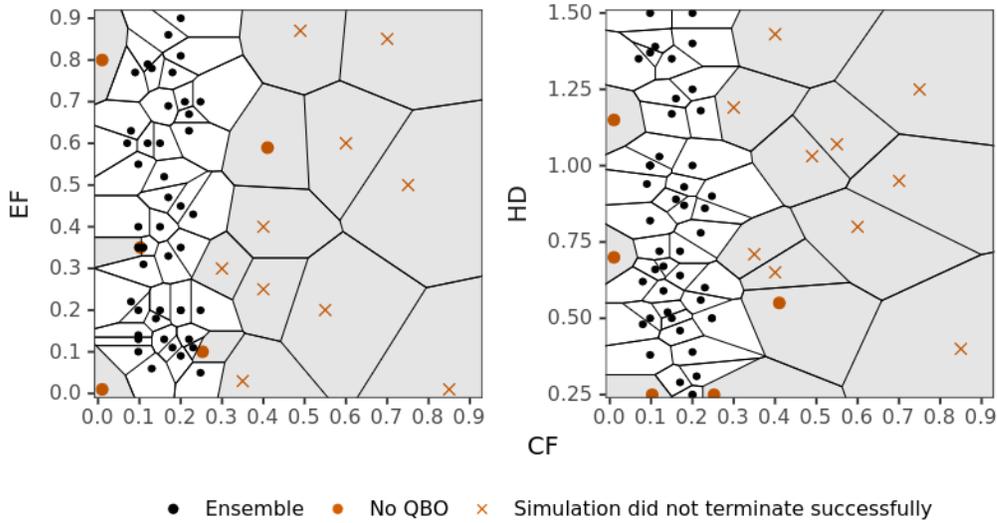

**Figure 4.** Locations in the E3SM physics parameter space where simulations were conducted. The black lines designates Voronoi cells (Lee & Schachter, 1980). Gray cells correspond to simulations that did not terminate successfully (numerically unstable, X marker) or did not show oscillating winds (nonphysical results, solid dot marker).

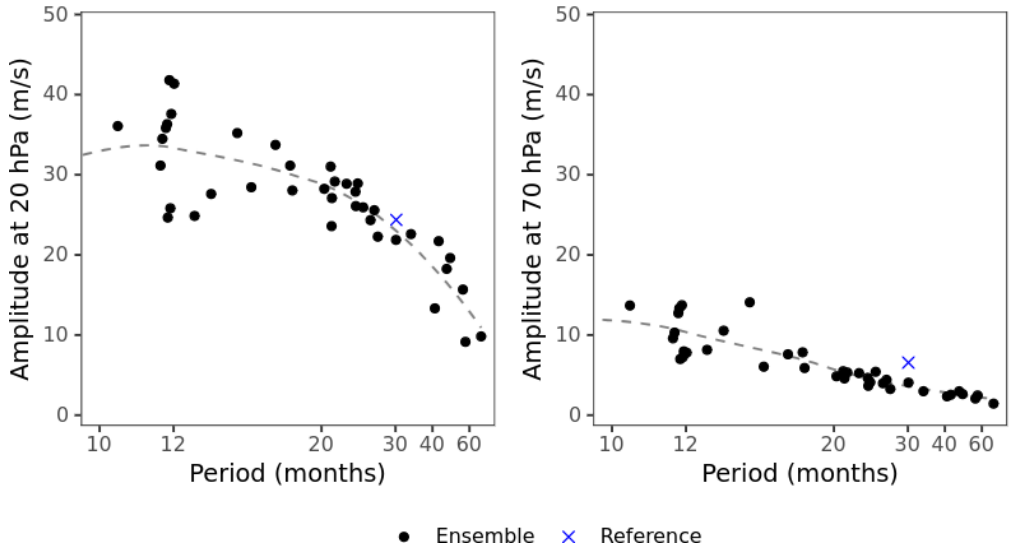

**Figure 5.** QBO period and amplitudes at 20 hPa (left) and 70 hPa (right) estimated by fitting the FFM to the 46 simulations under analysis (circle) and the reference data (cross). The dashed line, constructed using local smoothing (Cleveland & Loader, 1996), illustrates the approximate shape of the trend implied by E3SM.

We examine the sample correlation to further characterize the implicit structure of the QoIs. Given that the first three eigenmodes of the sample correlation matrix account for 97.3% of the energy and the estimated effective rank size (Roy & Vetterli, 2007) is approximately two, we find that the nominally seven quantities, one period and six amplitudes at different pressure levels, do not vary independently within a hypercube but instead reside in a highly constrained subspace.

The oscillation co-dynamics are captured by the biplots in Fig. 6, where the arrows indicate the direction and strength of the QoI's contribution to the eigenmodes (Gabriel, 1971). The first mode (81.2% variance explained) captures the tension between the period and the amplitudes, the second mode (10.1% variance explained) discriminates between lower- and upper-stratosphere amplitudes, and the third mode (6.0% variance explained) has a minuscule effect on the transition of the amplitude profile from the upper to the lower levels.

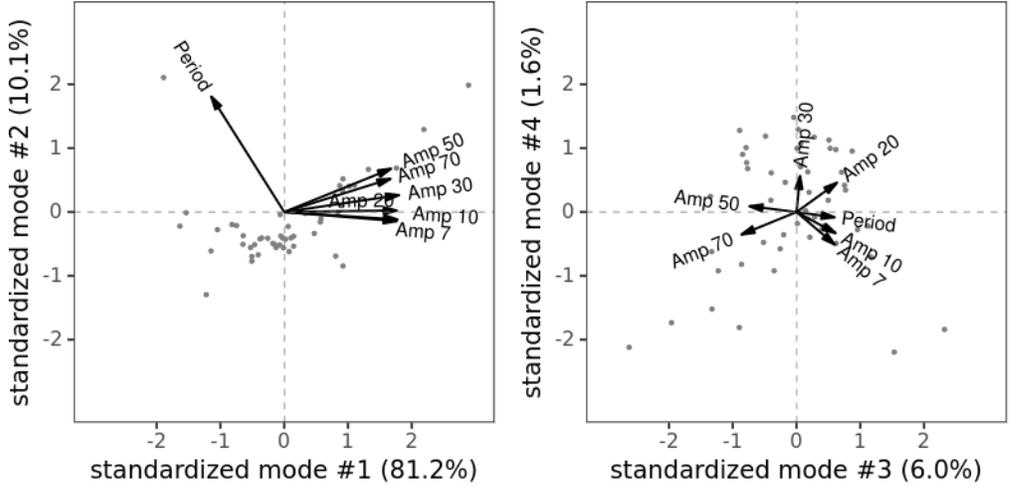

**Figure 6.** Biplots for the E3SM ensemble QoIs. The first mode (left plot horizontal axis) discriminates between QBO period and amplitude. The second mode (left plot vertical axis) discriminates between QBO amplitude at lower and upper levels.

The top three components explain 97.3% of the variance in the data and provide insight into the underlying physics, while the remaining components are negligible and provide no further insight into the QBO correlation structure. The first mode, which establishes a negative linear association between the QBO period and the simple mean of QBO amplitudes at all pressure levels, displays a strongly marked period-amplitude trade-off. Since amplitudes at all pressure levels contribute to the top mode of variation in approximately the same direction and magnitude, *calibrating the period will have a direct and opposite effect on the entire amplitude profile.* Consolidating the effects of the top two modes exposes a chain reaction in the calibration: adjusting the period will have a strong side-effect on the amplitude, and influencing the amplitude in the upper stratosphere will have a moderate side-effect on the lower stratosphere.

Zooming out to the MOO to improve the QBO, the correlation-based analysis of the E3SM ensemble data foreshadows a highly constrained system with limited capacity to target both period and amplitude simultaneously, and furthermore, to target amplitude at isolated pressure levels. Two complementary clustering analyses are reported in Section S4 of Supporting Information S1. The correlation- and distance-based analyses both reinforce the evidence that the QBO simulations are largely determined by two tensions among three sub-blocks. The QoI point estimates and 80% intervals are reported at an individual level in Section S5 of Supporting Information S1.

### 4.3 Surrogate modeling

Constructing the surrogate described in Section 3.2 requires specifying two components: the truncation constant $\tilde{J} \leq J$ and the form of the correlation function $r_j(\mathbf{x}, \mathbf{x}')$, as well as tuning several hyperparameters $\left\{ \sigma_{f_j}^2, \sigma_{\varepsilon_j}^2, \sigma_{x_{pj}}^2 \right\}_{j=1,p=1}^{J,P}$. The number of retained KLE modes introduces regularization in the predicted QoIs, while the choice of the correlation function influences the sensitivity of the predicted QoIs to input variations. We jointly select both components to balance smoothness and continuity in the data.

We begin by evaluating two complementary statistics related to surrogate training and goodness of fit. First, as discussed in Section 4.2, the first three modes capture up to 97.3% of the energy and correspond to physically interpretable rotations. In contrast, higher-order modes have a diminishing effect on QoI variance and offer little physical insight. Second, the surrogate-based signal-to-noise ratio $STN_j = \sigma_{f_j}/\sigma_{\varepsilon_j}$ shows that signal strength decreases drastically—by several orders of magnitude—indicating significant signal quality deterioration. From the fourth-order mode onward, error dominates the signal. To further assess the impact of mode selection, we perform a leave-one-out cross-validation study and compute the statistics defined in Section 3.2.3 to quantify the effect of increasing the number of modes on generalization error. The details are provided in Section S6 of Supporting Information S1.

Together, these analyses reveal that higher-order modes are neither explainable nor predictable by the proposed surrogate, and that the choice of covariance function has limited impact. We select the Matérn 3/2 correlation and retain the first three modes as a compromise among the points of maximum curvature in variance explained (strong energy compaction), signal-to-noise ratio (goodness of fit), and leave-one-out statistics (accuracy). This selection preserves most of the predictive accuracy while introducing regularization on the QoIs' random error. The selected model has an RMSE of 5 months for period and 4.5 m/s for amplitude at 20 hPa, equivalent to 18% of the reference values, setting a practical limit on the surrogate's efficiency in finding an optimal calibration.

### 4.4 Sensitivity analysis

We apply forward UQ techniques to assess the relative importance of the physics parameters for multiple QoIs, conducting a variance-based sensitivity analysis and estimating the global sensitivity Sobol' indices (Saltelli et al., 2006; Sobol , 2001). These indices do not depend on an initial set of model input values but instead consider the output across the entire domain of possible input parameters, using random sampling (Jansen, 1999; Saltelli et al., 2010). The definition and a numerical reporting of the estimates is provided in Section S7 of Supporting Information S1Although some indices vary across the QoIs, even after accounting for uncertainty in the index estimates, there is a clear ranking in the sensitivity of the QBO to the convection parameters. The oscillation is most responsive to CF, which accounts for 30-50% of the variance when considered alone and 65-75% when interactions are included, consistently across all QoIs. EF plays a secondary role, with nonzero main effects for the period and lower-stratospheric (higher pressure levels) amplitudes, and affects all QoIs approximately equally after accounting for interactions. HD has a tertiary role, marginally influencing QBO amplitude only in the middle-upper stratosphere (lower pressure levels). Considering that the sum of the total-effect indices for each QoI ranges between 1.2 and 1.5, and that CF and EF show the largest difference between total and main effects, we infer an implicit multiplicative structure for CF×EF in E3SM.

Since the physics parameters were normalized to $[-1, 1]$, the estimated length scales are nominally comparable and can serve as a proxy for feature importance. Notably, CF's length scales being close to zero across all modes indicate a fast reaction to small changes and the presence of local features. Since the first mode captures the majority of the variance and has coefficients of similar magnitude across all QoIs, we conclude that all QoIs exhibit local features as a function of CF.

### Main effect of convection in the simulated QBO

To further explore the response of the QoIs to changes in CF, we estimate the main effects $\hat{\mathbf{q}}(x_{CF}) = E_{CF}(\mathbf{q}|\mathbf{x})$ which represents the mean period and amplitudes at $x_{CF}$, averaged across all possible values of EF and HD. Figure 7 provides a parsimonious representation of an otherwise highly complex system and summarizes the essential relationship between the intensity of convection and the QBO: more intense convection is associated with faster and stronger oscillations. The QBO in E3SM can accordingly be reduced to its primary effects by noting that increases in CF are associated with approximately linear decreases in period and increases in amplitude, aside from three local deviations. This foretells the limitations of our calibration efforts.

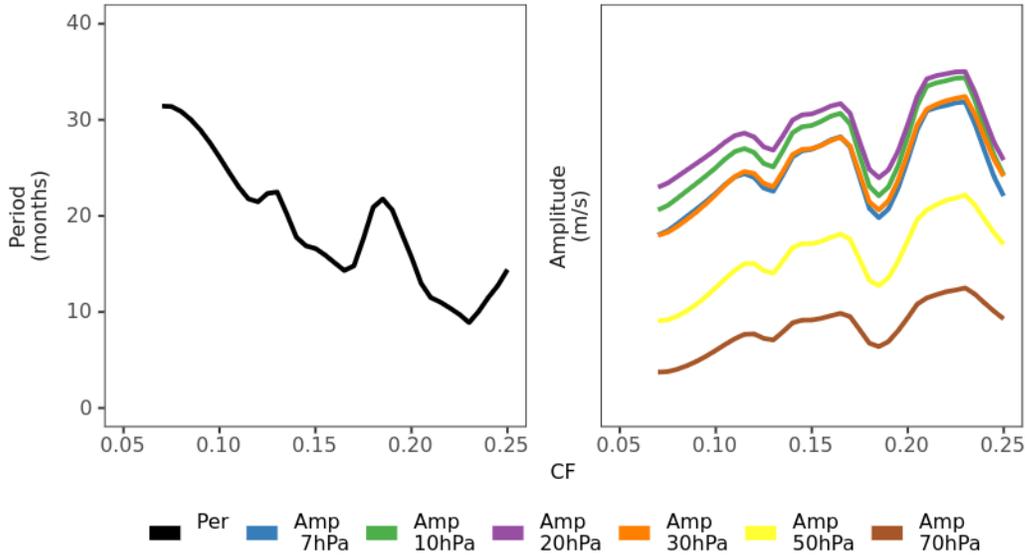

**Figure 7.** Main effects of CF on the simulated QBO period (left) and amplitude (right). An increase in CF is associated with a decrease in the period and an increase in the amplitudes that is approximately linear, with three local deviations from the trends. Different colors represent pressure levels, with the amplitude increasing with altitude.

### 4.5 Optimization

So far, we have formulated an efficient data reduction model to extract physically interpretable QoIs, built an inexpensive and sufficiently accurate surrogate, and identified the essential patterns linking convection and the QBO. We now proceed to set up a MOO problem to target parametric settings that best represent the fundamental dependence of equatorial winds on atmospheric deep convection.

By studying the simulation ensemble in Section 4.2 and the surrogate-based main effects in Section 4.4, we established that the simulated period and amplitudes belong to a highly constrained space dominated by a significant tension between the period and amplitude, and a lesser rigidity between amplitudes in the lower and upper stratosphere. Therefore, we consider it unlikely that a single combination of the physics parameters in E3SM version 2 will generate a QBO that satisfactorily reproduces both the reference period and the amplitude at every pressure level simultaneously.

To motivate this argument, we evaluate the integrated predicted QoIs $\hat{\mathbf{q}}(x_{CF}, x_{EF}) = E_{-HD}(\mathbf{q}|\mathbf{x})$, where we integrate over HD to marginalize over the parameter with the least reduction in the output uncertainty. In Fig. 8, red and blue represent integrated predicted values below and above the reference value, respectively, white regions represent small discrepancies, and black lines delineate the curves where a perfect match happens. Focusing solely on a single facet, the contour lines define the set of solutions where the QoI has a marginal expected value equal to the reference. The intersection of all the sets, which would result in a joint expectation matching all reference values, is empty.

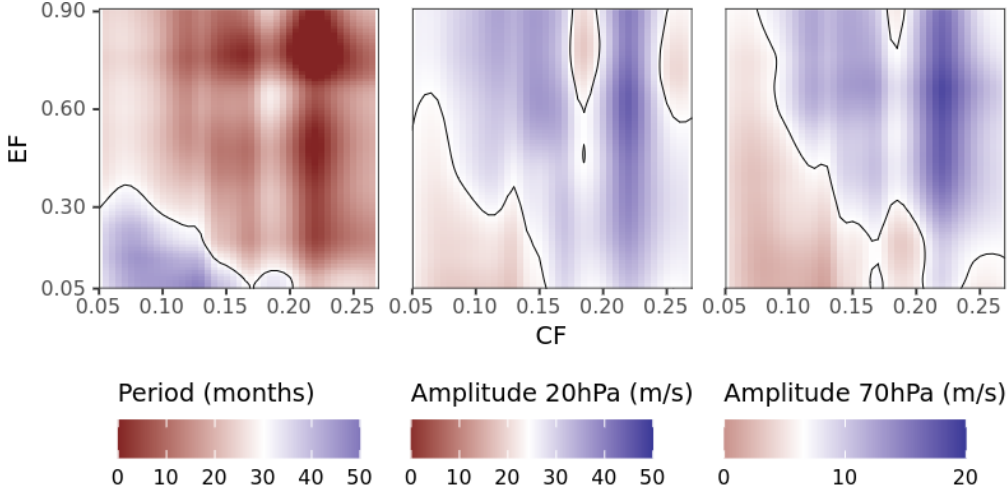

**Figure 8.** Predicted QBO period (left), amplitude at 20 hPa (middle) and 70 hPa (right) as a function of CF and EF after integrating out HD. Shading indicates whether the predicted value is above or below the reference. The contour delineates regions of zero difference, highlighting parameter combinations where predictions align with target values, but the empty intersection of the three contours indicates that no single point matches all reference values simultaneously.

We evaluate the surrogate and predict the QoIs over a fine grid with 200,000 elements, covering the input space. This grid is created using a tensor product of sequences chosen to provide higher resolution for HD, EF, and CF, with input prioritization based on their sensitivities. Adopting a no-preference approach, we approximate the Pareto efficient frontier to learn the trade-off between QBO period and amplitude. We opted against eliciting a preference function from the SMEs due to the absence of a natural hierarchy among the QoIs and the difficulty of combining variables with different physical meanings and measurement units. We compute the objective functions by substituting the predicted mean vector for each QoI in Eq. (10), with weights equal to the inverse of the reference values, $w_k = 1/\beta_{2k}^{\mathrm{REF}}$. Since the amplitude at 20 hPa is approximately four times larger than at 70 hPa, we balance the contribution of amplitudes across pressure levels by adjusting the weight of differences in the upper stratosphere and the lower stratosphere.

Figure 9 shows the surrogate-based objective functions, ensemble members, the origin representing perfect calibration, and the discrete approximation of the Pareto frontier. The Pareto frontier reflects a trade-off between QBO period and amplitude, representing the marginal rate of substitution between the two metrics. The gap between the frontier and the origin indicates that a near-perfect solution, aligning with reference amplitudes across all pressure levels, is unattainable. The vertical gap highlights the lower tension between amplitudes in the lower and upper stratosphere, as perfect alignment would cause the frontier to intersect the horizontal axis. The upper-left boundary contains solutions with improved amplitudes at the expense of longer periods, while the lower-right boundary shows solutions with improved periods, with minimal impact on amplitude. The "elbow" of the frontier represents the most cost-effective subset of solutions under a no-explicit-preference approach.

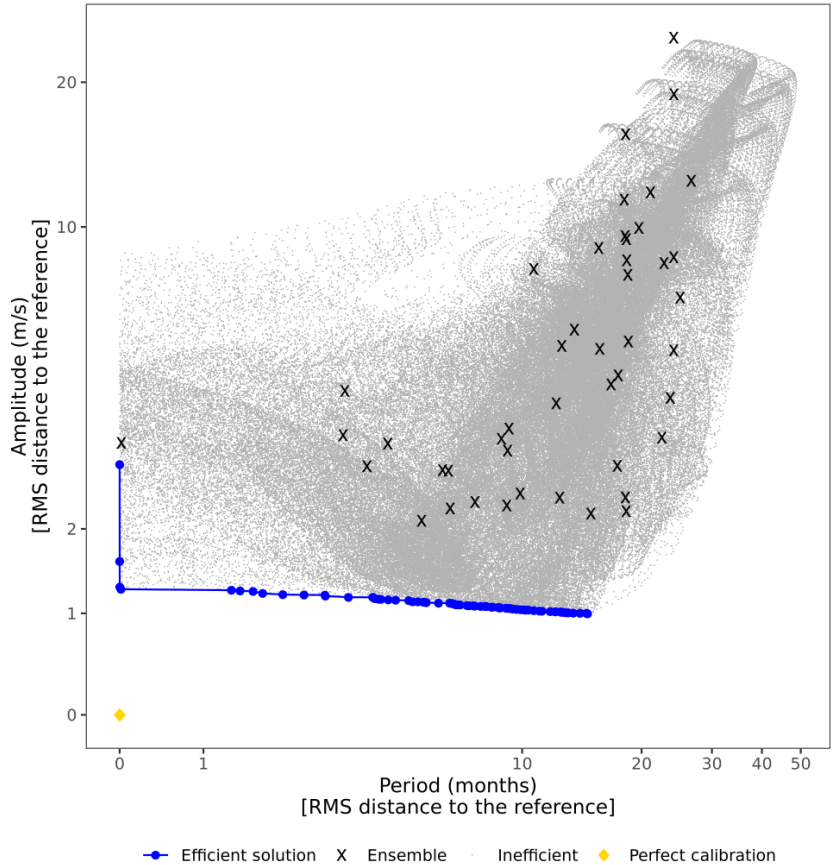

**Figure 9.** Surrogate-based objective functions (gray dots), E3SM ensemble members (black crosses), the origin representing perfect calibration (yellow diamond), and the discrete approximation of the Pareto frontier (blue circles). The axes are in log scale.

Figure 10 shows the physics parameters associated with the surrogate-based Pareto frontier. The range in CF covers the default value for E3SMv2, and the marginal distribution of HD shows a strong skew toward values smaller than 0.75. The efficient set reflects an internal structure that is ultimately learned from the E3SM ensemble data: to remain on the frontier, an increase in CF must primarily be compensated by a non-linear decrease in EF and secondarily coupled with an increase in HD. We separated the solutions into two groups for HD in [0.25, 0.75] and (0.75, 1.50]. Although there is a visually compelling argument that these two clusters display different correlation structures, we are unaware of a physical rationale behind this apparent change of regime.

### 4.6 Optimal configuration

We select a corner solution (EF = 0.12, CF = 0.16, HD = 0.48) from the Pareto frontier and run E3SM to validate this parameter set. The predicted, simulated, and reference QoIs are summarized in Figure 11. The simulated QBO period is approximately 3 months shorter than the predicted value, the simulated upper-stratospheric QBO amplitudes are close to expectations, and the simulated lower-stratospheric QBO amplitudes are smaller than predicted. This illustrates the two tensions governing the simulations: the trade-off between QBO period and amplitude, and the tension between lower- and upper-level amplitudes. It is difficult to find a solution that closely matches more than one element in the triad. The width of the period interval highlights the surrogate's limitation. The QoIs estimated from the validation simulation show some variability, attributed to the expectedly imperfect but sufficiently accurate surrogate. The simulated QBO period and amplitudes fall within the 80% predictive intervals, suggesting that surrogate accuracy was satisfactory.

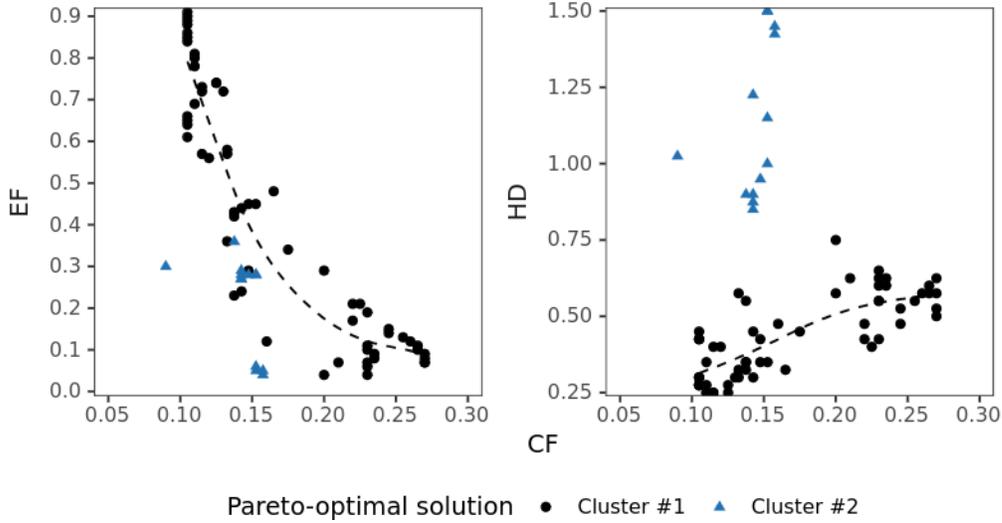

**Figure 10.** Physics parameter values along the discrete approximation to the surrogate-based Pareto frontier. Dashed lines indicate general trends in each panel. Left: Trade-off between CF and EF, where selecting optimal combinations of the QBO period and amplitudes requires increasing CF while decreasing EF. Right: The distribution of optimal CF and HD values reveals two distinct clusters. Cluster #1 (black circles) follows a positive trend, while Cluster #2 (blue triangles) exhibits elevated HD values, suggesting a different regime of parameter interactions.

**End-to-end calibration**

To conclude, we examine the model improvement sequence from E3SMv2 to the surrogate-based optimal configuration. This analysis includes two vertical grid configurations and three sets of physics parameter values. L72 and L80 refer to the prior and new vertical grids, composed of 72 and 80 atmospheric levels, respectively, as discussed in Yu et al. (2024). The prior default, new default, and surrogate-based optimal values are discussed in (Golaz et al., 2022), (Xie & Coauthors, 2025), and Section 4.6, respectively. The simulated wind fields are shown in Fig. 12, and the numerical results are provided in Section S8 of Supporting Information S1.

From (a) to (b) in the Fig. 12 panels, we observe changes in the tropical wind structure solely due to the new vertical grid with 80 levels. Increasing the vertical resolution around the lower stratosphere largely improved the tropical wind, which now displays a clear sequence of alternating winds. Despite these more defined cycles, the simulated QBO is weaker and approximately 10 months slower than the reference data. The differences between panels (b) and (c) resulted from several upgrades to the E3SM source code that took place during analyses in Yu et al. (2024) and this paper. The many updates to model physics, including to the parameterization of tropical convection but not to the parameterization of convectively generated gravity waves, were conducted as part of the model development cycle that led to the creation of E3SMv3. Changes in the atmospheric component model accelerated the oscillations, causing the same nominal values of the physics parameters to produce a faster oscillation with noticeably steeper stripes.

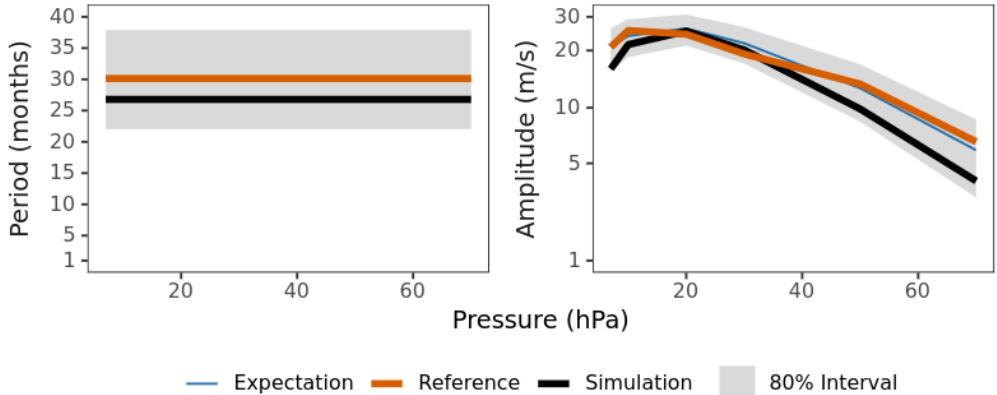

**Figure 11.** Predicted, reference, and simulated QoIs for the corner solution on the Pareto frontier. The expectation (blue) represents the predicted QoI, the reference (orange) denotes the target values, and the simulation (black) corresponds to the actual E3SM output. The shaded gray region indicates the 80% interval. The results demonstrate that the simulation closely follows the reference values, particularly for QBO amplitude, with minor deviations in the higher-pressure (lower-stratospheric) levels. The width in the prediction intervals for QBO period and higher-pressure (lower-stratospheric) amplitudes indirectly suggests a limit to the efficiency in the surrogate-based calibration.

In panel (d), we see a partial improvement selected for promotion to the v3 default values, stemming from our ongoing optimization work at the time of version freeze. Starting with E3SMv3 with the prior defaults, new parameter values based on the then-optimal process were selected, building upon an intermediate snapshot of the adaptive process. This coarse refinement only affected HD and led to improvements in the QBO period, amplitude, downward propagation, and east-west balance. Finally, in panel (e), we present the surrogate-based optimal solution highlighted in Fig. 11. The latter part of the process fine-tunes CF and EF, representing the trade-off illustrated in Figure 10, while keeping HD virtually unchanged.

In coordinating with related E3SM activities over the coarse of this study, the end-to-end surrogate-assisted MOO has progressed from (c) to (e), with the first stage (c) to (d) being a coarse refinement by applying a large correction in HD, and the second stage (d) to (e) as a more precise refinement with small corrections in CF and EF. We observe that E3SMv3 still simulates a slightly faster and weaker QBO. However, the jump from (d) to (e) is significantly smaller than from (c) to (e), suggesting that we have reached diminishing returns in our optimization. At this point, we concluded this phase of our QBO calibration process. We expect that continuing the optimization with surrogate refinement iterations would only yield marginal numerical improvements, having exhausted the majority of the potential in our current parametric formulation.

## 5 Discussion

We developed an end-to-end UQ workflow that calibrates the representation of convectively generated gravity waves in E3SM and yields a more realistic QBO through surrogate-accelerated MOO. We introduced the FFM to compress massive wind field data and extract physically interpretable QoIs. The FFM effectively translated established principles from atmospheric science into QoIs amenable to UQ, isolated the QBO signal from other oscillations and random error (Fig. 2), and reduced data dimensionality. A single sinusoidal wave captured the signal in the reference data, retaining 79.7% of the variance and achieving a 4:1 signal-to-noise ratio while reducing dimensionality by a factor of 50 and producing estimates consistent with existing literature (Fig. 3). The goodness of fit was acceptable, constraints were appropriate, estimates aligned with previous studies, free amplitudes exhibited smooth transitions despite the absence of an explicit smoothness mechanism, and local signal deviations effectively separated the QBO from other sources.

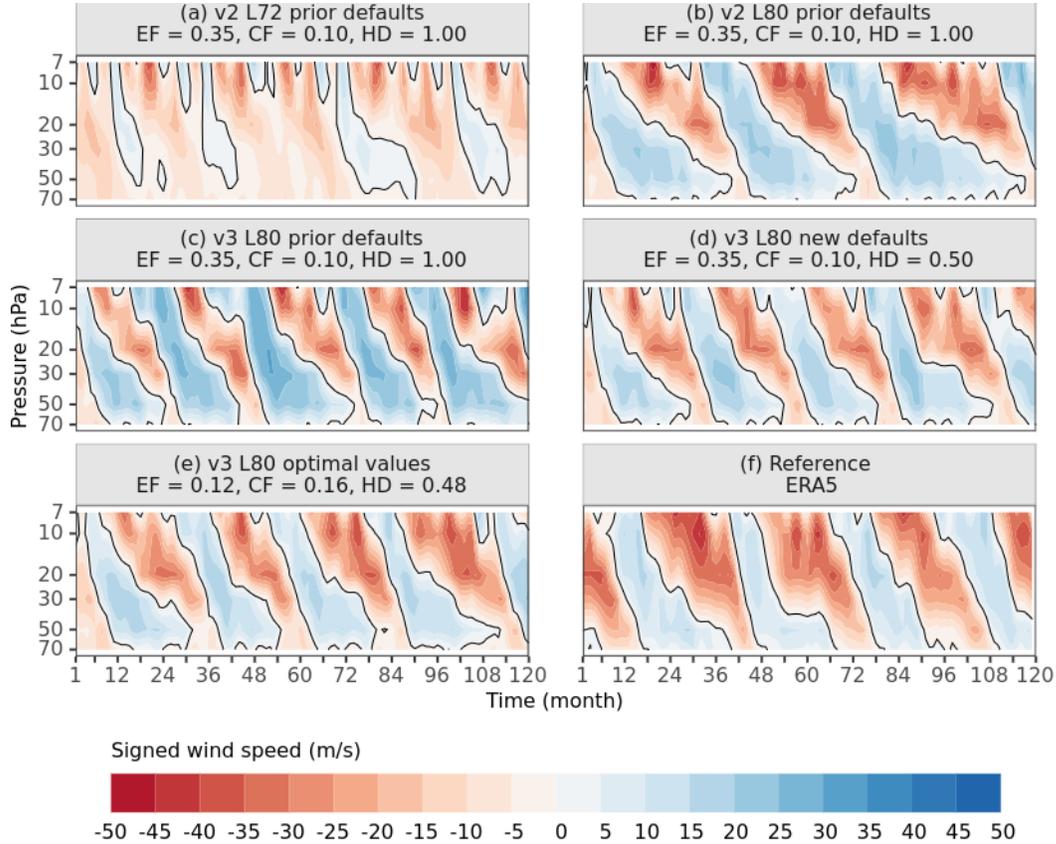

**Figure 12.** Improvement sequence of the wind fields from E3SMv2 to the surrogate-based optimal configuration applied to E3SMv3. The improvements resulting from the surrogate-assisted MOO are captured by the two-stage analysis (c) to (d) and (d) to (e), where (d) corresponds to an intermediate result during the adaptation loop.

We generated a simulation ensemble to explore the physics parameter space (Fig. 4). We identified two implicit forces in the simulated QBO: a nonlinear relationship between period and amplitude at 20 hPa, wherein longer periods were associated with lower amplitudes, and a negative log-linear trend between period and amplitude at 70 hPa (Fig. 5). The seven QoIs, comprising the period and amplitudes at six pressure levels in the stratosphere, exhibited a strong correlation structure. The first three eigenmodes of the sample correlation matrix accounted for 97.3% of the total variance (Fig. 6), yielding an estimated effective rank of approximately two. Higher-order modes were neither explainable nor predictable by the proposed surrogate. These constraints produced a highly restricted output space.

Next, we developed a statistical surrogate to predict E3SM-generated QBO behavior with satisfactory accuracy and at a fraction of the computational cost of a full simulation.

CF explained 30–50% of the variance when considered in isolation and 65–75% when interactions were included, consistently across all QoIs. Its near-zero length scales across all modes suggested a rapid response to small perturbations and the presence of localized features. At the core of a complex physical system, the QBO in E3SM could be characterized by noting that increases in CF were associated with approximately linear decreases in period and increases in amplitude, aside from three local deviations (Fig. 7). Furthermore, we identified an implicit upper bound on CF beyond which the model became numerically unstable.

Finally, we employed our cost-efficient surrogate to discretely approximate the Pareto frontier and quantify the trade-off between the QBO period and amplitude. A near-optimal solution aligning with reference amplitudes across all pressure levels was unattainable (Fig. 8). Even when disregarding the period, tensions in QBO amplitudes between the lower and upper stratosphere precluded fine-tuning at individual pressure levels (Fig. 9). The overall trend in the solution set shows that an increase in CF necessitates a primary non-linear decrease in EF, along with a secondary increase in HD (Fig. 10). We validated our workflow by running E3SM near the Pareto elbow and confirmed that the surrogate accuracy was satisfactory (Fig. 11). After observing diminishing returns in the optimization process, we concluded our calibration. Analysis of the end-to-end model improvement sequence, from E3SMv2 to the surrogate-based optimal configuration applied to E3SMv3, suggests that our workflow substantially contributed to an improved QBO (Fig. 12).

Our application also revealed several opportunities for future workflow enhancements. Although the FFM was both parsimonious and effective, it could benefit from further refinement. The standard error for the QBO period appeared optimistically narrow due to the FFM's simplistic formulation. Incorporating a time-aware correlation structure for the error could lead to more accurate uncertainty estimates (White, 1982). Furthermore, a secondary periodicity between 12 and 16 months emerged in the residuals, likely resulting from the extraction of a periodic signal from quasi-periodic data. Increasing the number of active frequencies within the FFM, analogous to adopting a polyphonic rather than monophonic approach, could address this issue. Also, mechanistic (Holton & Lindzen, 1972) and empirical (Baldwin et al., 2001) studies suggest that the QBO exhibits partially distinct characteristics during its easterly and westerly phases. This concept could be incorporated into the FFM by permitting certain coefficients to vary according to wind speed sign or magnitude. Said dependence of coefficients on wind direction might be modeled through structural changes or smooth transitions.

Enhancing surrogate accuracy would further benefit the workflow. The surrogate's RMSE, which amounts to 18% of the reference values, imposes a practical limit on optimization efficiency in this application. Exploring alternative predictive models, such as polynomial chaos expansion (Sargsyan et al., 2014) or neural networks (Diaz-Ibarra et al., 2025), may improve performance. Alternative surrogate modeling approaches that do not require explicit FFM construction warrant consideration. For instance, the surrogate modeling workflow for random fields proposed by Mueller et al. (2025) aligns well with the pressure–time data structure shown in Fig. 2. Spectral analysis techniques, such as wavelets (Torrence & Compo, 1998), can address quasi-periodicity via time localization, thereby reconciling phase differences between reference and ensemble data. A hybrid approach that integrates these components may yield an efficient and novel representation.

The final step in our workflow, MOO, offers several avenues for improvement. Distance- and similarity-based analyses of the ensemble revealed that QBO period and amplitude are primarily driven by tensions among period, lower-stratospheric, and upper-stratospheric amplitudes. A three-dimensional loss function could more effectively resolve the solution space. More radically, rather than targeting summary quantities in Eq. (10), estimating the Pareto frontier in a seven-dimensional space, where all QoIs are independently optimized, could resolve all partial trade-offs. High-dimensional Pareto optimization, however, poses increasing computational challenges (Sülflow et al., 2007).

Our loss function was constructed as a plug-in estimator that treated the predictive mean as a fixed known quantity. A natural extension would incorporate prediction uncertainty, following Bayesian model calibration (Kennedy & O'Hagan, 2001) or efficient global optimization (Jones et al., 1998). Although predictive surface uncertainty is one component, a comprehensive approach should integrate multiple sources of uncertainty, including aleatoric uncertainty in QoI estimates, structural uncertainty in the FFM, and surrogate construction uncertainties such as hyperparameter tuning and model selection. Moreover, reference values derived from reanalysis data are subject to data assimilation and random errors (Bosilovich et al., 2013; Parker, 2016), thereby introducing additional aleatoric and epistemic uncertainty. Although it is difficult to judge a priori which of these sources have the largest impact, their inclusion would enhance the robustness of the workflow.

Finally, we acknowledge that simulating the QBO remains a challenging problem. Our workflow reveals a strong tension between period and amplitude, which limits the degrees of freedom of the calibration. Modifying E3SM to alleviate this tension could boost the workflow's effectiveness and impact. Potential improvements include exposing currently hard-coded parameters, introducing new physics parameters, refining convection physics, further adjusting the vertical grid, or increasing geospatial resolution. Although atmospheric scientists are best suited to propose and evaluate such modifications, UQ can support these efforts through global sensitivity analysis (Iooss & Lemaître, 2015) to evaluate the effects of the additional parameters, surrogate-based optimization with multi-fidelity (Eldred & Dunlavy, 2006) to efficiently allocate simulations with varying degrees of spatial resolution and complexity, and embedded (Sargsyan et al., 2019) or external (Brynjarsdóttir & O Hagan, 2014) model discrepancy methods to further examine the trade-offs among the QoIs.

## Open Research Section

ERA5 data (Hersbach et al., 2017) is available from ECMWF MARS tape archive. Simulation data was generated from a version forked from E3SMv2 (Golaz et al., 2022; E3SM Project, 2023) that implements the redesigned vertical grid introduced in Yu et al. (2024). Data processing and analysis procedures were implemented in the R programming language (R Core Team, 2024) leveraging infrastructure (Barrett et al., 2024; Ooms, 2014; Pierce, 2023; Solymos & Zawadzki, 2023), statistical (Binois & Gramacy, 2021; Genz & Bretz, 2009; Iooss et al., 2024; D. Bates et al., 2024), visualization (Paradis & Schliep, 2019; Turner, 2024; Schloerke et al., 2024; Vu & Friendly, 2024; van den Brand, 2024; Wickham, 2016; Aphalo, 2024; Slowikowski, 2024; Henry et al., 2024; Pedersen, 2024; Neuwirth, 2022; Wickham et al., 2023), and table generation (Dahl et al., 2019) community packages.

## Acknowledgments

This research was supported through the U.S. Department of Energy Office of Science's Scientific Discovery through Advanced Computing (SciDAC) program, the Advanced Scientific Computing Research (ASCR) program, and the Biological and Environmental Research's (BER's) Earth System Model Development program area via the SciDAC project "Improving the quasi-biennial oscillation through surrogate-accelerated parameter optimization and vertical grid modification" (grant number: SCW1787). This research used resources of the National Energy Research Scientific Computing Center (NERSC), a Department of Energy Office of Science User Facility supported by the Office of Science under Contract No. DE-AC02-05CH11231, using NERSC award ASCR-ERCAP0031947. Los Alamos National Laboratory is operated by Triad National Security, LLC for the U.S. Department of Energy, National Nuclear Security Administration under Contract 89233218CNA000001. This work was performed under the auspices of the U.S. Department of Energy by Lawrence Livermore National Laboratory under Contract DE-AC52-07NA27344. This article has been authored by an employee of National Technology & Engineering Solutions of Sandia, LLC under Contract No. DE-NA0003525 with the U.S. Department of Energy (DOE). The employee owns all right, title and interest in and to the article and is solely responsible for its contents. The United States Government retains and the publisher, by accepting the article for publication, acknowledges that the United States Government retains a non-exclusive, paid-up, irrevocable, world-wide license to publish or reproduce the published form of this article or allow others to do so, for United States Government purposes. The DOE will provide public access to these results of federally sponsored research in accordance with the DOE Public Access Plan `https://www.energy.gov/downloads/doe-public-access-plan`. This paper describes objective technical results and analysis. Any subjective views or opinions that might be expressed in the paper do not necessarily represent the views of the U.S. Department of Energy or the United States Government.

## References

Ahlgrimm, M., & Forbes, R. (2014, January). Improving the representation of low clouds and drizzle in the ecmwf model based on arm observations from the azores. *Monthly Weather Review*, *142*(2), 668–685. Retrieved from `http://dx.doi.org/10.1175/MWR-D-13-00153.1` doi: 10.1175/mwr-d-13-00153.1

Alexander, M. J., & Holton, J. R. (1997, February). A model study of zonal forcing in the equatorial stratosphere by convectively induced gravity waves. *Journal of the Atmospheric Sciences*, *54*(3), 408–419. Retrieved from `http://dx.doi.org/10.1175/1520-0469(1997)054<0408:AMSOZF>2.0.CO;2` doi: 10.1175/1520-0469(1997)054<0408:amsozf>2.0.co;2

Ameli, S., & Shadden, S. C. (2022). *Noise estimation in gaussian process regression.* arXiv. Retrieved from `https://arxiv.org/abs/2206.09976` doi: 10.48550/ARXIV.2206.09976

Anstey, J. A., Butchart, N., Hamilton, K., & Osprey, S. M. (2020, July). The sparc quasi-biennial oscillation initiative. *Quarterly Journal of the Royal Meteorological Society*, *146*(744), 1455–1458. Retrieved from `http://dx.doi.org/10.1002/qj.3820` doi: 10.1002/qj.3820

Anstey, J. A., Osprey, S. M., Alexander, J., Baldwin, M. P., Butchart, N., Gray, L., … Richter, J. H. (2022, August). Impacts, processes and projections of the quasi-biennial oscillation. *Nature Reviews Earth & Environment*, *3*(9), 588–603. Retrieved from `http://dx.doi.org/10.1038/s43017-022-00323-7` doi: 10.1038/s43017-022-00323-7

Anstey, J. A., & Shepherd, T. G. (2013, March). High-latitude influence of the quasi-biennial oscillation. *Quarterly Journal of the Royal Meteorological Society*, *140*(678), 1–21. Retrieved from `http://dx.doi.org/10.1002/qj.2132` doi: 10.1002/qj.2132

Anstey, J. A., Simpson, I. R., Richter, J. H., Naoe, H., Taguchi, M., Serva, F., … Yukimoto, S. (2021, June). Teleconnections of the quasi-biennial oscillation in a multi-model ensemble of qbo-resolving models. *Quarterly Journal of the Royal Meteorological Society*, *148*(744), 1568–1592. Retrieved from `http://dx.doi.org/10.1002/qj.4048` doi: 10.1002/qj.4048

Aphalo, P. J. (2024). ggpp: Grammar extensions to 'ggplot2' [Computer software manual]. Retrieved from `https://CRAN.R-project.org/package=ggpp` (R package version 0.5.8-1)

Baldwin, M. P., & Dunkerton, T. J. (1998, September). Quasi-biennial modulation of the southern hemisphere stratospheric polar vortex. *Geophysical Research Letters*, *25*(17), 3343–3346. Retrieved from `http://dx.doi.org/10.1029/98GL02445` doi: 10.1029/98gl02445

Baldwin, M. P., & Dunkerton, T. J. (2001, October). Stratospheric harbingers of anomalous weather regimes. *Science*, *294*(5542), 581–584. Retrieved from `http://dx.doi.org/10.1126/science.1063315` doi: 10.1126/science.1063315

Baldwin, M. P., Gray, L. J., Dunkerton, T. J., Hamilton, K., Haynes, P. H., Randel, W. J., … Takahashi, M. (2001, May). The quasi-biennial oscillation. *Reviews of Geophysics*,

*39*(2), 179–229. Retrieved from `http://dx.doi.org/10.1029/1999RG000073` doi: 10.1029/1999rg000073

Barrett, T., Dowle, M., Srinivasan, A., Gorecki, J., Chirico, M., & Hocking, T. (2024). data.table: Extension of 'data.frame' [Computer software manual]. Retrieved from `https://CRAN.R-project.org/package=data.table` (R package version 1.15.4)

Bates, D., Maechler, M., & Jagan, M. (2024). Matrix: Sparse and dense matrix classes and methods [Computer software manual]. Retrieved from `https://CRAN.R-project.org/package=Matrix` (R package version 1.7-0)

Bates, D. M., & Watts, D. G. (1988). *Nonlinear regression analysis and its applications.* Wiley. Retrieved from `http://dx.doi.org/10.1002/9780470316757` doi: 10.1002/9780470316757

Bayarri, M. J., Berger, J. O., Paulo, R., Sacks, J., Cafeo, J. A., Cavendish, J., … Tu, J. (2007, May). A framework for validation of computer models. *Technometrics*, *49*(2), 138–154. Retrieved from `http://dx.doi.org/10.1198/004017007000000092` doi: 10.1198/004017007000000092

Bechtold, P., Köhler, M., Jung, T., Doblas-Reyes, F., Leutbecher, M., Rodwell, M. J., … Balsamo, G. (2008, July). Advances in simulating atmospheric variability with the ecmwf model: From synoptic to decadal time-scales. *Quarterly Journal of the Royal Meteorological Society*, *134*(634), 1337–1351. Retrieved from `http://dx.doi.org/10.1002/qj.289` doi: 10.1002/qj.289

Bechtold, P., Semane, N., Lopez, P., Chaboureau, J.-P., Beljaars, A., & Bormann, N. (2014, January). Representing equilibrium and nonequilibrium convection in large-scale models. *Journal of the Atmospheric Sciences*, *71*(2), 734–753. Retrieved from `http://dx.doi.org/10.1175/JAS-D-13-0163.1` doi: 10.1175/jas-d-13-0163.1

Berdahl, M., Leguy, G., Lipscomb, W. H., & Urban, N. M. (2021, June). Statistical emulation of a perturbed basal melt ensemble of an ice sheet model to better quantify antarctic sea level rise uncertainties. *The Cryosphere*, *15*(6), 2683–2699. Retrieved from `http://dx.doi.org/10.5194/tc-15-2683-2021` doi: 10.5194/tc-15-2683-2021

Beres, J. H., Alexander, M. J., & Holton, J. R. (2004, February). A method of specifying the gravity wave spectrum above convection based on latent heating properties and background wind. *Journal of the Atmospheric Sciences*, *61*(3), 324–337. Retrieved from `http://dx.doi.org/10.1175/1520-0469(2004)061<0324:amostg>2.0.co;2` doi: 10.1175/1520-0469(2004)061<0324:amostg>2.0.co;2

Beusch, L., Nicholls, Z., Gudmundsson, L., Hauser, M., Meinshausen, M., & Seneviratne, S. I. (2022, March). From emission scenarios to spatially resolved projections with a chain of computationally efficient emulators: coupling of magicc (v7.5.1) and mesmer (v0.8.3). *Geoscientific Model Development*, *15*(5), 2085–2103. Retrieved from `http://dx.doi.org/10.5194/gmd-15-2085-2022` doi: 10.5194/gmd-15-2085-2022

Binois, M., & Gramacy, R. B. (2021). hetGP: Heteroskedastic Gaussian process modeling and sequential design in R. *Journal of Statistical Software*, *98*(13), 1–44. doi: 10.18637/jss.v098.i13

Bishop, C. (1998). Bayesian pca. In M. Kearns, S. Solla, & D. Cohn (Eds.), *Advances in neural information processing systems* (Vol. 11). MIT Press. Retrieved from `https://proceedings.neurips.cc/paper_files/paper/1998/file/c88d8d0a6097754525e02c2246d8d27f-Paper.pdf`

Booker, J. R., & Bretherton, F. P. (1967, February). The critical layer for internal gravity waves in a shear flow. *Journal of Fluid Mechanics*, *27*(3), 513–539. Retrieved from `http://dx.doi.org/10.1017/S0022112067000515` doi: 10.1017/s0022112067000515

Bosilovich, M. G., Kennedy, J., Dee, D., Allan, R., & O'Neill, A. (2013). On the reprocessing and reanalysis of observations for climate. In *Climate science for serving society* (p. 51–71). Springer Netherlands. Retrieved from `http://dx.doi.org/10.1007/978-94-007-6692-1_3` doi: 10.1007/978-94-007-6692-1_3

Brooks, S., Gelman, A., Jones, G., & Meng, X.-L. (2011). *Handbook of markov chain monte carlo*. Chapman and Hall/CRC. Retrieved from `http://dx.doi.org/10.1201/b10905` doi: 10.1201/b10905

Brynjarsdóttir, J., & O Hagan, A. (2014, October). Learning about physical parameters: the importance of model discrepancy. *Inverse Problems*, *30*(11), 114007. Retrieved from `http://dx.doi.org/10.1088/0266-5611/30/11/114007` doi: 10.1088/0266-5611/30/11/114007

Bushell, A. C., Anstey, J. A., Butchart, N., Kawatani, Y., Osprey, S. M., Richter, J. H., … Yukimoto, S. (2020, March). Evaluation of the quasi-biennial oscillation in global climate models for the sparc qbo-initiative. *Quarterly Journal of the Royal Meteorological Society*, *148*(744), 1459–1489. Retrieved from `http://dx.doi.org/10.1002/qj.3765` doi: 10.1002/qj.3765

Cattell, R. B. (1966, April). The scree test for the number of factors. *Multivariate Behavioral Research*, *1*(2), 245–276. Retrieved from `http://dx.doi.org/10.1207/s15327906mbr0102_10` doi: 10.1207/s15327906mbr0102_10

Cheng, K., Lu, Z., Ling, C., & Zhou, S. (2020, January). Surrogate-assisted global sensitivity analysis: an overview. *Structural and Multidisciplinary Optimization*, *61*(3), 1187–1213. Retrieved from `http://dx.doi.org/10.1007/s00158-019-02413-5` doi: 10.1007/s00158-019-02413-5

Chinchuluun, A., & Pardalos, P. M. (2007, June). A survey of recent developments in multiobjective optimization. *Annals of Operations Research*, *154*(1), 29–50. Retrieved from `http://dx.doi.org/10.1007/s10479-007-0186-0` doi: 10.1007/s10479-007-0186-0

Chowdhary, K., Hoang, C., Lee, K., Ray, J., Weirs, V., & Carnes, B. (2022, November). Calibrating hypersonic turbulence flow models with the hifire-1 experiment using data-driven machine-learned models. *Computer Methods in Applied Mechanics and Engineering*, *401*, 115396. Retrieved from `http://dx.doi.org/10.1016/j.cma.2022.115396` doi: 10.1016/j.cma.2022.115396

Cleveland, W. S., & Loader, C. (1996). Smoothing by Local Regression: Principles and Methods. In W. Härdle & M. G. Schimek (Eds.), *Statistical Theory and Computational Aspects of Smoothing* (pp. 10–49). Heidelberg: Physica-Verlag HD. doi: 10.1007/978-3-642-48425-4_2

Collette, Y., & Siarry, P. (2004). *Multiobjective optimization*. Springer Berlin Heidelberg. Retrieved from `http://dx.doi.org/10.1007/978-3-662-08883-8` doi: 10.1007/978-3-662-08883-8

Coy, L., Newman, P. A., Strahan, S., & Pawson, S. (2020, September). Seasonal variation of the quasi-biennial oscillation descent. *Journal of Geophysical Research: Atmospheres*, *125*(18). Retrieved from `http://dx.doi.org/10.1029/2020JD033077` doi: 10.1029/2020jd033077

Crestaux, T., Le Maı̂tre, O., & Martinez, J.-M. (2009, July). Polynomial chaos expansion for sensitivity analysis. *Reliability Engineering & System Safety*, *94*(7), 1161–1172. Retrieved from `http://dx.doi.org/10.1016/j.ress.2008.10.008` doi: 10.1016/j.ress.2008.10.008

Dahl, D. B., Scott, D., Roosen, C., Magnusson, A., & Swinton, J. (2019). xtable: Export tables to latex or html [Computer software manual]. Retrieved from `https://CRAN.R-project.org/package=xtable` (R package version 1.8-4)

Dee, D. P., Uppala, S. M., Simmons, A. J., Berrisford, P., Poli, P., Kobayashi, S., … Vitart, F. (2011, April). The era-interim reanalysis: configuration and performance of the data assimilation system. *Quarterly Journal of the Royal Meteorological Society*, *137*(656), 553–597. Retrieved from `http://dx.doi.org/10.1002/qj.828` doi: 10.1002/qj.828

Diaz-Ibarra, O. H., Sargsyan, K., & Najm, H. N. (2025, January). Surrogate construction via weight parameterization of residual neural networks. *Computer Methods in Applied Mechanics and Engineering*, *433*, 117468. Retrieved from `http://dx.doi.org/10.1016/j.cma.2024.117468` doi: 10.1016/j.cma.2024.117468

Dunbar, O. R. A., Garbuno-Inigo, A., Schneider, T., & Stuart, A. M. (2021, September). Calibration and uncertainty quantification of convective parameters in an idealized gcm. *Journal of Advances in Modeling Earth Systems*, *13*(9). Retrieved from `http://dx.doi.org/10.1029/2020MS002454` doi: 10.1029/2020ms002454

E3SM Project, D. (2023). *Energy exascale earth system model v2.1.0.* [Computer Software] `https://doi.org/10.11578/E3SM/dc.20230110.5`. Retrieved from `https://www.osti.gov/doecode/biblio/98639` doi: 10.11578/E3SM/DC.20230110.5

Efstratiadis, A., & Koutsoyiannis, D. (2010, March). One decade of multi-objective calibration approaches in hydrological modelling: a review. *Hydrological Sciences Journal*, *55*(1), 58–78. Retrieved from `http://dx.doi.org/10.1080/02626660903526292` doi: 10.1080/02626660903526292

Eldred, M., & Dunlavy, D. (2006, September). Formulations for surrogate-based optimization with data fit, multifidelity, and reduced-order models. In *11th aiaa/issmo multidisciplinary analysis and optimization conference.* American Institute of Aeronautics and Astronautics. Retrieved from `http://dx.doi.org/10.2514/6.2006-7117` doi: 10.2514/6.2006-7117

Gabriel, K. R. (1971). The biplot graphic display of matrices with application to principal component analysis. *Biometrika*, *58*(3), 453–467. Retrieved from `http://dx.doi.org/10.1093/biomet/58.3.453` doi: 10.1093/biomet/58.3.453

Gan, Y., Duan, Q., Gong, W., Tong, C., Sun, Y., Chu, W., … Di, Z. (2014, January). A comprehensive evaluation of various sensitivity analysis methods: A case study with a hydrological model. *Environmental Modelling & Software*, *51*, 269–285. Retrieved from `http://dx.doi.org/10.1016/j.envsoft.2013.09.031` doi: 10.1016/j.envsoft.2013.09.031

Gelman, A., Carlin, J. B., Stern, H. S., Dunson, D. B., Vehtari, A., & Rubin, D. B. (2013). *Bayesian data analysis.* Chapman and Hall/CRC. Retrieved from `http://dx.doi.org/10.1201/b16018` doi: 10.1201/b16018

Genz, A., & Bretz, F. (2009). *Computation of multivariate normal and t probabilities.* Heidelberg: Springer-Verlag.

Ghanem, R. G., & Spanos, P. D. (2003). *Stochastic finite elements: A spectral approach* (Revised Edition ed.). Mineola, New York: Dover Publications.

Giorgetta, M. A., Manzini, E., & Roeckner, E. (2002, April). Forcing of the quasi-biennial oscillation from a broad spectrum of atmospheric waves. *Geophysical Research Letters*, *29*(8). Retrieved from `http://dx.doi.org/10.1029/2002GL014756` doi: 10.1029/2002gl014756

Golaz, J., Caldwell, P. M., Van Roekel, L. P., Petersen, M. R., Tang, Q., Wolfe, J. D., … Zhu, Q. (2019, July). The doe e3sm coupled model version 1: Overview and evaluation at standard resolution. *Journal of Advances in Modeling Earth Systems*, *11*(7), 2089–2129. Retrieved from `http://dx.doi.org/10.1029/2018MS001603` doi: 10.1029/2018ms001603

Golaz, J., Van Roekel, L. P., Zheng, X., Roberts, A. F., Wolfe, J. D., Lin, W., … Bader, D. C. (2022, December). The doe e3sm model version 2: Overview of the physical model and initial model evaluation. *Journal of Advances in Modeling Earth Systems*, *14*(12). Retrieved from `http://dx.doi.org/10.1029/2022MS003156` doi: 10.1029/2022ms003156

Gong, W., Duan, Q., Li, J., Wang, C., Di, Z., Ye, A., … Dai, Y. (2016, March). Multiobjective adaptive surrogate modeling-based optimization for parameter estimation of large, complex geophysical models. *Water Resources Research*, *52*(3), 1984–2008. Retrieved from `http://dx.doi.org/10.1002/2015WR018230` doi: 10.1002/2015wr018230

Gramacy, R. B. (2020). *Surrogates: Gaussian process modeling, design and optimization for the applied sciences.* Boca Raton, Florida: Chapman Hall/CRC. (`http://bobby.gramacy.com/surrogates/`)

Gray, W. M. (1984, September). Atlantic seasonal hurricane frequency. part i: El niño and 30 mb quasi-biennial oscillation influences. *Monthly Weather Review*, *112*(9), 1649–1668.

Retrieved from `http://dx.doi.org/10.1175/1520-0493(1984)112<1649:ASHFPI>2.0.CO;2` doi: 10.1175/1520-0493(1984)112<1649:ashfpi>2.0.co;2

Gunantara, N. (2018, January). A review of multi-objective optimization: Methods and its applications. *Cogent Engineering*, *5*(1), 1502242. Retrieved from `http://dx.doi.org/10.1080/23311916.2018.1502242` doi: 10.1080/23311916.2018.1502242

Hasebe, F. (1994, March). Quasi-biennial oscillations of ozone and diabatic circulation in the equatorial stratosphere. *Journal of the Atmospheric Sciences*, *51*(5), 729–745. Retrieved from `http://dx.doi.org/10.1175/1520-0469(1994)051<0729:QBOOOA>2.0.CO;2` doi: 10.1175/1520-0469(1994)051<0729:qboooa>2.0.co;2

Henry, L., Wickham, H., & Chang, W. (2024). ggstance: Horizontal 'ggplot2' components [Computer software manual]. Retrieved from `https://CRAN.R-project.org/package=ggstance` (R package version 0.3.7)

Hersbach, H., Bell, B., Berrisford, P., Hirahara, S., Horányi, A., Muñoz-Sabater, J., … others (2017). Complete era5 from 1940: Fifth generation of ecmwf atmospheric reanalyses of the global climate. *Copernicus Climate Change Service (C3S) Data Store (CDS)*. doi: 10.24381/cds.143582cf

Higdon, D., Gattiker, J., Williams, B., & Rightley, M. (2008, June). Computer model calibration using high-dimensional output. *Journal of the American Statistical Association*, *103*(482), 570–583. Retrieved from `http://dx.doi.org/10.1198/016214507000000888` doi: 10.1198/016214507000000888

Hirons, L. C., Inness, P., Vitart, F., & Bechtold, P. (2012, November). Understanding advances in the simulation of intraseasonal variability in the ecmwf model. part ii: The application of process-based diagnostics. *Quarterly Journal of the Royal Meteorological Society*, *139*(675), 1427–1444. Retrieved from `http://dx.doi.org/10.1002/qj.2059` doi: 10.1002/qj.2059

Holton, J. R., & Lindzen, R. S. (1972, September). An updated theory for the quasi-biennial cycle of the tropical stratosphere. *Journal of the Atmospheric Sciences*, *29*(6), 1076–1080. Retrieved from `http://dx.doi.org/10.1175/1520-0469(1972)029<1076:AUTFTQ>2.0.CO;2` doi: 10.1175/1520-0469(1972)029<1076:autftq>2.0.co;2

Hourdin, F., Mauritsen, T., Gettelman, A., Golaz, J.-C., Balaji, V., Duan, Q., … Williamson, D. (2017, March). The art and science of climate model tuning. *Bulletin of the American Meteorological Society*, *98*(3), 589–602. Retrieved from `http://dx.doi.org/10.1175/BAMS-D-15-00135.1` doi: 10.1175/bams-d-15-00135.1

Iooss, B., & Lemaître, P. (2015). A review on global sensitivity analysis methods. In *Uncertainty management in simulation-optimization of complex systems* (p. 101–122). Springer US. Retrieved from `http://dx.doi.org/10.1007/978-1-4899-7547-8_5` doi: 10.1007/978-1-4899-7547-8_5

Iooss, B., Veiga, S. D., Janon, A., Pujol, G., with contributions from Baptiste Broto, Boumhaout, K., … Weber, F. (2024). sensitivity: Global sensitivity analysis of model outputs and importance measures [Computer software manual]. Retrieved from `https://CRAN.R-project.org/package=sensitivity` (R package version 1.30.0)

Jansen, M. J. (1999, March). Analysis of variance designs for model output. *Computer Physics Communications*, *117*(1–2), 35–43. Retrieved from `http://dx.doi.org/10.1016/S0010-4655(98)00154-4` doi: 10.1016/s0010-4655(98)00154-4

Jones, D. R., Schonlau, M., & Welch, W. J. (1998). *Journal of Global Optimization*, *13*(4), 455–492. Retrieved from `http://dx.doi.org/10.1023/A:1008306431147` doi: 10.1023/a:1008306431147

Kang, S., Li, K., & Wang, R. (2024, September). A survey on pareto front learning for multi-objective optimization. *Journal of Membrane Computing*. Retrieved from `http://dx.doi.org/10.1007/s41965-024-00170-z` doi: 10.1007/s41965-024-00170-z

Karhunen, K. (1946). *Zur spektraltheorie stochastischer prozesse* (Vol. 34). Helsinki: Suomalainen tiedeakatemia.

Kennedy, M. C., & O'Hagan, A. (2001, September). Bayesian calibration of computer models. *Journal of the Royal Statistical Society Series B: Statistical Methodology*,

*63*(3), 425–464. Retrieved from `http://dx.doi.org/10.1111/1467-9868.00294` doi: 10.1111/1467-9868.00294

Langenbrunner, B., & Neelin, J. D. (2017, September). Multiobjective constraints for climate model parameter choices: Pragmatic<scp>p</scp>areto fronts in cesm1. *Journal of Advances in Modeling Earth Systems*, *9*(5), 2008–2026. Retrieved from `http://dx.doi.org/10.1002/2017MS000942` doi: 10.1002/2017ms000942

Lee, D. T., & Schachter, B. J. (1980, June). Two algorithms for constructing a delaunay triangulation. *International Journal of Computer 'I&' Information Sciences*, *9*(3), 219–242. Retrieved from `http://dx.doi.org/10.1007/BF00977785` doi: 10.1007/bf00977785

Li, C., & Yanai, M. (1996, February). The onset and interannual variability of the asian summer monsoon in relation to land–sea thermal contrast. *Journal of Climate*, *9*(2), 358–375. Retrieved from `http://dx.doi.org/10.1175/1520-0442(1996)009<0358:TOAIVO>2.0.CO;2` doi: 10.1175/1520-0442(1996)009<0358:toaivo>2.0.co;2

Li, Y., Chen, C., Benedict, J. J., Huang, K., Richter, J. H., & Bacmeister, J. (2025, January). Mechanisms in regulating the quasi-biennial oscillation in exascale earth system model version 2. *Journal of Geophysical Research: Atmospheres*, *130*(3). Retrieved from `http://dx.doi.org/10.1029/2024JD041868` doi: 10.1029/2024jd041868

Li, Y., Richter, J. H., Chen, C., & Tang, Q. (2023, August). A strengthened teleconnection of the quasi-biennial oscillation and tropical easterly jet in the past decades in e3smv1. *Geophysical Research Letters*, *50*(15). Retrieved from `http://dx.doi.org/10.1029/2023GL104517` doi: 10.1029/2023gl104517

Lindzen, R. S. (1987, April). On the development of the theory of the qbo. *Bulletin of the American Meteorological Society*, *68*(4), 329–337. Retrieved from `http://dx.doi.org/10.1175/1520-0477(1987)068<0329:OTDOTT>2.0.CO;2` doi: 10.1175/1520-0477(1987)068<0329:otdott>2.0.co;2

Lindzen, R. S., & Holton, J. R. (1968, November). A theory of the quasi-biennial oscillation. *Journal of the Atmospheric Sciences*, *25*(6), 1095–1107. Retrieved from `http://dx.doi.org/10.1175/1520-0469(1968)025<1095:ATOTQB>2.0.CO;2` doi: 10.1175/1520-0469(1968)025<1095:atotqb>2.0.co;2

Liu, H., Li, Y., Duan, Z., & Chen, C. (2020, November). A review on multi-objective optimization framework in wind energy forecasting techniques and applications. *Energy Conversion and Management*, *224*, 113324. Retrieved from `http://dx.doi.org/10.1016/j.enconman.2020.113324` doi: 10.1016/j.enconman.2020.113324

Liu, H., Ong, Y.-S., Shen, X., & Cai, J. (2020, November). When gaussian process meets big data: A review of scalable gps. *IEEE Transactions on Neural Networks and Learning Systems*, *31*(11), 4405–4423. Retrieved from `http://dx.doi.org/10.1109/TNNLS.2019.2957109` doi: 10.1109/tnnls.2019.2957109

Loève, M. (1963). *Probability theory.* New York: D. Van Nostrand Company Inc.

Mai, J. (2023, May). Ten strategies towards successful calibration of environmental models. *Journal of Hydrology*, *620*, 129414. Retrieved from `http://dx.doi.org/10.1016/j.jhydrol.2023.129414` doi: 10.1016/j.jhydrol.2023.129414

McKay, M. D., Beckman, R. J., & Conover, W. J. (1979, May). A comparison of three methods for selecting values of input variables in the analysis of output from a computer code. *Technometrics*, *21*(2), 239. Retrieved from `http://dx.doi.org/10.2307/1268522` doi: 10.2307/1268522

Mueller, J. N., Sargsyan, K., Daniels, C. J., & Najm, H. N. (2025, January). Polynomial chaos surrogate construction for random fields with parametric uncertainty. *SIAM/ASA Journal on Uncertainty Quantification*, *13*(1), 1–29. Retrieved from `http://dx.doi.org/10.1137/23M1613505` doi: 10.1137/23m1613505

Naoe, H., & Yoshida, K. (2019, July). Influence of quasi-biennial oscillation on the boreal winter extratropical stratosphere in qboi experiments. *Quarterly Journal of the Royal Meteorological Society*, *145*(723), 2755–2771. Retrieved from `http://dx.doi.org/10.1002/qj.3591` doi: 10.1002/qj.3591

Naujokat, B. (1986, September). An update of the observed quasi-biennial oscillation of the stratospheric winds over the tropics. *Journal of the Atmospheric Sciences*, *43*(17), 1873–1877. Retrieved from `http://dx.doi.org/10.1175/1520-0469(1986)043<1873:AUOTOQ>2.0.CO;2` doi: 10.1175/1520-0469(1986)043<1873:auotoq>2.0.co; 2

Neuwirth, E. (2022). Rcolorbrewer: Colorbrewer palettes [Computer software manual]. Retrieved from `https://CRAN.R-project.org/package=RColorBrewer` (R package version 1.1-3)

Nocedal, J., & Wright, S. J. (2006). *Numerical optimization*. Springer. Retrieved from `https://doi.org/10.1007/978-0-387-40065-5` doi: 10.1007/978-0-387-40065-5

Oberkampf, W. L., & Roy, C. J. (2010). *Verification and validation in scientific computing*. Cambridge University Press. Retrieved from `http://dx.doi.org/10.1017/CBO9780511760396` doi: 10.1017/cbo9780511760396

Ooms, J. (2014). The jsonlite package: A practical and consistent mapping between json data and r objects. *arXiv:1403.2805 [stat.CO]*. Retrieved from `https://arxiv.org/abs/1403.2805`

Osprey, S. M., Butchart, N., Knight, J. R., Scaife, A. A., Hamilton, K., Anstey, J. A., … Zhang, C. (2016, September). An unexpected disruption of the atmospheric quasi-biennial oscillation. *Science*, *353*(6306), 1424–1427. Retrieved from `http://dx.doi.org/10.1126/science.aah4156` doi: 10.1126/science.aah4156

Paradis, E., & Schliep, K. (2019). ape 5.0: an environment for modern phylogenetics and evolutionary analyses in R. *Bioinformatics*, *35*, 526-528. doi: 10.1093/bioinformatics/bty633

Parker, W. S. (2016, September). Reanalyses and observations: What's the difference? *Bulletin of the American Meteorological Society*, *97*(9), 1565–1572. Retrieved from `http://dx.doi.org/10.1175/BAMS-D-14-00226.1` doi: 10.1175/bams-d-14-00226 .1

Pascoe, C. L., Gray, L. J., Crooks, S. A., Juckes, M. N., & Baldwin, M. P. (2005, April). The quasi-biennial oscillation: Analysis using era-40 data. *Journal of Geophysical Research: Atmospheres*, *110*(D8). Retrieved from `http://dx.doi.org/10.1029/2004JD004941` doi: 10.1029/2004jd004941

Pedersen, T. L. (2024). patchwork: The composer of plots [Computer software manual]. Retrieved from `https://CRAN.R-project.org/package=patchwork` (R package version 1.2.0)

Pierce, D. (2023). ncdf4: Interface to unidata netcdf (version 4 or earlier) format data files [Computer software manual]. Retrieved from `https://CRAN.R-project.org/package=ncdf4` (R package version 1.22)

Piironen, J., & Vehtari, A. (2016, September). Projection predictive model selection for gaussian processes. In *2016 ieee 26th international workshop on machine learning for signal processing (mlsp)* (p. 1–6). IEEE. Retrieved from `http://dx.doi.org/10.1109/MLSP.2016.7738829` doi: 10.1109/mlsp.2016.7738829

Plumb, R. A. (1977, December). The interaction of two internal waves with the mean flow: Implications for the theory of the quasi-biennial oscillation. *Journal of the Atmospheric Sciences*, *34*(12), 1847–1858. Retrieved from `http://dx.doi.org/10.1175/1520-0469(1977)034<1847:TIOTIW>2.0.CO;2` doi: 10.1175/1520-0469(1977) 034<1847:tiotiw>2.0.co;2

R Core Team. (2024). R: A language and environment for statistical computing [Computer software manual]. Vienna, Austria. Retrieved from `https://www.R-project.org/`

Rasmussen, C. E., & Williams, C. K. I. (2005). *Gaussian processes for machine learning*. The MIT Press. Retrieved from `http://dx.doi.org/10.7551/mitpress/3206.001.0001` doi: 10.7551/mitpress/3206.001.0001

Richter, J. H., Anstey, J. A., Butchart, N., Kawatani, Y., Meehl, G. A., Osprey, S., & Simpson, I. R. (2020, April). Progress in simulating the quasi-biennial oscillation in cmip models. *Journal of Geophysical Research: Atmospheres*, *125*(8). Retrieved from `http://dx.doi.org/10.1029/2019JD032362` doi: 10.1029/2019jd032362

Richter, J. H., Chen, C., Tang, Q., Xie, S., & Rasch, P. J. (2019, November). Improved simulation of the qbo in e3smv1. *Journal of Advances in Modeling Earth Systems*, *11*(11), 3403–3418. Retrieved from `http://dx.doi.org/10.1029/2019MS001763` doi: 10.1029/2019ms001763

Roy, O., & Vetterli, M. (2007). The effective rank: A measure of effective dimensionality. In *2007 15th european signal processing conference* (p. 606-610).

Ruzika, S., & Wiecek, M. M. (2005, September). Approximation methods in multiobjective programming. *Journal of Optimization Theory and Applications*, *126*(3), 473–501. Retrieved from `http://dx.doi.org/10.1007/s10957-005-5494-4` doi: 10.1007/s10957-005-5494-4

Saltelli, A., Annoni, P., Azzini, I., Campolongo, F., Ratto, M., & Tarantola, S. (2010, February). Variance based sensitivity analysis of model output. design and estimator for the total sensitivity index. *Computer Physics Communications*, *181*(2), 259–270. Retrieved from `http://dx.doi.org/10.1016/j.cpc.2009.09.018` doi: 10.1016/j.cpc.2009.09.018

Saltelli, A., Ratto, M., Tarantola, S., & Campolongo, F. (2006, October). Sensitivity analysis practices: Strategies for model-based inference. *Reliability Engineering & System Safety*, *91*(10–11), 1109–1125. Retrieved from `http://dx.doi.org/10.1016/j.ress.2005.11.014` doi: 10.1016/j.ress.2005.11.014

Santner, T. J., Williams, B. J., & Notz, W. I. (2018). *The design and analysis of computer experiments*. Springer New York. Retrieved from `http://dx.doi.org/10.1007/978-1-4939-8847-1` doi: 10.1007/978-1-4939-8847-1

Sargsyan, K. (2015). Surrogate models for uncertainty propagation and sensitivity analysis. In *Handbook of uncertainty quantification* (p. 1–26). Springer International Publishing. Retrieved from `http://dx.doi.org/10.1007/978-3-319-11259-6_22-1` doi: 10.1007/978-3-319-11259-6_22-1

Sargsyan, K., Huan, X., & Najm, H. N. (2019). Embedded model error representation for bayesian model calibration. *International Journal for Uncertainty Quantification*, *9*(4), 365–394. Retrieved from `http://dx.doi.org/10.1615/Int.J.UncertaintyQuantification.2019027384` doi: 10.1615/int.j.uncertaintyquantification.2019027384

Sargsyan, K., Safta, C., Najm, H. N., Debusschere, B. J., Ricciuto, D., & Thornton, P. (2014). Dimensionality reduction for complex models via bayesian compressive sensing. *International Journal for Uncertainty Quantification*, *4*(1), 63–93. Retrieved from `http://dx.doi.org/10.1615/Int.J.UncertaintyQuantification.2013006821` doi: 10.1615/int.j.uncertaintyquantification.2013006821

Schloerke, B., Cook, D., Larmarange, J., Briatte, F., Marbach, M., Thoen, E., … Crowley, J. (2024). Ggally: Extension to 'ggplot2' [Computer software manual]. Retrieved from `https://CRAN.R-project.org/package=GGally` (R package version 2.2.1)

Schmidt, G. A., Bader, D., Donner, L. J., Elsaesser, G. S., Golaz, J.-C., Hannay, C., … Saha, S. (2017, September). Practice and philosophy of climate model tuning across six us modeling centers. *Geoscientific Model Development*, *10*(9), 3207–3223. Retrieved from `http://dx.doi.org/10.5194/gmd-10-3207-2017` doi: 10.5194/gmd-10-3207-2017

Shang, H. L. (2013, April). A survey of functional principal component analysis. *AStA Advances in Statistical Analysis*, *98*(2), 121–142. Retrieved from `http://dx.doi.org/10.1007/s10182-013-0213-1` doi: 10.1007/s10182-013-0213-1

Sharma, S., & Kumar, V. (2022, July). A comprehensive review on multi-objective optimization techniques: Past, present and future. *Archives of Computational Methods in Engineering*, *29*(7), 5605–5633. Retrieved from `http://dx.doi.org/10.1007/s11831-022-09778-9` doi: 10.1007/s11831-022-09778-9

Slowikowski, K. (2024). ggrepel: Automatically position non-overlapping text labels with 'ggplot2' [Computer software manual]. Retrieved from `https://CRAN.R-project.org/package=ggrepel` (R package version 0.9.5)

Smith, A. K., Holt, L. A., Garcia, R. R., Anstey, J. A., Serva, F., Butchart, N., … Yoshida, K. (2020, January). The equatorial stratospheric semiannual oscillation and time-mean winds in qboi models. *Quarterly Journal of the Royal Meteorological Society*, *148*(744), 1593–1609. Retrieved from `http://dx.doi.org/10.1002/qj.3690` doi: 10.1002/qj.3690

Smith, R. C. (2024). *Uncertainty quantification: Theory, implementation, and applications, second edition.* Society for Industrial and Applied Mathematics. Retrieved from `http://dx.doi.org/10.1137/1.9781611977844` doi: 10.1137/1.9781611977844

Sobol , I. (2001, February). Global sensitivity indices for nonlinear mathematical models and their monte carlo estimates. *Mathematics and Computers in Simulation*, *55*(1–3), 271–280. Retrieved from `http://dx.doi.org/10.1016/S0378-4754(00)00270-6` doi: 10.1016/s0378-4754(00)00270-6

Solymos, P., & Zawadzki, Z. (2023). pbapply: Adding progress bar to '*apply' functions [Computer software manual]. Retrieved from `https://CRAN.R-project.org/package=pbapply` (R package version 1.7-2)

Stein, M. (1987, May). Large sample properties of simulations using latin hypercube sampling. *Technometrics*, *29*(2), 143–151. Retrieved from `http://dx.doi.org/10.1080/00401706.1987.10488205` doi: 10.1080/00401706.1987.10488205

Sudret, B. (2008, July). Global sensitivity analysis using polynomial chaos expansions. *Reliability Engineering & System Safety*, *93*(7), 964–979. Retrieved from `http://dx.doi.org/10.1016/j.ress.2007.04.002` doi: 10.1016/j.ress.2007.04.002

Sülflow, A., Drechsler, N., & Drechsler, R. (2007). Robust multi-objective optimization in high dimensional spaces. In S. Obayashi, K. Deb, C. Poloni, T. Hiroyasu, & T. Murata (Eds.), *Evolutionary multi-criterion optimization* (pp. 715–726). Berlin, Heidelberg: Springer Berlin Heidelberg. Retrieved from `http://dx.doi.org/10.1007/978-3-540-70928-2_54` doi: 10.1007/978-3-540-70928-2_54

Sun, R., Duan, Q., & Huo, X. (2021, November). Multi-objective adaptive surrogate modeling-based optimization for distributed environmental models based on grid sampling. *Water Resources Research*, *57*(11). Retrieved from `http://dx.doi.org/10.1029/2020WR028740` doi: 10.1029/2020wr028740

Sundararajan, S., & Keerthi, S. S. (2001, May). Predictive approaches for choosing hyperparameters in gaussian processes. *Neural Computation*, *13*(5), 1103–1118. Retrieved from `http://dx.doi.org/10.1162/08997660151134343` doi: 10.1162/08997660151134343

Tiedtke, M. (1989, August). A comprehensive mass flux scheme for cumulus parameterization in large-scale models. *Monthly Weather Review*, *117*(8), 1779–1800. Retrieved from `http://dx.doi.org/10.1175/1520-0493(1989)117<1779:ACMFSF>2.0.CO;2` doi: 10.1175/1520-0493(1989)117<1779:acmfsf>2.0.co;2

Timmermann, A., An, S.-I., Kug, J.-S., Jin, F.-F., Cai, W., Capotondi, A., … Zhang, X. (2018, July). El niño–southern oscillation complexity. *Nature*, *559*(7715), 535–545. Retrieved from `http://dx.doi.org/10.1038/s41586-018-0252-6` doi: 10.1038/s41586-018-0252-6

Torrence, C., & Compo, G. P. (1998, January). A practical guide to wavelet analysis. *Bulletin of the American Meteorological Society*, *79*(1), 61–78. Retrieved from `http://dx.doi.org/10.1175/1520-0477(1998)079<0061:APGTWA>2.0.CO;2` doi: 10.1175/1520-0477(1998)079<0061:apgtwa>2.0.co;2

Turner, R. (2024). deldir: Delaunay triangulation and dirichlet (voronoi) tessellation [Computer software manual]. Retrieved from `https://CRAN.R-project.org/package=deldir` (R package version 2.0-4)

United States Committee on Extension to the Standard Atmosphere. (1976). *U.s. standard atmosphere, 1976* (Tech. Rep.). National Oceanic and Atmospheric Administration (NOAA). Retrieved from `https://ntrs.nasa.gov/citations/19770009539` (NOAA-S/T-76-1562)

van den Brand, T. (2024). ggh4x: Hacks for 'ggplot2' [Computer software manual]. Retrieved from `https://CRAN.R-project.org/package=ggh4x` (R package version 0.2.8)

van der Maaten, L., & Hinton, G. (2008). Visualizing data using t-sne. *Journal of Machine Learning Research*, *9*(86), 2579–2605.

Vehtari, A., Mononen, T., Tolvanen, V., Sivula, T., & Winther, O. (2016, January). Bayesian leave-one-out cross-validation approximations for gaussian latent variable models. *J. Mach. Learn. Res.*, *17*(1), 3581–3618.

Vu, V. Q., & Friendly, M. (2024). ggbiplot: A grammar of graphics implementation of biplots [Computer software manual]. Retrieved from `https://CRAN.R-project.org/package=ggbiplot` (R package version 0.6.2)

Wahba, G. (1990). *Spline models for observational data.* Society for Industrial and Applied Mathematics. Retrieved from `http://dx.doi.org/10.1137/1.9781611970128` doi: 10.1137/1.9781611970128

Wallace, J. M., & Kousky, V. E. (1968, September). Observational evidence of kelvin waves in the tropical stratosphere. *Journal of the Atmospheric Sciences*, *25*(5), 900–907. Retrieved from `http://dx.doi.org/10.1175/1520-0469(1968)025<0900:OEOKWI>2.0.CO;2` doi: 10.1175/1520-0469(1968)025<0900:oeokwi>2.0.co;2

Wang, K. A., Pleiss, G., Gardner, J. R., Tyree, S., Weinberger, K. Q., & Wilson, A. G. (2019). Exact gaussian processes on a million data points. In *Proceedings of the 33rd international conference on neural information processing systems.* Red Hook, NY, USA: Curran Associates Inc. Retrieved from `https://dl.acm.org/doi/10.5555/3454287.3455599`

Wang, Y., Rao, J., Lu, Y., Ju, Z., Yang, J., & Luo, J. (2023, October). A revisit and comparison of the quasi-biennial oscillation (qbo) disruption events in 2015/16 and 2019/20. *Atmospheric Research*, *294*, 106970. Retrieved from `http://dx.doi.org/10.1016/j.atmosres.2023.106970` doi: 10.1016/j.atmosres.2023.106970

Watanabe, S., Hamilton, K., Osprey, S., Kawatani, Y., & Nishimoto, E. (2018, February). First successful hindcasts of the 2016 disruption of the stratospheric quasi-biennial oscillation. *Geophysical Research Letters*, *45*(3), 1602–1610. Retrieved from `http://dx.doi.org/10.1002/2017GL076406` doi: 10.1002/2017gl076406

White, H. (1982, January). Maximum likelihood estimation of misspecified models. *Econometrica*, *50*(1), 1. Retrieved from `http://dx.doi.org/10.2307/1912526` doi: 10.2307/1912526

Wickham, H. (2016). *ggplot2: Elegant graphics for data analysis.* Springer-Verlag New York. Retrieved from `https://ggplot2.tidyverse.org`

Wickham, H., Pedersen, T. L., & Seidel, D. (2023). scales: Scale functions for visualization [Computer software manual]. Retrieved from `https://CRAN.R-project.org/package=scales` (R package version 1.3.0)

Wu, L., Su, H., Zeng, X., Posselt, D. J., Wong, S., Chen, S., & Stoffelen, A. (2024, February). Uncertainty of atmospheric winds in three widely used global reanalysis datasets. *Journal of Applied Meteorology and Climatology*, *63*(2), 165–180. Retrieved from `http://dx.doi.org/10.1175/JAMC-D-22-0198.1` doi: 10.1175/jamc-d-22-0198.1

Xie, S., & Coauthors. (2025). The energy exascale earth system model version 3. part i: Overview of the atmospheric component. *Journal of Advances in Modeling Earth Systems.*

Xie, S., Lin, W., Rasch, P. J., Ma, P., Neale, R., Larson, V. E., … Zhang, Y. (2018, October). Understanding cloud and convective characteristics in version 1 of the e3sm atmosphere model. *Journal of Advances in Modeling Earth Systems*, *10*(10), 2618–2644. Retrieved from `http://dx.doi.org/10.1029/2018MS001350` doi: 10.1029/2018ms001350

Yarger, D., Wagman, B. M., Chowdhary, K., & Shand, L. (2024, April). Autocalibration of the e3sm version 2 atmosphere model using a pca-based surrogate for spatial fields. *Journal of Advances in Modeling Earth Systems*, *16*(4). Retrieved from `http://dx.doi.org/10.1029/2023MS003961` doi: 10.1029/2023ms003961

Yu, W., Hannah, W. M., Benedict, J. J., Chen, C.-C., & Richter, J. H. (2024, June). Improving the qbo forcing by resolved waves with vertical grid refinement in e3smv2. *ESS Open Archive.* Retrieved from `http://dx.doi.org/10.22541/essoar.171781265.56480322/v1` doi: 10.22541/essoar.171781265.56480322/v1

Zhang, C., Golaz, J.-C., Forsyth, R., Vo, T., Xie, S., Shaheen, Z., … Ullrich, P. A. (2022, December). The e3sm diagnostics package (e3sm diags v2.7): a python-based diagnostics package for earth system model evaluation. *Geoscientific Model Development*, *15*(24), 9031–9056. Retrieved from `http://dx.doi.org/10.5194/gmd-15-9031-2022`  doi: 10.5194/gmd-15-9031-2022

Zhang, G., & McFarlane, N. A. (1995, September). Sensitivity of climate simulations to the parameterization of cumulus convection in the canadian climate centre general circulation model. *Atmosphere-Ocean*, *33*(3), 407–446. Retrieved from `http://dx.doi.org/10.1080/07055900.1995.9649539`  doi: 10.1080/07055900.1995.9649539

# Supporting Information for "Improving the quasi-biennial oscillation via a surrogate-accelerated multi-objective optimization"

Luis Damiano[1], Walter Hannah[2], Chih-Chieh Chen[3], James J. Benedict[4],

Khachik Sargsyan[5], Bert J. Debusschere[5], Michael S. Eldred[1]

[1]Optimization & Uncertainty Quantification, Sandia National Laboratories, NM, USA

[2]Atmospheric, Earth & Energy Science, Lawrence Livermore National Laboratory, CA, USA

[3]Climate and Global Dynamics Laboratory, National Center for Atmospheric Research, CO, USA

[4]Fluid Dynamics and Solid Mechanics Group, Los Alamos National Laboratory, NM, USA

[5]Plasma & Reacting Flow Science, Sandia National Laboratories, CA, USA

## 1.  Surrogate correlation functions

The correlation function provides us with a mechanism to explicitly state the prior assumptions regarding the smoothness of $f_j : \mathcal{X} \to \mathcal{Z}_j$, the unknown mapping from the physics parameters to the spectral features. Numerous functional forms for the correlation, which can further be combined to create new forms, have been proposed (Duvenaud, 2014). In our application, we select a form for $r_j(\mathbf{x}, \mathbf{x}')$ from a small set of candidates

---

Corresponding author: Luis Damiano (ladamia@sandia.gov)

via a leave-one-out cross-validation study. Namely, we consider the squared exponential (SE), Matérn 5/2, and Matérn 3/2 forms:

$$r_{\text{SE}}(\mathbf{x}, \mathbf{x}') = \prod_{p=1}^{P} \exp\left(-\frac{(x_p - x'_p)^2}{2\sigma_{x_{pj}}{}^2}\right),$$

$$r_{\text{M(5/2)}}(\mathbf{x}, \mathbf{x}') = \prod_{p=1}^{P} \left(1 + \frac{\sqrt{5}\,(x_p - x'_p)}{\sigma_{x_{pj}}} + \frac{5(x_p - x'_p)^2}{3\sigma_{x_{pj}}{}^2}\right) \exp\left(-\frac{\sqrt{5}(x_p - x'_p)}{\sigma_{x_{pj}}}\right),$$

$$r_{\text{M(3/2)}}(\mathbf{x}, \mathbf{x}') = \prod_{p=1}^{P} \left(1 + \frac{\sqrt{3}\,(x_p - x'_p)}{\sigma_{x_{pj}}}\right) \exp\left(-\frac{\sqrt{3}\,(x_p - x'_p)}{\sigma_{x_{pj}}}\right),$$

where $\mathbf{x}, \mathbf{x}' \in \mathcal{X}$ are two locations in the design space, $x_p, x'_p$ are the values for the $p$-th parameter at said locations, and $\sigma_{x_{pj}} > 0$ is the length-scale hyperparameter for the the $p$-th physics parameter in the $j$-th correlation function.

The forms were selected due to their distinct characteristics and flexibility in accommodating different degrees of smoothness of $f_j$. The SE is infinite differentiability, indicating that functions sampled from the GP exhibit a high degree of smoothness. The Matérn 5/2 possesses continuous first and second derivatives, thus providing a balance between smoothness and flexibility. Finally, the Matérn 3/2 has continuous first derivatives but lacks continuous second derivatives, making it suitable for functions that exhibit some degree of roughness or discontinuity.

The candidates belong to the family of the often called automatic relevance determination (ARD) priors (Neal, 1996), where separate length-scales $\sigma_{x_{pj}}$ enable the latent function $f_j$ to vary at different speeds with respect to different inputs $x_p$. ARD introduces multiple regularizations as the marginal likelihood favors solutions with large length-scales for those inputs along which the latent function is flat, a mechanism for Bayesian Occam's razor put in place to prevent significant overfitting by pruning multi-dimensional inputs toward sparse representations (MacKay, 1994). Length scales are sometimes known as correlation lengths (Santner et al., 2018) or ranges (Cressie, 1993), and their inverse as roughness (Kennedy & O'Hagan, 2001).

The inverse of the length-scale has sometimes been informally interpreted as a measure of feature relevance. A smaller length-scale indicates that points must be in close proximity to exhibit significant correlation, the predicted output responds more sensitively to small changes in the input variables, and thus the feature is considered more relevant. The converse also follows. However, it has been warned that the length-scales are also affected by other intrinsic characteristics of the data such as the output responding linearly or non-linearly to changes in the input variable (Piironen & Vehtari, 2016). In our application, we mitigate the scale-effect by normalizing the physics parameters to $[-1, 1]^P$, rendering the length-scales comparable across the physics parameters.

## 2. Surrogate-predicted QoIs

We present a step-by-step guide to obtain the predictive distribution of the QoI vector $\hat{\mathbf{q}}$ for a set of parameter values $\mathbf{x}_*$. Recall that $\mathbf{X}$, $\mathbf{Q}$, and $\mathbf{Z}$ are, respectively, the matrices with the physics parameters, QoIs, and the spectral features for the simulations in the ensemble. Define $\mathbf{c}$ and $\mathbf{s}$ the vectors with the means and standard deviations across the columns of $\mathbf{Q}$, and $\mathbf{A} = \mathbf{V}^{-1}$ where the columns of $\mathbf{V}$ correspond to the eigenvectors of the sample correlation matrix. Denote by $S(\mathbf{x}, \mathbf{x}')$ the covariance matrix.

The recipe can be summarized in three steps. First, calculate the prediction mean $\mathrm{E}\langle \hat{z}_j \rangle$ and variance $\mathrm{V}\langle \hat{z}_j \rangle$ for the $\tilde{J} \leq J$ spectral features (independent normal). Second, calculate the prediction mean vector $\mathbf{m}_{\hat{\mathbf{q}}_0}$ and covariance matrix $\mathbf{S}_{\hat{\mathbf{q}}_0}$ for the zero-mean, unit-variance quantities $\hat{\mathbf{q}}_0$ by reversing the KLE (multivariate normal). Third, calculate the prediction mean vector $\mathbf{m}_{\hat{\mathbf{q}}}$ and covariance matrix $\mathbf{S}_{\hat{\mathbf{q}}}$ for the QoIs $\hat{\mathbf{q}}$ by reversing the scaling across columns (multivariate normal).

**Step 1**. The predictive distributions for the $\tilde{J}$ spectral features are independent normal given by

$$\hat{z}_j | \mathbf{x}_* \sim \mathrm{N}(\mathrm{E}\langle \hat{z}_j \rangle, \mathrm{V}\langle \hat{z}_j \rangle) \tag{1}$$

$$\mathrm{E}\langle \hat{z}_j \rangle = S(\mathbf{x}_*, \mathbf{X})\Big[ S(\mathbf{X}, \mathbf{X}) + \sigma_{\varepsilon_j}^2 \mathbf{I} \Big]^{-1} \mathbf{z}_j \tag{2}$$

$$\mathrm{V}\langle \hat{z}_j \rangle = S(\mathbf{x}_*, \mathbf{x}_*) - S(\mathbf{x}_*, \mathbf{X})\Big[ S(\mathbf{X}, \mathbf{X}) + \sigma_{\varepsilon_j}^2 \mathbf{I} \Big]^{-1} S(\mathbf{X}, \mathbf{x}_*), \tag{3}$$

and the joint predictive distributions for the $\tilde{J}$ spectral features is multivariate normal with mean vector and diagonal covariance matrix as follows

$$\hat{\mathbf{z}} | \mathbf{x}_* \sim \mathrm{MVN}(\mathbf{m}_{\hat{z}}, \mathbf{S}_{\hat{z}}) \tag{4}$$

$$\mathbf{m}_{\hat{z}} = (\mathrm{E}\langle \hat{z}_1 \rangle, \ldots, \mathrm{E}\langle \hat{z}_J \rangle) \tag{5}$$

$$\mathbf{S}_{\hat{z}} = \mathrm{diag}\left( \mathrm{V}\langle \hat{z}_1 \rangle, \ldots, \mathrm{V}\langle \hat{z}_J \rangle \right). \tag{6}$$

**Step 2**. The zero-mean, unit-variance quantities $\hat{\mathbf{q}}_0$ are also multivariate normal with mean vector and covariance matrix as follows

$$\hat{\mathbf{q}}_0 | \mathbf{x}_* \sim \mathrm{MVN}(\mathbf{m}_{\hat{\mathbf{q}}_0}, \mathbf{S}_{\hat{\mathbf{q}}_0}) \tag{7}$$

$$\mathbf{m}_{\hat{\mathbf{q}}_0} = \mathbf{A}\mathbf{m}_{\hat{z}} \tag{8}$$

$$\mathbf{S}_{\hat{\mathbf{q}}_0} = \mathbf{A}\mathbf{S}_{\hat{z}}\mathbf{A}^\top. \tag{9}$$

**Step 3**. The QoIs $\hat{\mathbf{q}}$ can be retrieved by column-wise rescaling using the empirical mean $\mathbf{c}$ and scale $\mathbf{s}$

$$\hat{\mathbf{q}} | \mathbf{x}_* \sim \mathrm{MVN}(\mathbf{m}_{\hat{\mathbf{q}}}, \mathbf{S}_{\hat{\mathbf{q}}}) \tag{10}$$

$$\mathbf{m}_{\hat{\mathbf{q}}} = (\mathbf{m}_{\hat{\mathbf{q}}_0} + \mathbf{c}) \odot \mathbf{s} \tag{11}$$

$$\mathbf{S}_{\hat{\mathbf{q}}} = \mathrm{diag}(\mathbf{s})\mathbf{S}_{\hat{\mathbf{q}}_0}\,\mathrm{diag}(\mathbf{s}). \tag{12}$$

## 3.    FFM fit to the reference data

Figure S1 shows the reference data, the signal, and the 80% interval for the noisy observables. The sinusoidal waves capture the global pattern, while local deviations from the signal correspond to high-frequency patterns that we want to separate from the QBO. Allowing for a free amplitude and east-west (E-W) wind biases are reasonable assumptions, as the vertical dilation and shift in the observations and fitted lines vary across the facets. We observe that some fitted lines might temporarily desynchronize due to the constant period. Although the periods vary slightly locally, the FFM fits a global period across all pressure levels and cycles, seemingly finding an average period across all 6×4 cycles.

Figure S2 shows the estimated power spectral density. The data display significant spectral densities at 30 months consistently across all pressure levels. Removing the signal spectrum, which is analogous to a pulse function around $\hat{\tau} = 30.1$, uncovers several significant estimates at higher frequencies. Enhanced spectral power below 8, near 12, and around 30 months are attributed to the semi-annual oscillation, the annual seasonality, and the QBO, respectively. The residual spectrum includes some significant activity between 20 and 70 hPa between 14 and 16 months. Recalling that we fit a periodic model to quasi-periodic data, the active spectrum could be attributed to a secondary mode of variability in the QBO that could be captured by extending the number of harmonics.

## 4.    FFM fitted to the ensemble data

Figure S3 shows the coefficient of determination ($R^2$) for each simulation in the E3SM ensemble, measuring the proportion of variance explained by the FFM. The explained variance ranges from 11% to 91% with a median of 71%. This variation suggests that while the FFM captures a significant portion of the dynamics in most simulations, some runs did not produce a dominant mode of variability, which contradicts the definition of the QBO. Simulations with higher $R^2$ values (e.g., 39, 40, and 15) likely exhibit dominant frequency components aligned with the FFM's design, resulting in more predictable behavior. In contrast, those with lower values (e.g., 48, 43, and 34) may contain substantial variance unaccounted for by the FFM, possibly due to short-term or high-frequency fluctuations and higher random error. The reference observational dataset also has a relatively high $R^2$, suggesting it behaves similarly to the more predictable ensemble members.

From Fig. S4 (left), two key characteristics emerge. First, there is a strong trade-off between QBO amplitude and period. Second, amplitude exhibits a block-like variation with two sub-blocks: 7–30 hPa (upper stratosphere) and 50–70 hPa (lower stratosphere). Smaller off-block amplitude correlations suggest an implicit secondary trade-off between amplitudes in the upper and lower stratosphere. The phylogram in Fig. S4 (right) was constructed using hierarchical clustering on the standardized QoIs, employing Ward's minimum variance method to identify compact clusters (Murtagh & Legendre, 2014). The results reveal three distinct clusters: period, lower stratospheric amplitudes, and upper stratospheric amplitudes.

## 5. Physics parameters and QoIs

A total of 61 simulations were conducted to explore the physics parameter space. E3SM failed to terminate successfully for 10 parameter combinations with relatively large CF values, suggesting an implicit upper bound on the convective fraction beyond which numerical instability occurs. Additionally, for 5 parameter sets near the sampling space boundaries, the model ran to completion but did not produce a QBO-like pattern. The remaining 46 simulations are retained for downstream analysis. The parameter values and the QoIs are listed in Table S1 and visualized in Figs. S5 and S6.

## 6.    Statistics for surrogate selection

Figure S7 illustrates the effect of the covariance function and the number of modes on surrogate accuracy in predicting the QoIs, as measured by the PPLD. Numerical results are reported in Table S2. The PPLD statistic represents the log probability of the observed values under the model's predictive distribution, given that the model is trained on all other data points, with higher values indicating more consistent predictability of a mode. While all tested configurations with at least two modes nearly achieve maximal accuracy for amplitude at 20 hPa and 70 hPa, the third mode significantly improves period prediction. The narrow spread in the curves suggests that the choice of correlation function has a secondary effect on predictive performance, with the Matérn 3/2 consistently yielding the best results. This provides moderate evidence that the QoIs respond relatively roughly to variations in the physics parameters. The optimal number of modes remains consistent across all correlation functions.

In Table S3, we further examine the predictive accuracy of spectral features using the Matérn 3/2 covariance function. The observed decreasing trend in leave-one-out accuracy statistics indicates that the surrogate model's predictive performance declines with increasing mode order. Specifically, negative values in these statistics suggest that beyond the third mode, the surrogate's predictions are less accurate than those generated by a simple standard normal model. This decline in accuracy underscores the diminishing returns of incorporating higher-order modes into the surrogate model, emphasizing the need to select an optimal truncation point where additional modes become detrimental.

## 7. Surrogate-based global sensitivity index

The total-effect sensitivity index for the $j$-th QoI on the $p$-th physics parameter is given by,

$$T_{pj} = 1 - \frac{Var_{-p}(E_p(Q_j|\mathbf{X}_{-p}))}{Var(Q_j)}, \tag{13}$$

where $\mathbf{X}_{-p}$ indicates all input variables except the $p$-th one. This statistic provides a comprehensive measure of the parameter's influence on the QoIs, with values closer to 0 and 1 indicating lower and higher influence, respectively. Because Sobol's decomposition is linear and additive, $T_{pj}$ can exceed 1 in cases of nonlinear systems with multiplicative interactions among the parameters.

The estimates are reported in Fig. S8. The oscillation is most responsive to CF, which accounts for 30-50% of the variance when considered alone and 65-75% when interactions are included, consistently across all QoIs. EF plays a secondary role, with nonzero main effects for the period and lower-stratospheric (higher pressure levels) amplitudes, and affects all QoIs approximately equally after accounting for interactions. HD has a tertiary role, marginally influencing QBO amplitude only in the middle-upper stratosphere (lower pressure levels).

The main-effect sensitivity index for the $j$-th QoI on the $p$-th physics parameter is defined as,

$$S_{pj} = \frac{Var_p(E_{-p}(Q_j|\mathbf{X}_p))}{Var(Q_j)}, \tag{14}$$

where $\mathbf{X}_p$ represents the $p$-th input variable. This index quantifies the direct contribution of the $p$-th parameter to the variance of the QoI, with values ranging from 0 to 1, where a value closer to 1 indicates a higher influence of the parameter on the QoI. The main-effect sensitivity index quantifies the proportion of the total variance in the $j$-th QoI attributable to the $p$-th parameter alone and is particularly useful for identifying the primary drivers of variability in the model output.

In Fig. S9, we observe that approximately 45% of the total variance in the QBO period and the lower-stratospheric (higher pressure)amplitude, and 35% of the upper-stratospheric (lower pressure) amplitude, is attributed to CF. This provides strong evidence that CF is largely the main driver across all the QoIs. On the contrary, the main effects of EF and HD are lesser and localized to some QoIs: EF has a moderate influence on the QBO period and the lower-stratospheric amplitudes, whereas HD has only a minor direct effect on the upper-stratospheric amplitudes. Notably, HD exerts no influence on the period. Recalling that their total effects influence all the QoIs with some degree of consistency, we deduce that EF and HD affect the upper- and lower-stratospheric amplitudes, respectively, only via interactions.

## 8. End-to-end calibration

Table S4 shows the estimated QoIs and objective functions for the sequence of models in the end-to-end calibration analysis.

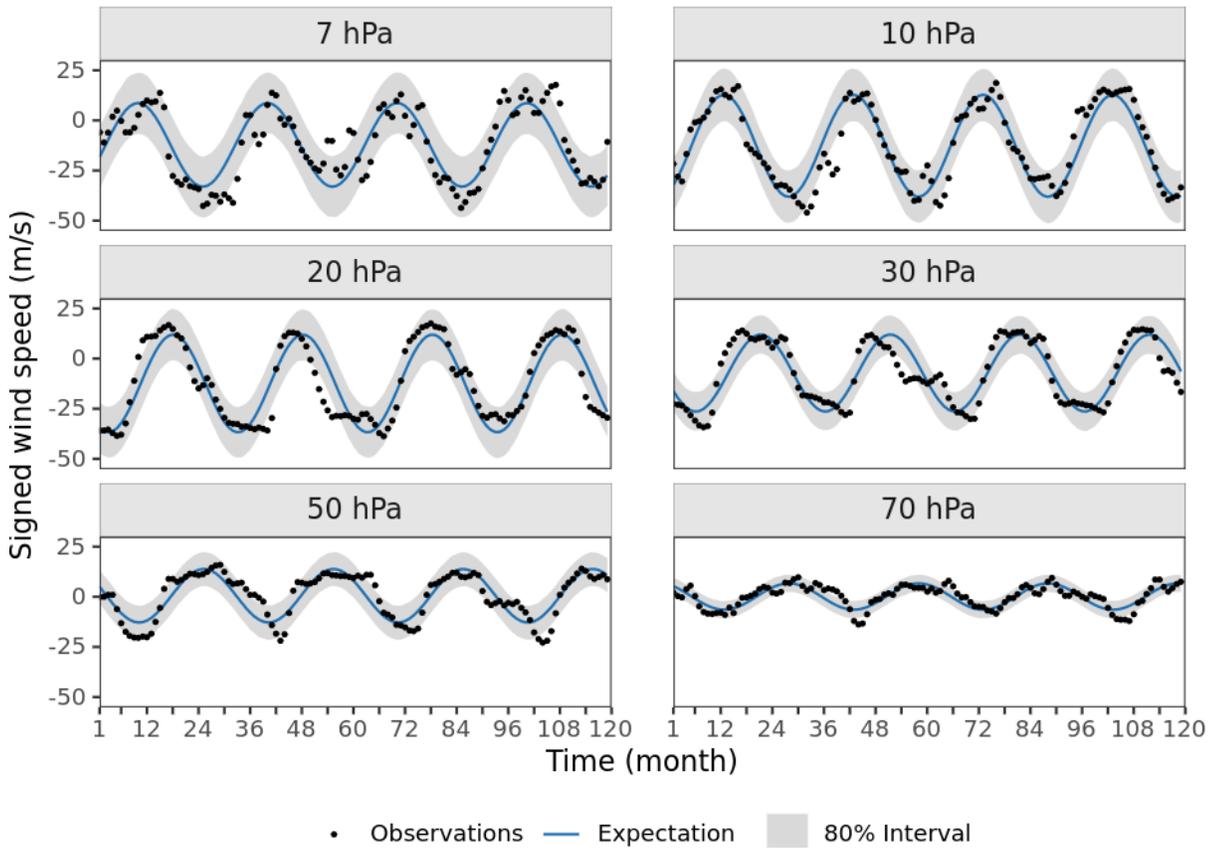

**Figure S1.**    Wind time series for the reference data set (ERA5). Each facet isolates a pressure level and corresponds to a horizontal slice in the time-pressure cross-sections. The expectation (blue line) shows the sine wave that best fits the raw data (black circles) and the interval (gray ribbon) shows the central 80% interval for a normal distribution calibrated with the sample residual variance.

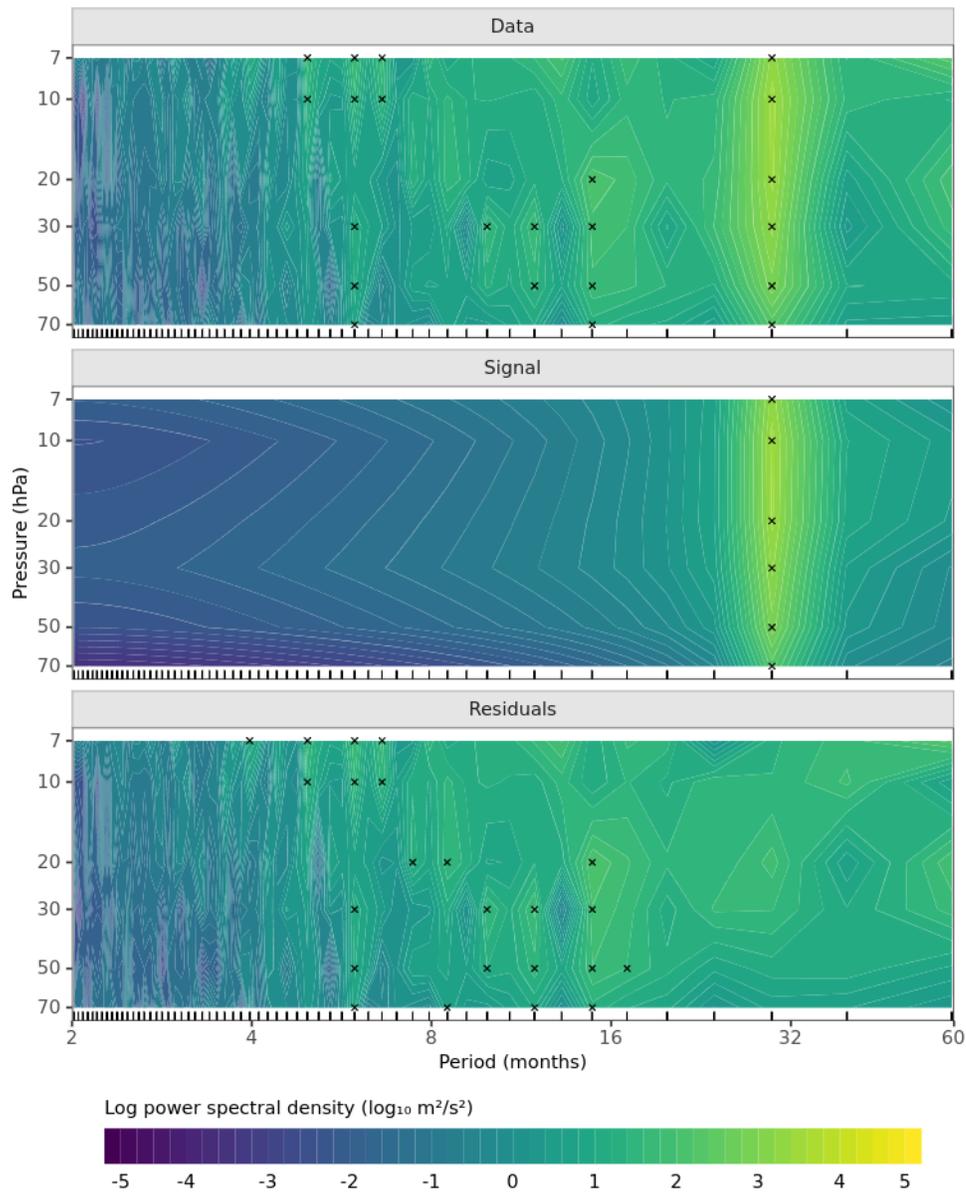

**Figure S2.** Sampling spectrum density for the reference data set. Top: sampling spectrum from raw data. Middle: sampling spectrum from the model mean. Bottom: difference between the sampling spectra from the data and the model mean (unexplained variance at secondary frequencies).

March 11, 2025, 12:27pm

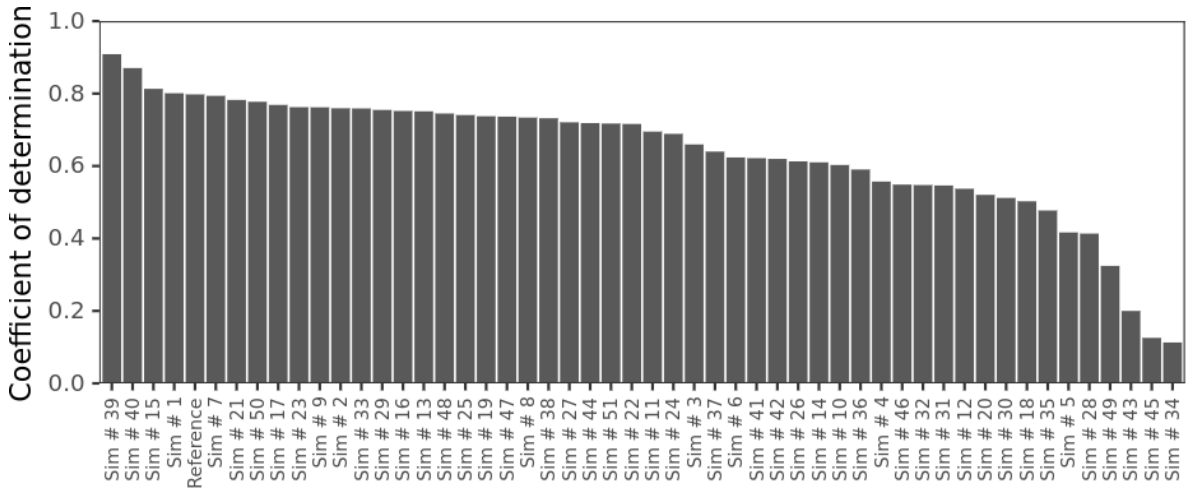

**Figure S3.**    Percentage of variance explained by the FFM for every E3SM ensemble run. The remaining variance is associated with short-term (high-frequency) dynamics or random error.

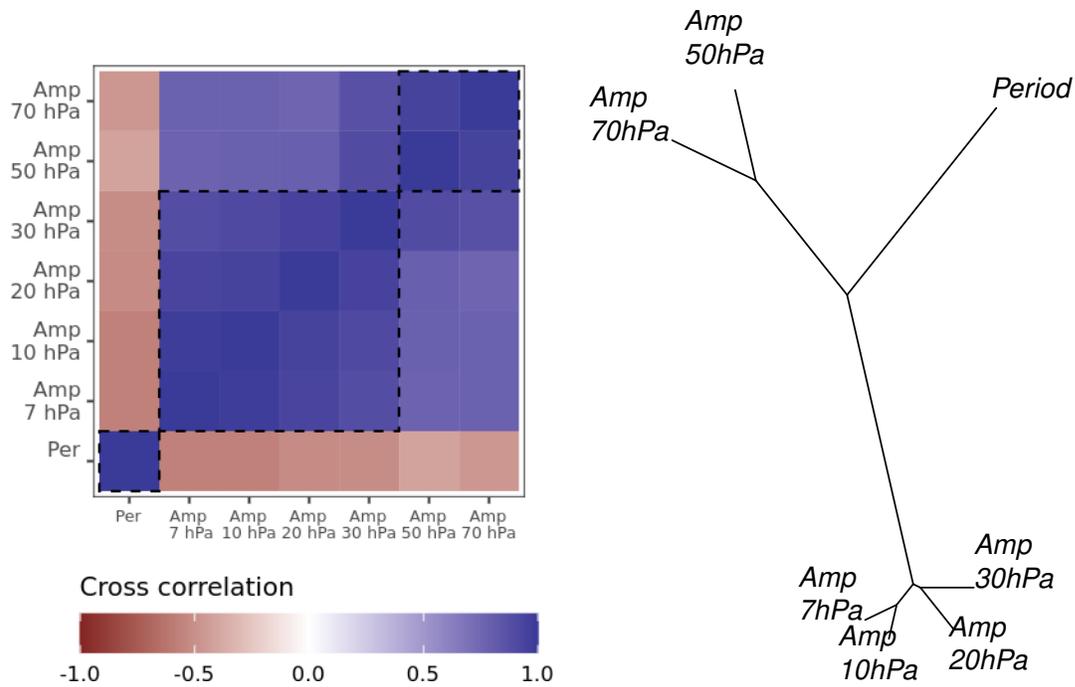

**Figure S4.** Similarity analysis among the ensemble QoIs. Left: sample correlation plot. Right: phylogenetic tree from hierarchical clustering on the euclidean distance between the standardized QoIs.

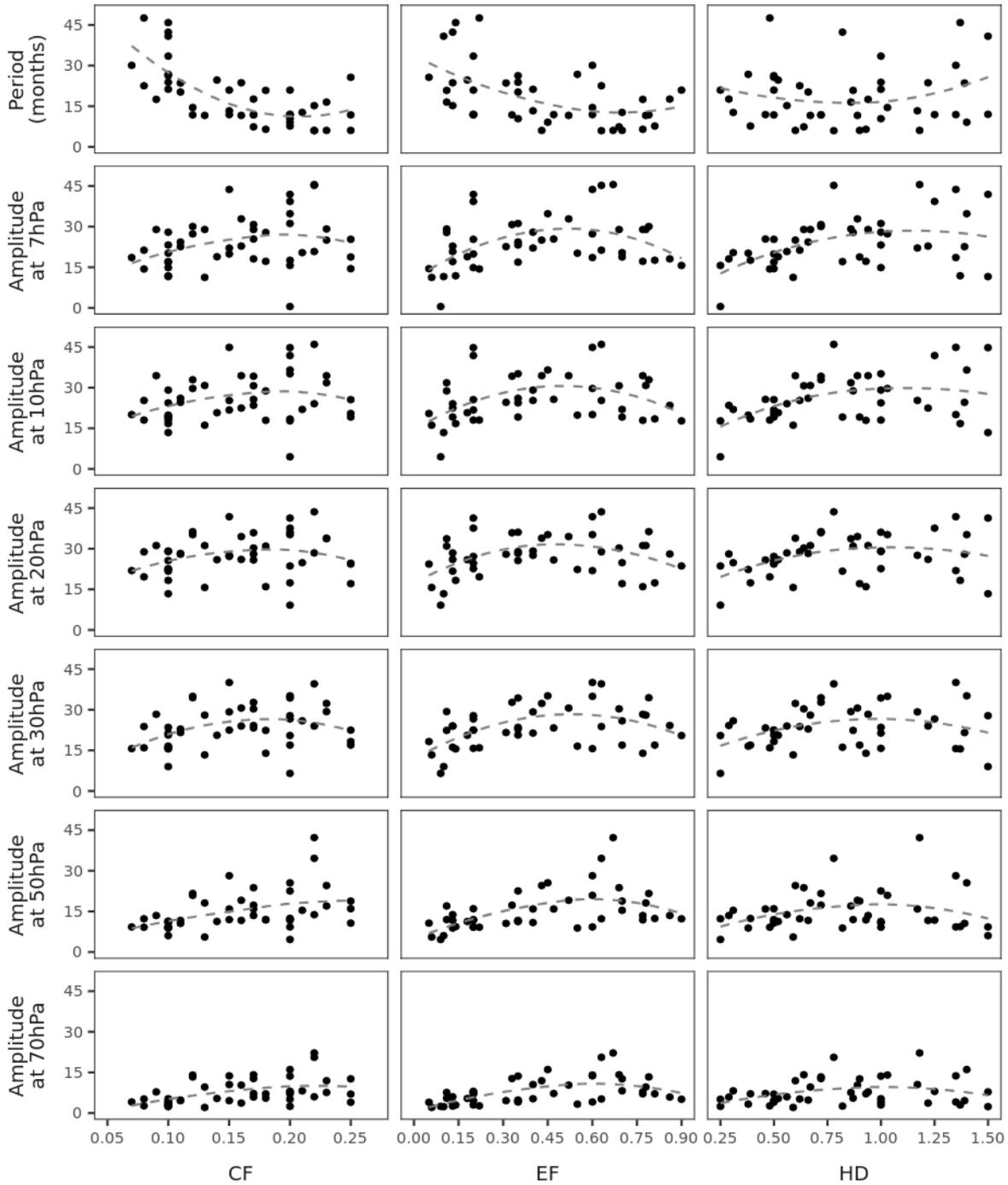

**Figure S5.**    Ensemble QoIs (circle) as a function of the physics parameters. The line, constructed using local smoothing (Cleveland & Loader, 1996), illustrates the overall trend in the E3SM ensemble.

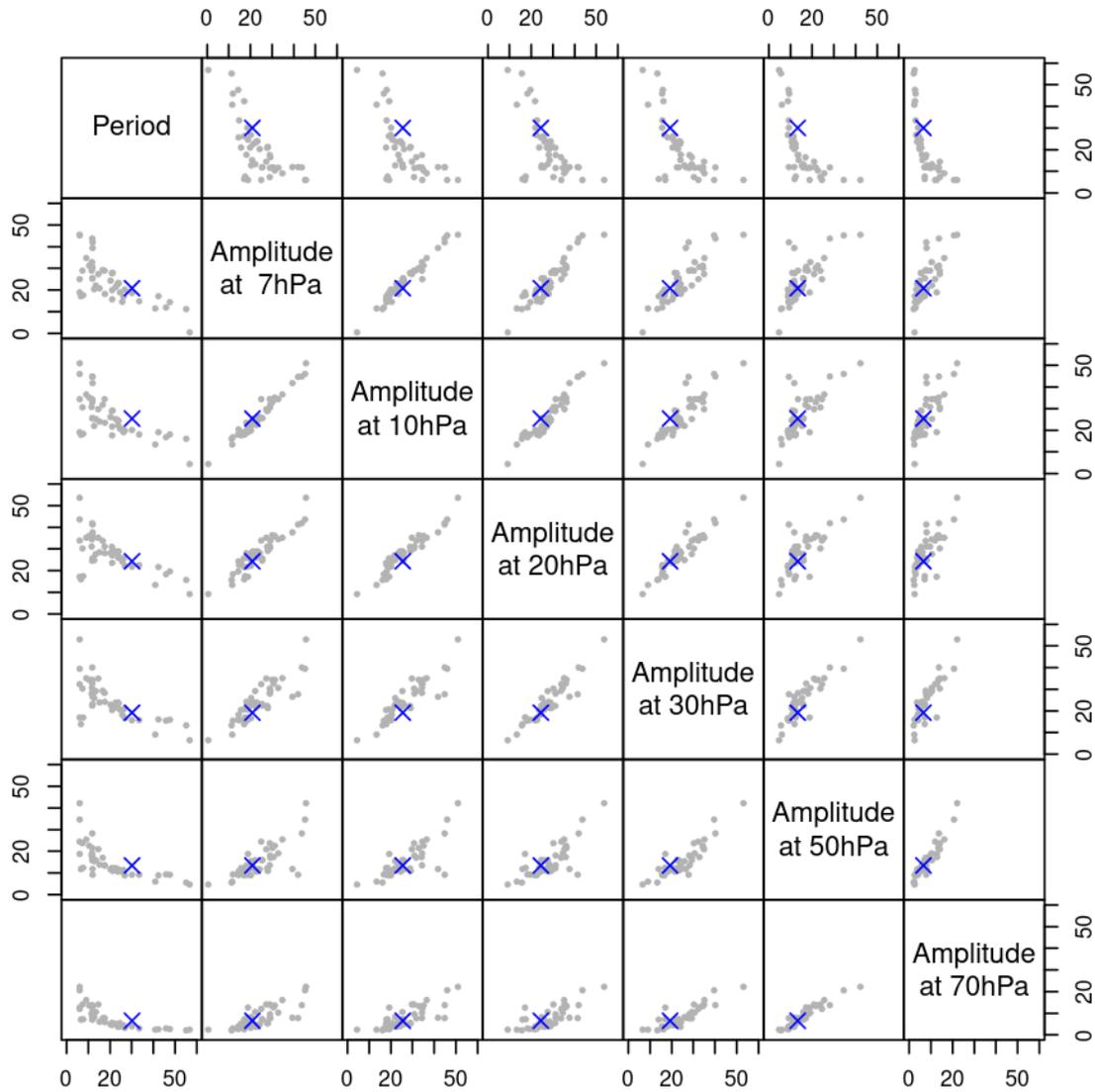

**Figure S6.**    Pair plots for the QoIs in the E3SM ensemble (circle) and the reference (cross).

| CF | EF | HD | Period | Amp 7hPa | Amp 10hPa | Amp 20hPa | Amp 30hPa | Amp 50hPa | Amp 70hPa |
|---|---|---|---|---|---|---|---|---|---|
|  |  |  | 30.1 (30.0, 30.3) | 20.9 (18.9, 22.8) | 25.5 (23.8, 27.2) | 24.4 (22.8, 26.0) | 19.2 (18.0, 20.4) | 13.4 (12.3, 14.4) | 6.6 (6.0, 7.1) |
| 0.07 | 0.60 | 1.35 | 30.1 (29.9, 30.3) | 18.6 (14.9, 22.2) | 20.1 (17.0, 23.2) | 21.9 (19.7, 24.0) | 15.6 (14.1, 17.2) | 9.2 (8.2, 10.3) | 4.1 (3.4, 4.7) |
| 0.08 | 0.22 | 0.48 | 47.6 (47.3, 47.9) | 14.4 (12.5, 16.3) | 18.1 (16.7, 19.4) | 19.6 (18.6, 20.6) | 15.9 (14.7, 17.2) | 9.1 (8.3, 10.0) | 2.6 (2.2, 3.1) |
| 0.08 | 0.63 | 0.62 | 22.6 (22.5, 22.6) | 21.3 (17.8, 24.8) | 25.3 (22.1, 28.4) | 28.9 (27.1, 30.7) | 23.8 (22.4, 25.2) | 12.3 (10.9, 13.6) | 5.2 (4.5, 5.9) |
| 0.09 | 0.77 | 0.94 | 17.5 (17.5, 17.6) | 28.9 (26.0, 31.9) | 34.4 (31.5, 37.3) | 31.2 (29.3, 33.0) | 28.3 (26.4, 30.2) | 13.5 (12.2, 14.7) | 7.8 (7.1, 8.6) |
| 0.10 | 0.10 | 1.50 | 40.8 (39.8, 41.8) | 11.5 (9.1, 13.9) | 13.4 (11.2, 15.6) | 13.3 (11.5, 15.2) | 9.0 (7.3, 10.7) | 6.0 (5.2, 6.8) | 2.4 (1.9, 2.8) |
| 0.10 | 0.13 | 0.82 | 42.3 (42.1, 42.6) | 17.1 (15.1, 19.1) | 19.2 (17.6, 20.7) | 21.7 (20.9, 22.5) | 16.1 (15.0, 17.3) | 8.9 (8.0, 9.7) | 2.6 (2.0, 3.1) |
| 0.10 | 0.14 | 1.37 | 45.8 (45.5, 46.2) | 11.8 (8.5, 15.2) | 16.8 (13.9, 19.7) | 18.3 (16.4, 20.1) | 15.5 (14.5, 16.5) | 9.4 (8.8, 10.0) | 3.0 (2.4, 3.5) |
| 0.10 | 0.20 | 1.00 | 33.5 (33.3, 33.7) | 14.8 (11.8, 17.8) | 18.0 (15.6, 20.4) | 22.6 (21.2, 24.1) | 15.7 (14.4, 17.1) | 9.2 (8.3, 10.0) | 3.0 (2.3, 3.6) |
| 0.10 | 0.35 | 0.50 | 26.2 (26.1, 26.3) | 16.9 (15.0, 18.8) | 19.1 (17.7, 20.5) | 25.6 (24.1, 27.1) | 20.7 (19.2, 22.1) | 11.4 (10.1, 12.7) | 4.4 (3.8, 5.0) |
| 0.10 | 0.35 | 1.00 | 23.9 (23.8, 24.0) | 23.2 (20.2, 26.3) | 24.4 (21.9, 26.8) | 28.9 (27.2, 30.7) | 23.4 (21.5, 25.4) | 11.3 (10.0, 12.6) | 4.1 (3.4, 4.7) |
| 0.10 | 0.40 | 1.00 | 21.3 (21.2, 21.3) | 28.0 (25.5, 30.5) | 29.1 (26.9, 31.4) | 29.2 (27.3, 31.0) | 21.3 (19.6, 23.0) | 10.9 (9.7, 12.0) | 5.3 (4.5, 6.1) |
| 0.10 | 0.55 | 0.38 | 26.8 (26.7, 26.9) | 20.2 (17.6, 22.8) | 19.9 (17.4, 22.4) | 22.3 (20.8, 23.8) | 16.6 (15.2, 18.0) | 8.9 (7.7, 10.0) | 3.2 (2.7, 3.8) |
| 0.11 | 0.31 | 1.39 | 23.6 (23.5, 23.7) | 22.7 (19.0, 26.3) | 24.6 (21.2, 27.9) | 27.9 (25.2, 30.6) | 21.5 (19.3, 23.8) | 10.6 (9.3, 11.9) | 4.6 (3.8, 5.5) |
| 0.11 | 0.35 | 0.66 | 20.2 (20.2, 20.3) | 24.3 (21.7, 26.9) | 26.1 (23.8, 28.3) | 28.3 (26.5, 30.0) | 22.9 (21.1, 24.7) | 11.7 (10.5, 12.8) | 4.8 (4.0, 5.6) |
| 0.12 | 0.60 | 1.03 | 14.5 (14.5, 14.5) | 27.3 (23.5, 31.2) | 29.7 (26.1, 33.3) | 35.2 (32.0, 38.4) | 34.9 (32.1, 37.8) | 20.9 (19.0, 22.8) | 14.1 (13.0, 15.1) |
| 0.12 | 0.79 | 0.72 | 11.8 (11.8, 11.8) | 30.0 (27.0, 33.1) | 32.9 (30.2, 35.7) | 36.3 (33.6, 39.1) | 34.5 (31.5, 37.4) | 21.6 (19.6, 23.6) | 13.3 (12.2, 14.5) |
| 0.13 | 0.06 | 0.59 | 55.1 (54.4, 55.9) | 11.2 (9.2, 13.3) | 16.1 (14.7, 17.5) | 15.7 (14.5, 16.9) | 13.3 (12.1, 14.5) | 5.6 (4.8, 6.4) | 2.1 (1.6, 2.6) |
| 0.13 | 0.78 | 0.67 | 11.6 (11.6, 11.6) | 29.0 (25.8, 32.1) | 30.8 (27.5, 34.1) | 31.2 (27.6, 34.7) | 28.0 (24.6, 31.5) | 18.1 (15.1, 21.0) | 9.6 (7.8, 11.4) |
| 0.14 | 0.18 | 0.52 | 24.6 (24.5, 24.7) | 18.9 (14.6, 23.2) | 20.7 (18.5, 23.0) | 25.9 (24.7, 27.2) | 20.6 (18.8, 22.5) | 11.3 (9.7, 12.8) | 5.4 (4.8, 6.0) |
| 0.15 | 0.20 | 0.50 | 21.0 (20.9, 21.0) | 19.9 (18.0, 21.7) | 21.8 (20.2, 23.3) | 27.1 (25.4, 28.7) | 22.5 (21.1, 23.8) | 11.9 (11.0, 12.8) | 4.6 (4.0, 5.1) |
| 0.15 | 0.40 | 1.17 | 13.4 (13.3, 13.4) | 22.1 (17.6, 26.6) | 25.3 (20.8, 29.7) | 27.6 (23.2, 31.9) | 29.2 (26.1, 32.3) | 15.9 (14.1, 17.7) | 10.5 (9.2, 11.8) |
| 0.15 | 0.60 | 1.35 | 11.9 (11.8, 11.9) | 43.8 (38.4, 49.1) | 44.9 (39.9, 49.8) | 41.8 (37.1, 46.5) | 40.1 (35.8, 44.4) | 28.2 (25.0, 31.3) | 13.7 (11.9, 15.6) |
| 0.16 | 0.13 | 1.22 | 23.7 (23.5, 23.8) | 22.9 (19.5, 26.3) | 22.5 (19.2, 25.8) | 26.1 (24.0, 28.2) | 24.0 (21.9, 26.0) | 11.7 (10.5, 12.9) | 3.7 (2.8, 4.5) |
| 0.16 | 0.52 | 0.89 | 11.6 (11.6, 11.7) | 32.9 (28.9, 36.9) | 34.4 (30.5, 38.3) | 34.5 (30.8, 38.2) | 30.6 (27.0, 34.3) | 19.1 (15.9, 22.2) | 10.3 (8.2, 12.4) |
| 0.17 | 0.33 | 0.72 | 11.8 (11.7, 11.8) | 30.7 (27.7, 33.7) | 34.3 (31.4, 37.1) | 35.8 (32.9, 38.7) | 32.7 (29.6, 35.8) | 17.3 (14.9, 19.7) | 12.7 (11.1, 14.4) |
| 0.17 | 0.47 | 0.46 | 11.9 (11.8, 11.9) | 25.4 (22.3, 28.5) | 25.6 (22.7, 28.6) | 25.8 (22.9, 28.8) | 23.2 (20.2, 26.3) | 15.8 (13.3, 18.4) | 7.3 (5.9, 8.6) |
| 0.17 | 0.69 | 0.64 | 7.3 (7.3, 7.3) | 28.9 (25.1, 32.7) | 30.7 (26.9, 34.4) | 30.3 (26.6, 33.9) | 30.3 (26.6, 33.9) | 23.8 (21.2, 26.4) | 14.1 (12.5, 15.7) |
| 0.17 | 0.86 | 0.29 | 17.7 (17.6, 17.7) | 18.1 (15.3, 20.9) | 23.4 (20.8, 26.1) | 28.1 (26.4, 29.8) | 24.2 (22.5, 25.8) | 13.4 (12.5, 14.4) | 5.9 (5.2, 6.5) |
| 0.18 | 0.11 | 0.87 | 20.9 (20.8, 20.9) | 27.8 (25.5, 30.1) | 28.8 (26.8, 30.8) | 31.0 (29.4, 32.6) | 22.3 (20.6, 24.0) | 12.0 (10.6, 13.4) | 5.5 (4.8, 6.2) |
| 0.18 | 0.77 | 0.93 | 6.5 (6.5, 6.5) | 17.2 (12.2, 22.2) | 17.9 (13.0, 22.9) | 16.0 (11.0, 21.0) | 14.0 (9.0, 18.9) | 11.8 (8.2, 15.5) | 7.1 (4.7, 9.5) |
| 0.20 | 0.09 | 0.25 | 56.8 (55.0, 58.6) | 0.5 (-1.9, 2.8) | 4.4 (2.1, 6.8) | 9.4 (5.9, 12.9) | 7.4 (4.0, 10.7) | 4.7 (4.0, 5.4) | 2.5 (2.0, 2.9) |
| 0.20 | 0.20 | 1.25 | 11.9 (11.9, 12.0) | 39.3 (36.3, 42.4) | 41.8 (38.6, 45.0) | 37.6 (34.1, 41.1) | 26.6 (23.1, 30.0) | 11.8 (9.6, 13.9) | 8.0 (6.4, 9.5) |
| 0.20 | 0.20 | 1.50 | 12.0 (12.0, 12.0) | 42.0 (39.9, 44.0) | 44.7 (42.7, 46.8) | 41.4 (38.6, 44.1) | 27.8 (24.9, 30.6) | 9.2 (7.2, 11.2) | 7.8 (6.3, 9.3) |
| 0.20 | 0.35 | 1.00 | 10.4 (10.4, 10.4) | 31.2 (27.0, 35.4) | 35.1 (31.6, 38.7) | 36.1 (32.8, 39.3) | 34.4 (31.3, 37.4) | 22.6 (20.0, 25.1) | 13.7 (12.0, 15.3) |
| 0.20 | 0.45 | 1.40 | 9.1 (9.0, 9.1) | 34.8 (29.7, 39.8) | 36.5 (31.5, 41.5) | 35.1 (30.1, 40.1) | 25.5 (22.4, 28.7) | 16.1 (14.1, 18.1) | 7.2 (5.9, 8.4) |
| 0.20 | 0.81 | 0.39 | 7.6 (7.6, 7.7) | 17.6 (14.0, 21.2) | 18.4 (14.7, 22.1) | 17.4 (13.8, 21.0) | 16.9 (13.6, 20.3) | 12.4 (10.1, 14.6) | 7.2 (5.9, 8.4) |
| 0.20 | 0.90 | 0.25 | 20.9 (20.9, 21.0) | 15.6 (13.9, 17.4) | 17.7 (16.1, 19.3) | 23.6 (21.7, 25.5) | 20.4 (18.9, 22.0) | 12.2 (11.2, 13.3) | 5.1 (4.6, 5.7) |
| 0.21 | 0.70 | 0.31 | 12.7 (12.7, 12.8) | 20.4 (18.0, 22.8) | 22.0 (19.6, 24.4) | 24.9 (22.5, 27.3) | 25.9 (23.4, 28.5) | 15.4 (14.0, 16.8) | 8.2 (7.4, 8.9) |
| 0.22 | 0.13 | 0.56 | 15.2 (15.2, 15.2) | 20.8 (17.8, 23.8) | 24.1 (21.0, 27.2) | 28.5 (25.8, 31.2) | 24.0 (22.1, 25.8) | 13.8 (12.7, 14.9) | 6.0 (5.4, 6.6) |
| 0.22 | 0.63 | 0.78 | 6.0 (6.0, 6.0) | 45.2 (41.3, 49.1) | 46.0 (42.3, 49.6) | 43.6 (39.7, 47.5) | 39.6 (35.3, 43.8) | 34.6 (30.9, 38.3) | 20.6 (18.3, 22.9) |
| 0.22 | 0.67 | 1.18 | 6.0 (6.0, 6.0) | 45.6 (40.6, 50.5) | 51.0 (46.3, 55.7) | 53.6 (50.9, 56.3) | 53.1 (49.4, 56.7) | 42.3 (38.6, 45.9) | 22.2 (19.8, 24.6) |
| 0.23 | 0.11 | 0.86 | 16.6 (16.5, 16.6) | 29.2 (26.8, 31.5) | 31.8 (29.6, 34.0) | 33.7 (31.8, 35.7) | 29.3 (26.9, 31.7) | 17.0 (15.0, 19.0) | 7.6 (6.7, 8.4) |
| 0.23 | 0.43 | 0.60 | 6.0 (6.0, 6.0) | 25.0 (21.0, 29.0) | 34.4 (30.7, 38.2) | 33.9 (32.1, 35.6) | 32.3 (30.2, 34.4) | 24.5 (22.3, 26.7) | 10.5 (10.0, 13.4) |
| 0.25 | 0.05 | 0.50 | 25.7 (25.6, 25.8) | 14.5 (12.7, 16.3) | 20.4 (18.8, 22.1) | 24.3 (23.0, 25.7) | 18.3 (16.9, 19.6) | 10.7 (9.8, 11.6) | 4.0 (3.4, 4.5) |
| 0.25 | 0.20 | 0.50 | 11.8 (11.8, 11.9) | 25.3 (22.0, 28.7) | 25.6 (22.3, 28.9) | 26.4 (23.1, 28.0) | 22.4 (19.3, 25.6) | 16.0 (13.7, 18.3) | 7.0 (5.7, 8.4) |
| 0.25 | 0.70 | 0.90 | 6.0 (6.0, 6.1) | 18.8 (13.8, 23.9) | 19.2 (14.2, 24.1) | 17.1 (12.2, 22.0) | 17.0 (12.3, 21.6) | 18.8 (15.0, 22.7) | 12.7 (10.2, 15.1) |

**Table S1.** Estimated QoIs for the reference data (first gray row) and the E3SM ensemble simulations (remaining rows). Each cell shows the MLE point estimate and the corresponding 80% interval. The second gray row highlights the ensemble member with lowest mean square distance to the reference value.

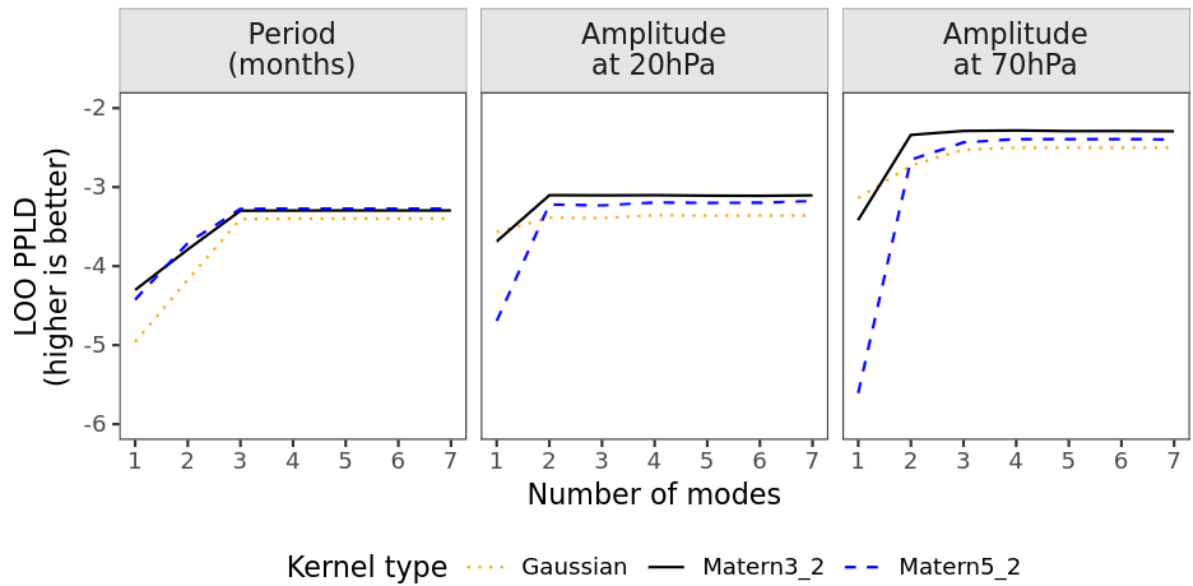

**Figure S7.** Posterior predictive log density evaluated at the test sample from a leave-one-out cross-validation study. Higher values are associated with higher accuracy in the probabilistic predictions of the QoIs.

| Correlation function | Modes | Period | Amplitude | | | | | | |
|---|---|---|---|---|---|---|---|---|---|
| | | | Mean | 7 | 10 | 20 | 30 | 50 | 70 |
| Squared exponential | 1 | -5.0 | -3.5 | -3.6 | -3.5 | -3.6 | -3.4 | -3.8 | -3.1 |
| | 2 | -4.2 | -3.3 | -3.5 | -3.4 | -3.4 | -3.4 | -3.3 | -2.7 |
| | 3 | <u>-3.4</u> | -3.2 | -3.4 | -3.3 | -3.4 | -3.3 | -2.9 | -2.5 |
| | 4 | <u>-3.4</u> | <u>-3.1</u> | -3.4 | -3.3 | -3.4 | -3.3 | -2.9 | -2.5 |
| | 5 | -3.4 | -3.1 | -3.4 | -3.3 | -3.4 | -3.3 | -2.8 | -2.5 |
| | 6 | -3.4 | -3.1 | -3.4 | -3.3 | -3.4 | -3.3 | -2.8 | -2.5 |
| | 7 | -3.4 | -3.1 | -3.4 | -3.3 | -3.4 | -3.3 | -2.8 | -2.5 |
| Matérn 5/2 | 1 | -4.4 | -4.6 | -4.8 | -3.8 | -4.7 | -3.8 | -5.0 | -5.6 |
| | 2 | -3.7 | -3.3 | -3.5 | -3.2 | -3.2 | -3.8 | -3.3 | -2.7 |
| | 3 | <u>-3.3</u> | -3.1 | -3.3 | -3.2 | -3.2 | -3.4 | -2.9 | -2.4 |
| | 4 | <u>-3.3</u> | <u>-3.0</u> | -3.3 | -3.2 | -3.2 | -3.3 | -2.9 | -2.4 |
| | 5 | -3.3 | <u>-3.0</u> | -3.3 | -3.2 | -3.2 | -3.3 | -2.8 | -2.4 |
| | 6 | -3.3 | -3.0 | -3.3 | -3.2 | -3.2 | -3.3 | -2.8 | -2.4 |
| | 7 | -3.3 | -3.0 | -3.3 | -3.2 | -3.2 | -3.3 | -2.8 | -2.4 |
| Matérn 3/2 | 1 | -4.3 | -3.6 | -3.8 | -3.4 | -3.7 | -3.3 | -3.8 | -3.4 |
| | 2 | -3.8 | -3.0 | -3.2 | -3.1 | -3.1 | -3.3 | -2.9 | -2.3 |
| | 3 | -3.3 | -2.9 | -3.1 | -3.1 | -3.1 | -3.2 | -2.8 | -2.3 |
| | 4 | -3.3 | -2.9 | -3.1 | -3.1 | -3.1 | -3.2 | -2.8 | -2.3 |
| | 5 | -3.3 | -2.9 | -3.1 | -3.1 | -3.1 | -3.2 | -2.7 | -2.3 |
| | 6 | -3.3 | -2.9 | -3.1 | -3.1 | -3.1 | -3.2 | -2.7 | -2.3 |
| | 7 | -3.3 | -2.9 | -3.1 | -3.1 | -3.1 | -3.2 | -2.7 | -2.3 |

**Table S2.** Posterior predictive log density (PPLD) evaluated at the test sample from a leave-one-out cross-validation study. Higher values are associated with higher accuracy in the probabilistic predictions of the QoIs. We highlight the smallest number of modes needed to achieve maximum PPLD at a single-digit significance in the period and the mean across amplitudes for each covariance function (underlined) and the selected correlation function and number of modes for the surrogate (gray row). The numeric values in the 3rd, 7th, and 10th columns correspond to the values plotted in Figure S7.

| Mode | Cum. energy | Variance | | | Leave one out | | Length scales | | |
|------|------|--------|-------|-----|-----|-------|------|------|------|
| | | Signal | Error | STN | LLR | $R^2$ | EF | CF | HD |
| 1 | 0.84 | 1.64 | $2.4 \times 10^{-8}$ | $8 \times 10^3$ | 68 | 0.62 | 1.40 | 0.06 | 1.35 |
| 2 | 0.08 | 1.01 | $3.7 \times 10^{-7}$ | $2 \times 10^3$ | 38 | 0.45 | 0.29 | 0.06 | 1.32 |
| 3 | 0.06 | 0.93 | $2.3 \times 10^{-1}$ | 2 | 13 | 0.34 | 0.17 | 0.14 | 1.65 |
| 4 | 0.02 | 0.29 | $7.1 \times 10^{-1}$ | $6 \times 10^{-1}$ | -105 | 0.08 | 0.83 | 0.47 | 0.20 |
| 5 | 0.01 | 0.69 | $3.7 \times 10^{-1}$ | 1 | -6 | -0.02 | 0.18 | 0.04 | 2.59 |
| 6 | < 0.01 | 0.19 | $7.9 \times 10^{-1}$ | $5 \times 10^{-1}$ | -208 | 0.06 | 2.22 | 0.05 | 1.46 |
| 7 | < 0.01 | 1.02 | $7.6 \times 10^{-7}$ | $1 \times 10^3$ | 1 | -0.02 | 0.17 | 0.04 | 0.20 |

**Table S3.** Statistics from GPs with Matérn 3/2 correlation function fitted separately to the spectral features. The cumulative energy (second column) and the learned variances (first group) evidence a decrease in relevance and surrogate's goodness of fit after the third-order mode. The leave-one-out validation statistics (second group) evidence a sharp decrease in the surrogate's accuracy to predict the forth-order mode and higher, at which point a simple standard normal model has better predictive performance. The length scales (third group) were estimated by ML.

March 11, 2025, 12:27pm

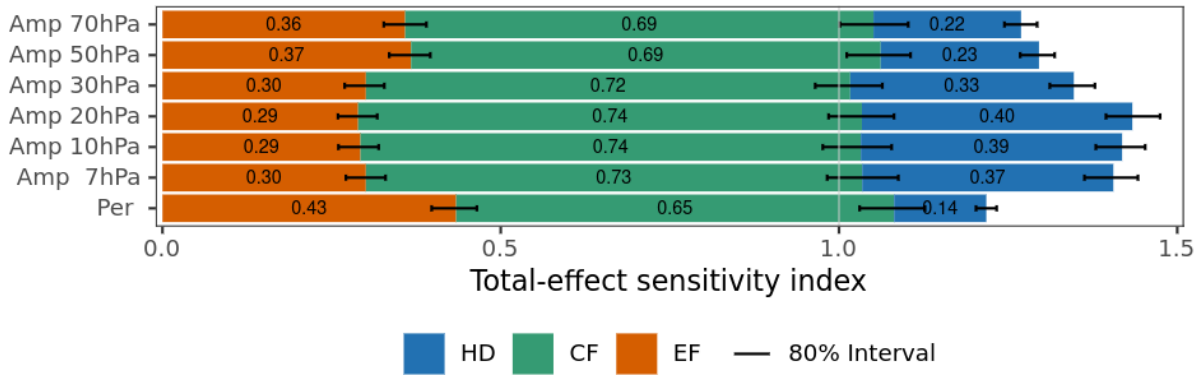

**Figure S8.** Surrogate-based total-effect index quantifying the contributions of the physics parameters to the QoIs. Color bars and text reflect the point estimate, whereas error bars represent the 80% intervals estimated via random sampling. CF dominates QBO period and amplitude across pressure levels, whereas EF has a secondary contribution, particularly for the period. HD's contribution is limited to the middle-upper stratosphere (lower pressure levels). The sums over all parameters being greater than 1 hint, to some extent, that the QBO originates from a nonlinear system with multiplicative interactions among the parameters.

March 11, 2025, 12:27pm

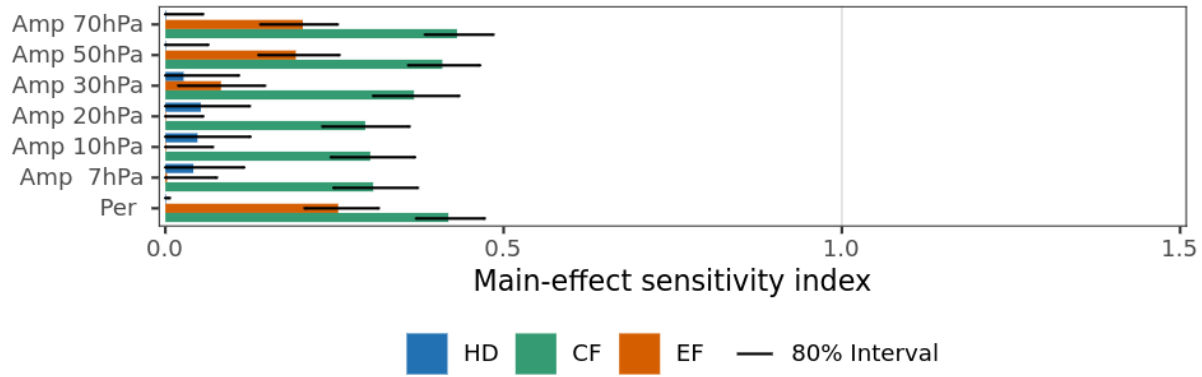

**Figure S9.** Surrogate-based main-effect index quantifying the direct influence of the physics parameters on each QoI. Color bars and text reflect the point estimate, whereas error bars represent the 80% intervals estimated via random sampling. CF is the main driver across all the QoIs. EF has a moderate influence on the QBO period and the lower-stratospheric amplitudes, whereas HD has only a minor direct effect on the upper-stratospheric amplitudes.

| Model configuration / observational data | Period | Amplitude | | | | | | Loss | |
|---|---|---|---|---|---|---|---|---|---|
| | | 7 | 10 | 20 | 30 | 50 | 70 | Period | Amplitude |
| (a) v2 L72 prior defaults | 30.9 | 6.4 | 6.4 | 7.3 | 6.8 | 4.1 | 1.7 | 0.81 | 11.55 |
| (b) v2 L80 prior defaults | 37.1 | 23.0 | 26.2 | 23.0 | 17.0 | 9.3 | 4.2 | 6.99 | 2.55 |
| (c) v3 L80 prior defaults | 23.9 | 23.2 | 24.4 | 28.9 | 23.4 | 11.3 | 4.1 | 6.24 | 2.88 |
| (d) v3 L80 new defaults | 26.2 | 16.9 | 19.1 | 25.6 | 20.7 | 11.4 | 4.4 | 3.90 | 2.94 |
| (e) v3 L80 optimal values | 26.8 | 16.1 | 21.5 | 25.3 | 20.1 | 9.9 | 4.0 | 3.34 | 3.04 |
| (f) Reference ERA5 | 30.1 | 20.9 | 25.5 | 24.4 | 19.2 | 13.4 | 6.6 | - | - |

**Table S4.**     Estimated QoIs and objective functions for the E3SM simulations in the improvement sequence from E3SMv2 to the surrogate-based optimal configuration applied to E3SMv3. Additionally, we include the QoIs estimated from the reference data (gray).

# References

Cleveland, W. S., & Loader, C. (1996). Smoothing by Local Regression: Principles and Methods. In W. Härdle & M. G. Schimek (Eds.), *Statistical Theory and Computational Aspects of Smoothing* (pp. 10–49). Heidelberg: Physica-Verlag HD. doi: 10.1007/978-3-642-48425-4_2

Cressie, N. A. C. (1993). *Statistics for spatial data.* Wiley. Retrieved from `http://dx.doi.org/10.1002/9781119115151`  doi: 10.1002/9781119115151

Duvenaud, D. (2014). *Automatic model construction with gaussian processes* (Doctoral dissertation, University of Cambridge). doi: 10.17863/CAM.14087

Kennedy, M. C., & O'Hagan, A. (2001, September). Bayesian calibration of computer models.

*Journal of the Royal Statistical Society Series B: Statistical Methodology*, *63*(3), 425–464. Retrieved from `http://dx.doi.org/10.1111/1467-9868.00294`  doi: 10.1111/1467-9868 .00294

MacKay, D. J.  (1994, 12).  Bayesian nonlinear modeling for the prediction competition. ASHRACE.  Retrieved from `https://www.osti.gov/biblio/33309`

Murtagh, F., & Legendre, P.  (2014, October).  Ward's hierarchical agglomerative clustering method: Which algorithms implement ward's criterion?  *Journal of Classification*, *31*(3), 274–295. Retrieved from `http://dx.doi.org/10.1007/s00357-014-9161-z` doi: 10.1007/ s00357-014-9161-z

Neal, R. M. (1996). *Bayesian learning for neural networks*. Springer New York. Retrieved from `http://dx.doi.org/10.1007/978-1-4612-0745-0`  doi: 10.1007/978-1-4612-0745-0

Piironen, J., & Vehtari, A. (2016, September). Projection predictive model selection for gaussian processes. In *2016 ieee 26th international workshop on machine learning for signal processing (mlsp)* (p. 1–6). IEEE. Retrieved from `http://dx.doi.org/10.1109/MLSP.2016.7738829` doi: 10.1109/mlsp.2016.7738829

Santner, T. J., Williams, B. J., & Notz, W. I. (2018).  *The design and analysis of computer experiments*.  Springer New York.  Retrieved from `http://dx.doi.org/10.1007/978-1 -4939-8847-1`  doi: 10.1007/978-1-4939-8847-1